\documentclass[fleqn,usenatbib]{mnras}

\usepackage{newtxtext,newtxmath}

\usepackage[T1]{fontenc}
\usepackage{ae,aecompl}

\usepackage{graphicx}	
\usepackage{amsmath}	
\usepackage{amssymb}	

\usepackage{soul} 
\usepackage{xargs} 
\usepackage[pdftex,dvipsnames]{xcolor}  
\usepackage{amsmath}

\title[Binary pop synth models for core-collapse GRBs]{Binary population synthesis models for core-collapse gamma-ray burst progenitors}

\author[A. A. Chrimes et al.]{A. A. Chrimes,$^{1}$\thanks{E-mail: A.Chrimes@warwick.ac.uk}
E. R. Stanway,$^{1}$
and J. J. Eldridge$^{2}$
\\
$^{1}$Department of Physics,  University of Warwick, Gibbet Hill Road, Coventry, CV4 7AL, UK\\
$^{2}$Department of Physics, University of Auckland, Private Bag 92019, Auckland, New Zealand\\
}

\date{Accepted XXX. Received YYY; in original form ZZZ}

\pubyear{2019}

\begin{document}
\label{firstpage}
\pagerange{\pageref{firstpage}--\pageref{lastpage}}
\maketitle

\begin{abstract}
Long-duration gamma-ray bursts (GRBs) are understood to be the final fate for a subset of massive, stripped envelope, rapidly rotating stars. Beyond this, our knowledge of the progenitor systems is limited. Using the BPASS (Binary Population and Spectral Synthesis) stellar evolution models, we investigate the possibility that some massive stars in binaries can maintain the angular momentum required for jet production, while still loosing their outer envelope through winds or binary interactions. We find that a total hydrogen mass of $M_\mathrm{H}<5\times10^{-4}$\,M$_\odot$ and a helium ejecta mass fraction of $F_\mathrm{He}<0.20$ provide the best thresholds for the supernova II/Ibc and Ib/Ic divisions respectively. Tidal interactions in binaries are accounted for by applying a tidal algorithm to post-process the stellar evolution models output by BPASS. We show that the observed volumetric gamma-ray burst rate evolution can be recreated using two distinct pathways and plausible distributions for burst parameters. In the first pathway, stars are spun up by mass accretion into a quasi-homogeneous state. In the second, tides maintain rotation where otherwise the star would spin down. Both lead to type Ic supernova progenitors, and a metallicity distribution consistent with the GRB host galaxy population. The inferred core angular momentum threshold for jet production is consistent with theoretical requirements for collapsars, given the assumptions made in our model. We can therefore reproduce several aspects of core collapse supernova/GRB observation and theory simultaneously. We discuss the predicted observable properties of GRB progenitors and their surviving companions.
\end{abstract}




\section{Introduction}
Long-duration gamma-ray bursts (GRBs) mark the evolutionary end points for a subset of massive, rapidly rotating, stripped envelope stars. Their preference for low metallicity, star-forming host galaxies \citep{2006Natur.441..463F,2009ApJ...691..182S,2010AJ....140.1557L,2016ApJ...817....8P,2019arXiv190100872M,2019A&A...623A..26P}, association with energetic hydrogen and helium-poor type Ic-BL (broad line) supernovae \citep{1998Natur.395..670G,1998Natur.395..672I,1999ApJ...516..788W,2001ApJ...550..991N,2003ApJ...591L..17S,2003Natur.423..847H,2004ASSL..302..277N,2004PThPS.155..299N,2008AJ....135.1136M,2016ApJ...832..108M,2017AdAst2017E...5C}, and theoretical considerations, all point to a core-collapse scenario. This could be a `collapsar', \citep{1993ApJ...405..273W,1999A&AS..138..499W}, in which a collapsing star powers a rapidly spinning black hole central engine, which launches relativistic jets. If the jets produced can escape the envelope \citep[a requirement which might explain the association with stripped envelope supernovae, e.g.][]{2016ApJ...832..108M}, and we are aligned along the jet axis, the event is observable as a relativistically-beamed GRB \citep{1999ApJ...524..262M,2001ApJ...550..410M,2006ARA&A..44..507W,2013RSPTA.37120275H}. Alternatively, the central engine could be a rapidly spinning neutron star with a strong magnetic field \citep[a magnetar,][]{2014MNRAS.443...67M}, which powers relativistic jets as it spins down. 

Although the requirement for a stripped envelope progenitor with a large reserve of angular momentum at core-collapse is understood, the exact nature of the progenitors and the evolutionary pathways which lead to long-duration (core-collapse) GRBs are not well constrained \citep[see][and others for reviews]{2016SSRv..202...33L,2019EPJA...55..132F}. Models invoking single stars struggle to maintain enough angular momentum over the stellar lifetime, while simultaneously losing the outer envelope through winds \citep{2001A&A...369..574V,2005A&A...443..581H,2005A&A...442..587V,2006ARA&A..44..507W,2011A&A...536L..10V,2016ApJ...832..108M}.
 
One promising pathway is through chemically homogeneous evolution. In the later stages of binary evolution, the primary can expand and fill its Roche lobe, triggering accretion onto the secondary star. This can spin up the secondary, and if sufficient angular momentum is transferred, rotational mixing can occur within the star \citep{2007A&A...465L..29C,2009A&A...497..243D,2011MNRAS.414.3501E,2013ApJ...764..166D}. If the star becomes fully mixed, it evolves chemically homogeneously, becoming smaller, hotter and hydrogen deficient, evolving blue-wards on the Hertzsprung-Russell diagram. If this process occurs at low metallicity, the secondary can retain the angular momentum gained from mass transfer, and end its life as a rapidly spinning type Ic progenitor - a good candidate for producing a GRB. \citet{2019MNRAS.482..870E} calculated the volumetric event rate of GRBs as a function of redshift, assuming that they arise solely from a quasi-homogeneous evolution (QHE) pathway \citep{2006A&A...460..199Y,2007A&A...465L..29C}, using the BPASS stellar evolution and population synthesis code \citep{2017PASA...34...58E}. They found that although QHE stars could account for the observed GRB rate, there are significant uncertainties arising from luminosity function and beaming corrections. Additionally, BPASS only implements QHE pathways at metallicities $Z<0.004$. GRBs are not exclusively found at such low metallicity, with observed examples arising in environments exceeding Solar abundance \citep[e.g.][]{2009AIPC.1133..269G,2010ApJ...712L..26L,2015A&A...579A.126S,2018A&A...616A.169M}. QHE therefore cannot be the sole contributing pathway. 

In this paper, we explore the possibility that massive stars in binaries, without undergoing QHE, can also produce GRBs given the right evolutionary history, even at moderate to high metallicities \citep[see e.g.][]{2004ApJ...607L..17P,2004MNRAS.348.1215I,2005A&A...435..247P,2008A&A...484..831D,2015ApJ...802..103T}. This route is particularly promising as binary interactions are likely responsible for the rapid rotation of massive stars \citep{2013ApJ...764..166D}. To investigate, we introduce a tidal interaction post-processing algorithm to the BPASS binary stellar evolution models in order to account for angular momentum transfer between orbital and stellar rotation. We investigate subsets of the resultant high mass, rapidly spinning, stripped envelope population, and explore the consequences of matching synthetic GRB rates arising from our selection criteria to observations. The paper is structured as follows: In Section \ref{sec:CCSNe}, we outline the categorisation of models into different classes of core-collapse progenitor, producing updated BPASS predictions for the properties of supernova progenitors. Section \ref{sec:tBPASS} describes our implementation of tides within BPASS, and the effect of their inclusion. A comparison to observations is presented in Section \ref{sec:eventrates}, with a Bayesian analysis used to infer the most likely parameter values in our two-progenitor pathway model. Section \ref{sec:results} details the results, including predictions for progenitor properties such as their position on the Hertzsprung-Russell diagram, and the interior angular momentum at core-collapse. A discussion of the physical interpretation of these models follows in Section \ref{sec:discussion}, with conclusions presented in Section \ref{sec:conc}. Where required, a flat $\Lambda$CDM cosmology with $h$ = 0.7, ${\Omega}_\mathrm{M}$ = 0.3 and ${\Omega}_{\Lambda}$ = 0.7 is used.

\section{Classification of Core-Collapse Progenitor Models}\label{sec:CCSNe}
GRBs, when close enough (typically within ${\sim}$1\,Gpc) and followed up with prompt and deep optical-NIR observations, typically show evidence for an associated broad-lined type-Ic supernova (Ic-BL SN). The preference for stripped-envelope supernovae of this type is likely a jet escape requirement, as an envelope of helium or hydrogen around the star may prove an insurmountable barrier to the ejection of fast-moving material and hence stifle any incipient jet. It appears as though not every Ic-BL SN has an accompanying GRB, although in some cases the jet might be choked, as may have been the case in SN\,2008D \citep[][]{2008Sci...321.1185M,2013RSPTA.37120273P,2016ApJ...832..108M,2017MNRAS.472..616S,2018ApJ...860...38B}. Nevertheless, the minimum requirement for a stripped envelope is an observational constraint on the progenitor stars of GRBs. Using the binary stellar evolution models of BPASS \citep[v2.2.1,][]{2018MNRAS.479...75S}, the first step in identifying GRB pathways is to find models which likely produce type Ic-SNe. 

At each step in the BPASS models, stellar properties such as mass and chemical composition are calculated. When core carbon burning ends, the stellar model is assumed to progress rapidly through the final stages of core burning (if massive) before ending its life \citep{2002RvMP...74.1015W}. If a supernova occurs, the stellar model is then replaced by a neutron star or black hole based on the final stellar mass and the amount of mass that could be removed by energy injection from a supernova of $10^{51}$\,erg. Because supernovae are observationally classified according to their spectral properties, which reflect the chemical composition of the ejecta and circumstellar medium \citep{doi:10.1146/annurev.astro.35.1.309}, we can predict which supernova type each model will produce based on its chemical properties immediately before core-collapse. 

Our first task is to identify stars which produce supernovae. The fate of stars with final masses $<2M_{\odot}$ is uncertain. They might produce weak, faint and fast core-collapse supernovae \citep{2016MNRAS.461.2155M}, or, depending on the central carbon abundance, type Ia supernovae. Alternatively, they could form white dwarfs without a supernova. Supernovae from these stars would in any case be rare events \citep{2017PASA...34...58E}.

We find models that likely experience a supernova by two methods. First, if at the end point of our model the CO core mass is greater than $3M_{\odot}$, a supernova occurs. Second, at lower masses, due to the complexity of second dredge-up and carbon burning in super-AGB stars, we require that the final total mass exceeds  $2M_{\odot}$, the CO core mass is greater than $1.38M_{\odot}$ and core carbon burning has occurred (when the ONe core is great than $>0.1M_{\odot}$). Then, if the remnant mass is in the range ${\sim}$1.4-3$M_{\odot}$, a neutron star results. Above this remnant mass, black holes are created, and supernovae in this regime are uncertain, with `islands of explodability' in mass \citep[e.g., ][]{2016ApJ...821...38S,2019ApJ...871...64A,2019arXiv190100215W,2019arXiv190500474S}. It is believed that most stars above a zero-age main sequence mass of $20M_{\odot}$ will create a black hole remnant and have their explosion engulfed within the Schwarzschild radius, thus producing no visible display. This picture is increasingly backed up by a lack of high mass stars seen in pre-explosion imaging \citep{2009ARA&A..47...63S,2015PASA...32...16S}, possible tension between star formation and SN rates \citep{2011ApJ...738..154H}, and the direct observation of a quietly disappearing massive star \citep{2017MNRAS.468.4968A}. As such, we initially disregard any model which produces a remnant mass greater than $3M_{\odot}$ (i.e. a black hole) as a vanishing event \citep[see][for a description of how remnant and ejecta masses are calculated]{2004MNRAS.353...87E}.

\begin{figure}
\centering
\includegraphics[width=0.95\columnwidth]{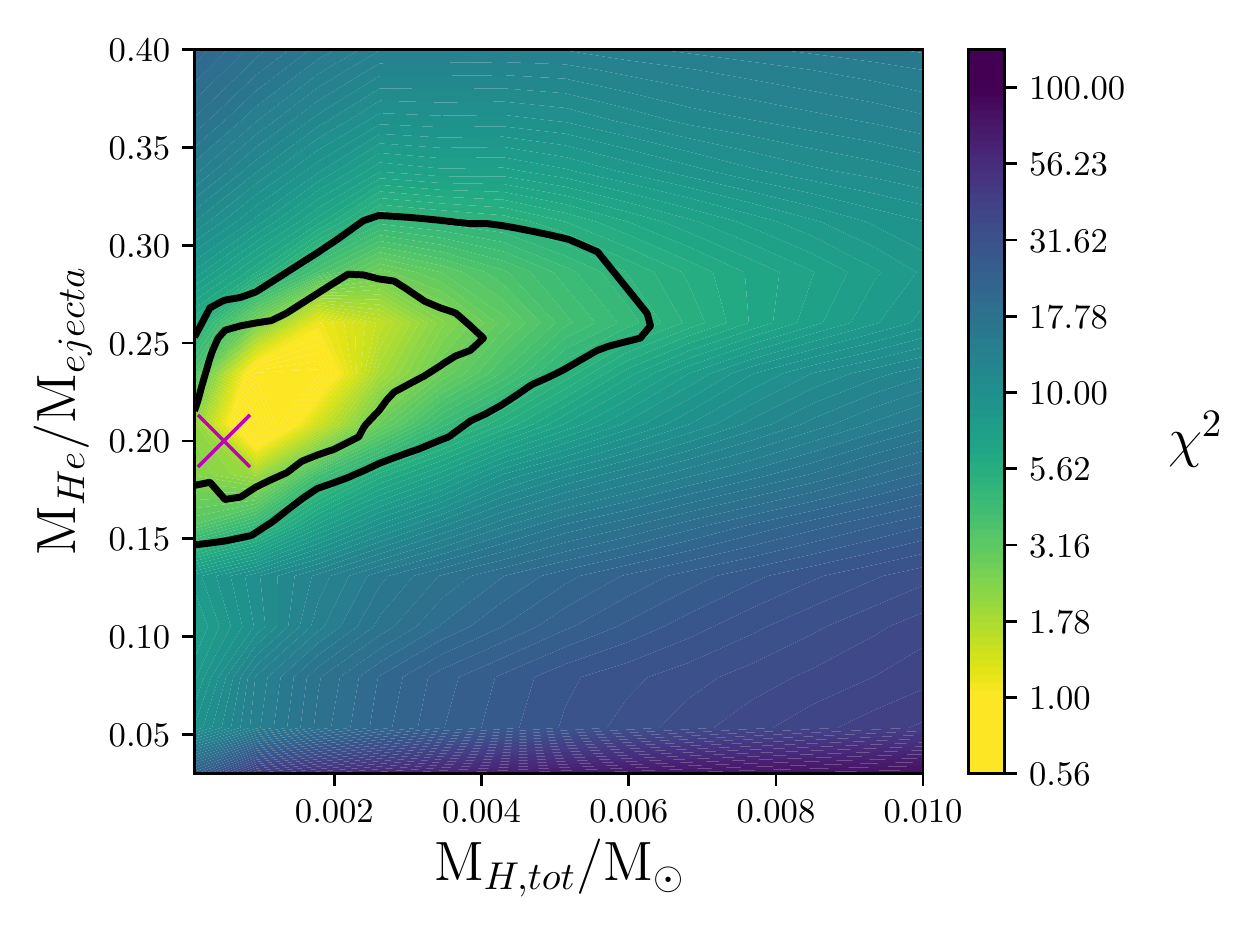}
\caption{The result of matching type II, Ib and Ic SN relative fractions to observations, by minimising ${\chi}^{2}$ over $M_\mathrm{H,tot}$ and $F_\mathrm{He,ejecta}$ parameter space. The best fit classification thresholds, assuming that black hole producing events do not go supernova, are marked by a cross. Solid contours represent the 68 and 90 per cent confidence regions for one degree of freedom.}
\label{fig:bestvalues}
\end{figure}

Otherwise, if the stellar structure at the end of the model still satisfies the carbon-oxygen, oxygen-neon and total mass conditions listed above, we assume that a visible supernova will be produced. We then split these models into three categories, types Ib, Ic, and II, based on their chemical composition \citep{2011MNRAS.414.3501E,2013MNRAS.436..774E,2017PASA...34...58E}. 

The type I/type II divide is defined by the total mass of hydrogen in the star; observationally this is probed by the presence of hydrogen lines in the SN spectrum, or lack thereof. Theoretically, a small residue of hydrogen may still be present without leaving observational evidence, as long as it is below some threshold. We vary this threshold, labelling models with a total hydrogen mass less than $M_\mathrm{H,tot}$ as type I and those with more as type II. The type Is are further divided into Ibs (which have visible helium lines) and Ics (with no helium or hydrogen in their spectra). This is decided by the fraction of helium in the ejecta, $F_\mathrm{He,ejecta}$, and this threshold is also varied. The total hydrogen mass is used, rather than the fraction of hydrogen in the ejecta, because (unless QHE is underway) the majority of the hydrogen should be on the surface of the star and therefore mixed into the ejecta upon supernova.

\begin{figure}
\centering
\includegraphics[width=0.95\columnwidth]{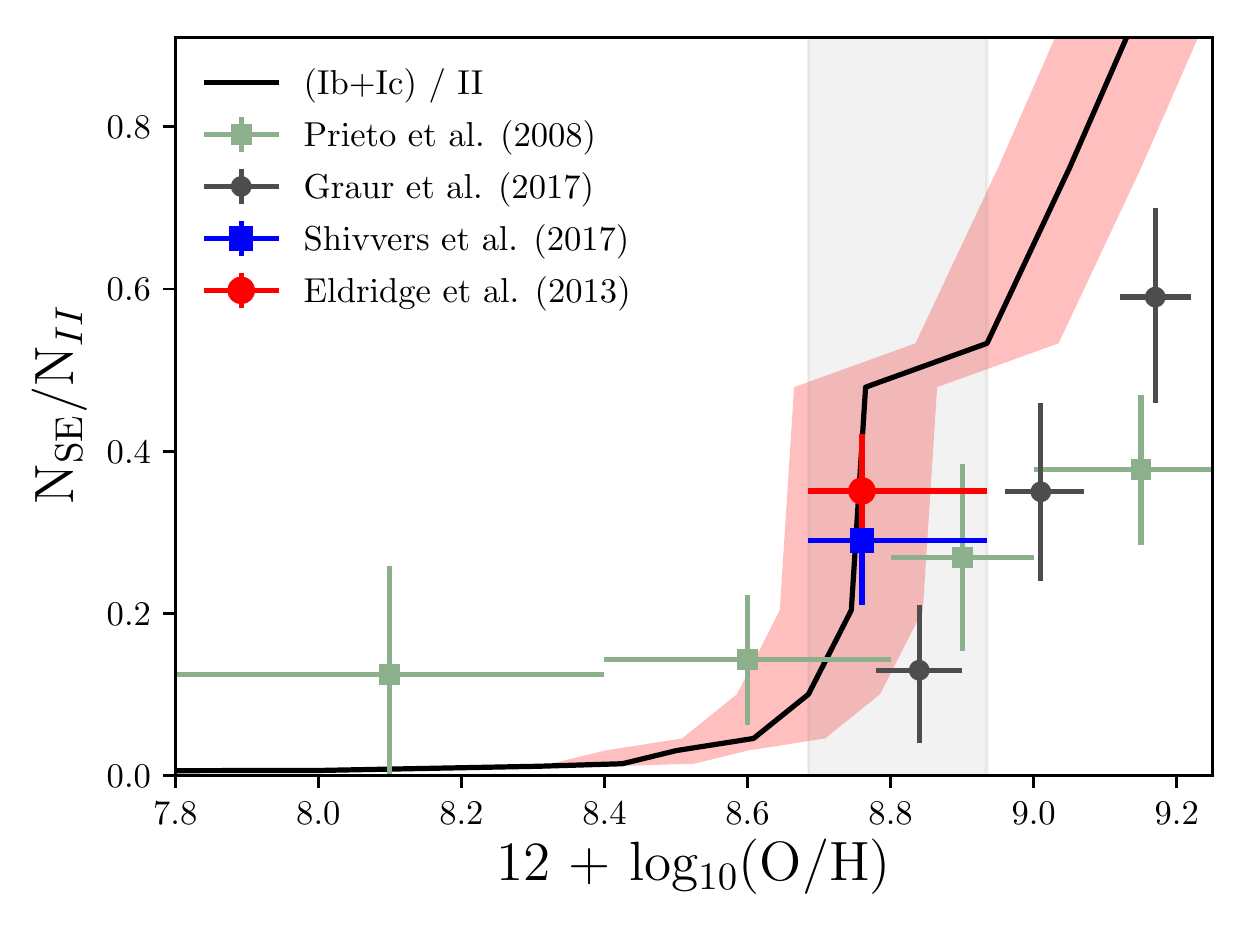}
\caption{The ratio of stripped envelope (SE) to hydrogen-rich (type II) supernovae as a function of metallicity, using our best fit $M_\mathrm{H,tot}$ and $F_\mathrm{He,ejecta}$ thresholds (black line), assuming that supernovae only occur if a black hole is not produced. Metallicites in the range 0.008 to 0.02 by mass fraction are indicated by a grey shaded band. The red shaded region is a 0.1 dex uncertainty, intended to give an indication of the discrepancies between different metallicity scales \citep{2017PASA...34...58E}. Where a sample metallicity is not explicitly defined, we assume that it samples the local metal mass fraction spread of 0.008 to 0.02.}
\label{fig:relfrac}
\end{figure}

\begin{figure*}
\centering
\includegraphics[width=0.95\textwidth]{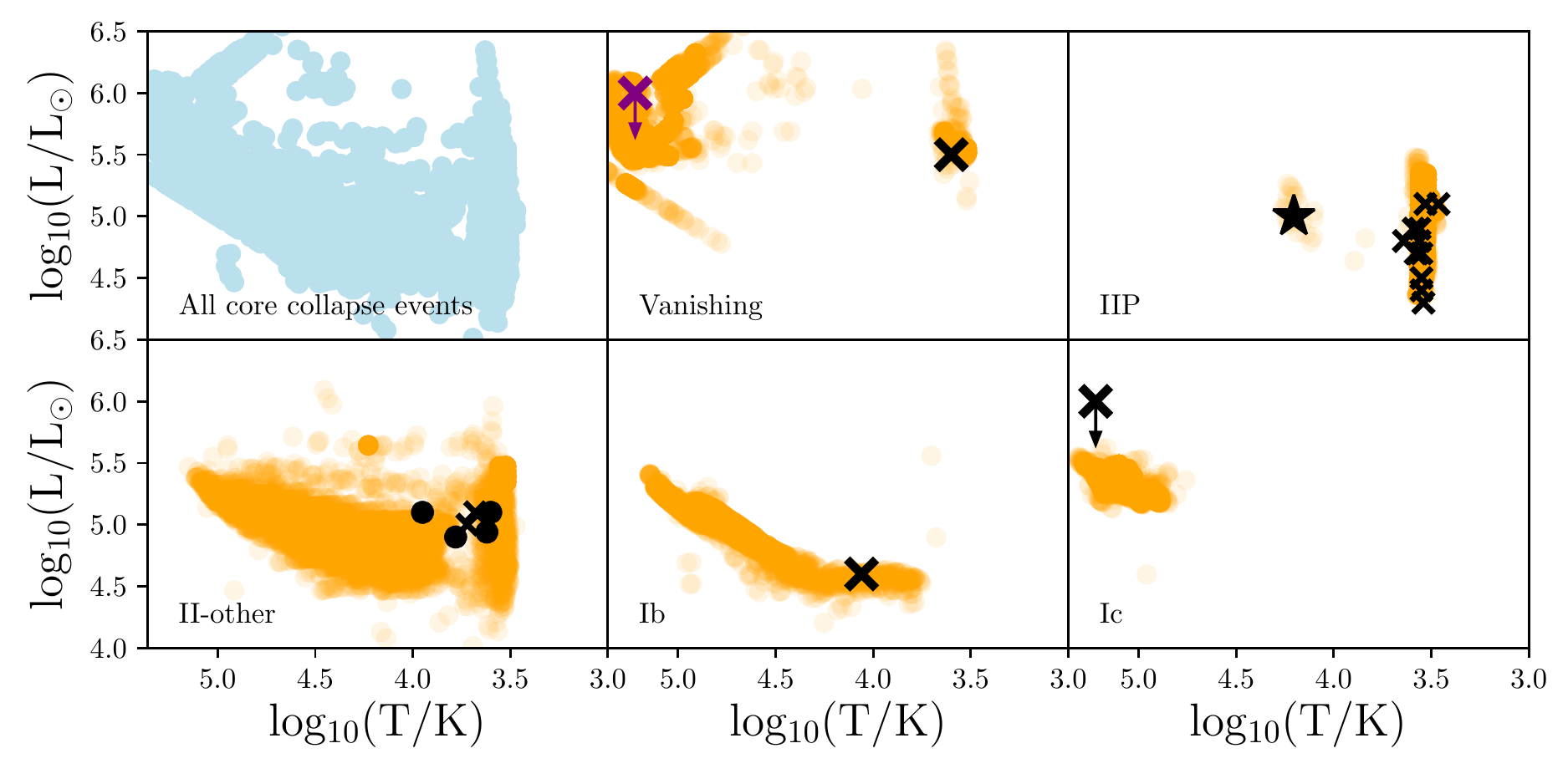}
\caption{Following the categorisation of BPASS model such that the observed supernova relative rates are reproduced, we place these models on the HR diagram (in orange) according to core-collapse type. The shading represents the number density of models. Observed progenitors (in black) include the vanishing star N6946-BH1 \citep[marked by a cross,][]{2017MNRAS.468.4968A}, for which we have adjusted the luminosity to reflect an updated distance estimate \citep{2019MNRAS.485L..58E}, SN\,1987A \citep[marked by a star in the IIP section,][]{1987ApJ...321L..41W}, several IIP SNe from \citet{2015PASA...32...16S}, type IIb and IIL SNe \citep[both classified as II-other, and marked by dots and crosses respectively,][]{2015PASA...32...16S}, SN\,1993J \citep[also a IIb event,][]{1994AJ....107..662A}, the Ib progenitor of SN\,iPTF13bvn \citep{2016MNRAS.461L.117E} and finally the candidate Ic progenitor of SN\,2017ein \citep{2018ApJ...860...90V}. The marked luminosity for 2017ein is an upper estimate, assuming that it was a single star \citep{2018MNRAS.480.2072K}. This progenitor system is also indicated on the 'vanishing' panel (in purple) for reference. }
\label{fig:HR}
\end{figure*}

The models in each category are then assigned a weighting, which corresponds to the estimated occurrence frequency of each model in a star formation episode, assuming $10^{6}M_{\odot}$ of metal enriched gas were allowed to collapse and form stars with 100 per cent efficiency. The weightings are informed by observations of stars in terms of binary fractions, mass distributions, binary orbital periods and mass ratios  \citep{2017ApJS..230...15M} and correspond to the weightings from v2.2 of the BPASS spectral synthesis models \citep{2018MNRAS.479...75S}. We use the standard BPASS initial mass function (IMF). This is parameterised as a power-law in $\frac{dN}{dm}$, and has a slope of -1 in the range 0.1-0.5M$_{\odot}$, and -2.35 from 0.5 to a maximum of 300M$_{\odot}$. 

For a $10^{6}M_{\odot}$ stellar population, at each metallicity $Z$ (12 over the range $10^{-4}$ to 0.04 by metal mass fraction) we calculate the relative contribution to the total number of supernovae from each type (II, Ib and Ic), summed over 10\,Gyr from $t=0$. The average of the fractions over the $Z=$0.008, 0.010, 0.014 and 0.020 metallicities are then compared to observed fractions from volume limited supernova studies. The metallicity range used for comparison is broadly representative of the range seen in the local Universe. For the type I/type II ratio we refer to \citet{2013MNRAS.436..774E}, which is a volume-limited sample and therefore representative of the intrinsic ratio. \citet{2017PASP..129e4201S} is used for the Ib/Ic ratio, due to their careful reclassification of a number of stripped envelope SNe from the Lick Observatory Supernova Search (LOSS).

In using the LOSS data, we have classified type IIb events as Hydrogen-rich (so that they are in the type II category), and we have grouped regular, peculiar and broad-line Ic SNe together. We also subdivide the type IIs into IIP (plateau) and II-other classes, but do not use this for constraining the pre-core-collapse properties as we would have too many degrees of freedom. Instead, we use the selection criteria of \citet{2017PASA...34...58E}, where type IIPs are defined to have $M_\mathrm{H,tot} > 1.5M_{\odot}$ and a hydrogen to helium ratio greater than 1.05. 

We use the difference between the BPASS weighted fractions and the observed fractions to calculate ${\chi}^{2}$, which we then minimise by varying $M_\mathrm{H,tot}$ and $F_\mathrm{He,ejecta}$ across parameter space. We show the results of this process in Figure \ref{fig:bestvalues}. We find that the best fitting values are $M_\mathrm{H,tot}$ = $5\times10^{-4}M_{\odot}$ and $F_\mathrm{He,ejecta} = 0.20$, with SN types II, Ib and Ic making up 75, 14 and 10 per cent of the local Universe supernova rate respectively. This compares to the \citet{2013MNRAS.436..774E} and \citet{2017PASP..129e4201S} combined percentages of $74\pm7.6$, $14.5\pm2.6$ and $11.5\pm2.0$. The differences correspond to a ${\chi}^{2}$, equivalent to a reduced ${\chi}^{2}$ in this case, of 0.27. The large uncertainties on the observed fractions, taken together with the large number of free parameters, allows for a range of values above these best fitting thresholds. Hydrogen masses up to 0.0022, and $F_\mathrm{He,ejecta}$ values in the range 0.17-0.26, are permitted within the 68 per cent confidence interval. Nonetheless, there is a clear preference for lower values, with the upper end of the explored parameter space firmly excluded.

\citet{2011MNRAS.414.2985D} predicted the supernova spectra from Wolf-Rayet stars, and found that hydrogen lines are initially visible if more than $10^{-3}M_{\odot}$ of hydrogen is present, likely making such events type IIb. This is consistent with our $M_\mathrm{H,tot}$ = $5\times10^{-4}M_{\odot}$ type I/II threshold. \citet{2012MNRAS.424.2139D} then showed that helium ejecta fractions as low as 0.2-0.3 can produce visible helium lines in the spectrum, again consistent with our results. 

\begin{table}
\centering 
\caption{The BPASS metallicity scaling used in this work, adapted from \citet{2018MNRAS.477..904X}. For a discussion of metallicity scales and uncertainties, we refer the reader to \citet{2017PASA...34...58E}.}
\begin{tabular}{c c c c} 
\hline 
Metal fraction & Previous & New & \\
\newline
by mass & 12$+$log(O$/$H) & 12$+$log(O$/$H) &  12$+$log(Fe$/$H) \\ 
\hline 
0.0001 & 6.60 & 7.00 & 5.23 \\
0.001 & 7.61 & 8.00 & 6.24 \\
0.002 & 7.91 & 8.31 & 6.54 \\
0.003 & 8.09 & 8.43 & 6.72 \\
0.004 & 8.21 & 8.51 & 6.84 \\
0.006 & 8.39 & 8.61 & 7.02 \\
0.008 & 8.52 & 8.69 & 7.15 \\
0.010 & 8.62 & 8.75 & 7.25 \\
0.014 & 8.77 & 8.77 & 7.40 \\
0.020 & 8.93 & 8.93 & 7.57 \\
0.030 & 9.13 & 9.05 & 7.76 \\
0.040 & 9.27 & 9.13 & 7.90 \\
\end{tabular}
\label{tab:metals}
\end{table}

Using the selection criteria determined from comparison to the local Universe supernova rates, we extend these cuts over the entire BPASS metallicity range. The predicted stripped envelope to hydrogen-rich ratio as a function of metallicity is shown in Figure \ref{fig:relfrac}, with the same ratio as reported by various observational studies also shown. The BPASS metallicity scaling used in this paper includes modifications to the previous scale used \citep{2018MNRAS.477..904X}, to improve agreement with the data at low metallicities \citep[see Figure 44 of][ and Table \ref{tab:metals}]{2017PASA...34...58E}. We note that the stellar evolution is driven primarily by iron abundance and bulk metallicity mass fraction, rather than the oxygen abundance. The BPASS relative rates generally agree with observation, including with those studies listed in \citep{2009ARA&A..47...63S}, which are not shown on the Figure for clarity. However, at the highest metallicities, there are discrepancies. Our models appear to over-predict the rate of stripped envelope events relative to Hydrogen-rich events at significantly super-Solar metallicities. We caution that in this regime the remnant mass and ejecta composition are strongly sensitive to the assumed mass loss rates on the asymptotic giant branch and immediately preceding the explosion. Small adjustments in the assumed wind prescription may have a significant effect on the supernova type ratio and further work is required to explore this further.

Using our classification criteria, we place every pre-core-collapse model (in the metallicity range 0.08-0.020) on a Hertzsprung-Russell diagram in Figure \ref{fig:HR}. We also show a range of known or candidate progenitors, identified from pre-explosion imaging. In most cases, there is good agreement. The only candidate which could be in tension is that of the type Ic progenitor, SN 2017ein \citep{2018ApJ...860...90V,2018MNRAS.480.2072K}, although it is currently unclear how much of the pre-explosion stellar light was from the progenitor, rather than a possible binary companion, and so this data represents an upper limit. We note that increasing the remnant threshold for supernova production (i.e. assuming lower supernova mass-ejection energies or less efficient black-hole stifling of any core-collapse event) would improve the agreement here. The physical properties of observed Wolf-Rayet stars are similarly more typical of the most luminous Ic progenitor models in our classification, with luminosities around log$_{10}(L/L_{\odot}){\sim}5.6$ \citep{2019arXiv190806238N}, although we have no constraint on whether these stars will go supernova or not. 

A possible resolution to the type Ic tension is that 2017ein may have produced a supernova {\it and} a black hole remnant. Modelling of pre core-collapse stellar structure shows that some black-hole producing events can successfully explode if the core has a low compactness parameter \citep{2016ApJ...821...38S}, although this is just one of several relevant quantities \citep{2019arXiv190901371M}. Furthermore, work in this area has typically focused on red supergiants, since these are the class of stars which we would have expected to see in pre-explosion imaging - stripped envelope supernovae are typically too distant to expect to see the progenitor. Therefore, we have few constraints on the explodability of such stars, and the generic vanishing threshold we have applied might not be applicable to stripped envelope stars. Another possibility is that supernovae can occur in stars that would otherwise implode thanks to energy injection from a rotationally powered central engine. This could be from disc winds \citep[e.g.][]{1999ApJ...524..262M} or a jet \citep[for example, type Ic-BL SNe might harbour choked jets,][]{2008Sci...321.1185M,2013RSPTA.37120273P,2016ApJ...832..108M,2019Natur.565..324I}. We refer the reader to \citet{2019ApJ...871L..25P} for a review of these ideas.

Regardless of the mechanism, it is likely that some black hole producing events cause visible supernovae, and that these have been classified as vanishing in our analysis so far. One example are pair-instability supernovae \citep[PISNe,][]{2002ApJ...567..532H}, which we classify as occurring in stars with helium cores between 64 and 133 $M_{\odot}$ at core-collapse. However, these particular events likely disrupt the star leaving no remnant. They are also exceptionally rare and do not significantly impact the analysis thus far. 

It is likely that GRBs do indeed require such a black hole central engine, but these events are also luminous and associated with supernovae. In order to decide which BPASS models end their lives as GRBs, we therefore need to consider the rotation of stars which are Ic-like in their chemical composition, but which also produce black holes at core-collapse. We note that these additional black hole producing SNe are rare, given that they have higher mass progenitors. They make a small difference to the relative SN rates in the local Universe, therefore having a minimal impact on our supernova categorisation - this is demonstrated later, see for example Figure \ref{fig:grb_rate}.

\section{Introducing Complex Tidal Interactions to BPASS}\label{sec:tBPASS}
Surface composition and structure of the pre-explosion progenitor allow us to identify the subset of BPASS models which produce Ic SNe and black holes. In order to produce a GRB, however, rapid rotation is also required. BPASS accounts for the first order effects of mass loss and mass transfer on the spins of stars in binaries, but we must also consider the transfer of angular momentum between the orbit and stellar spins due to tidal interactions. BPASS currently only invokes tidal interactions when Roche lobe overflow occurs \citep{2017PASA...34...58E}, however tides may have a substantial impact at all stages of stellar evolution. In this section we describe our first-order application of tides to the BPASS output models.

\subsection{Radiative damping of the dynamical tide}
There are two mechanisms for tidal dissipation within stars. These are convective damping of the equilibrium tide, and radiative damping of the dynamical tide \citep{1975A&A....41..329Z,1977A&A....57..383Z,1981A&A....99..126H,1989ApJ...342.1079G,2002MNRAS.329..897H}. Tidal forces are strongly dependent on radius, whatever the dissipation mechanism, and so the nature of the stellar envelope (rather than the core) determines which mechanism is dominant. Given that we are interested in stars with masses greater than $2M_{\odot}$, which have radiative envelopes for the majority of their lives, we employ radiative damping of the dynamical tide and assume that convective damping is negligible.

In the radiative regime, the mass and orbit of a companion star introduces a time-varying external gravitational potential. This variation couples to $g$-modes (where buoyancy is the restoring force) within the radiative envelope. The density within this zone is not constant, and the frequencies corresponding to orbital motion preferentially drive $g$-modes deeper within the radiative zone, near the convective core, because the Brunt-V\"{a}is\"{a}l\"{a} oscillation frequency is density dependent. The induced $g$-modes distort the star, allowing gravitational torques to act. The excited waves would be standing waves if they were reflected at the stellar surface, but because the radiative time-scale there is short, they are only partially reflected, undergoing a phase shift. Angular momentum is therefore transported from the core to the surface, or vice versa, by the induced $g$-modes \citep{1975A&A....41..329Z,1989ApJ...342.1079G}.

\subsection{Implementation of tides}
We do not recalculate the BPASS models to include tides at each time-step, as this would be extremely computationally expensive. Instead, our approach is analogous to the rapid population synthesis models of \citet{2002MNRAS.329..897H}. We choose a subset of the BPASS output models and implement a post-processing algorithm as described below.  First, we identify every primary and secondary star model which produces a black hole remnant and has the chemical properties of a Ic progenitor immediately before core-collapse. In this model subset, the lowest initial mass, at any metallicity, is $10M_{\odot}$. Using this to inform our strategy, we opt to consider every primary and secondary model in BPASS with Zero Age Main Sequence (ZAMS) masses greater than $7M_{\odot}$, expecting that binary interaction, enhanced by tides, might cause some slightly lower mass stars to move onto black hole/type-Ic pathways. 

For each model, over each time-step, we calculate the expected change in the orbital semi-major axis ${\Delta}a_\mathrm{tides}$ due to tides. The change in $a$ and corresponding change in the stellar rotational angular velocity ${\Omega}$ are given by,
\begin{equation}
{\Delta}{a_\mathrm{tides}}=\frac{6ka}{t_\mathrm{damp}}q(1+q)\Big(\frac{R}{a}\Big)^{8}\Big(\frac{\Omega}{\omega}-1\Big),
\end{equation}
and
\begin{equation}
{\Delta}{\Omega}=-\frac{3k}{t_\mathrm{damp}}\frac{q^{2}}{r_{g}^{2}}\Big(\frac{R}{a}\Big)^{6}(\Omega-\omega)
\end{equation}
where $k$ is an apsidal motion constant \citep[typically in the range 0.01-0.1, we adopt 0.05,][]{1975A&A....41..329Z}, $q$ is the ratio of secondary mass $M$ to primary mass, $R$ is the stellar radius, ${\omega}$ is the orbital angular velocity and $r_{g} = I^{0.5}M^{-0.5}R^{-1}$ \citep{1977A&A....57..383Z,1981A&A....99..126H,2002MNRAS.329..897H,2019ses..book.....E}. The moment of inertia $I$ is assumed to be that of a solid sphere, $\frac{2}{5}MR^{2}$, using the total stellar mass $M$ (and assuming solid body rotation). This approximation is made for the current study as detailed calculations using stellar structure models would be computationally expensive. Most stars show only a small differential rotation gradient while still on the main sequence, although post-main sequence stars may show a significant disconnect between envelope and core rotation \citep{2000ApJ...528..368H}.  
The final term left undefined is the damping time-scale, $t_\mathrm{damp}$. This is the time-scale on which tidally induced distortions dissipate. For radiative damping, it can be shown that the damping time-scale is,
\begin{equation}
t_\mathrm{damp} = \frac{ka^{5}}{(1.9782\times10^{4})\,MR^{2}\,(1+q)^{5/6}\,E_{2}}\,\mathrm{yrs}
\end{equation}
where $M$, $R$ and $a$ are in Solar units \citep{2002MNRAS.329..897H}, and $E_{2}$ is a second-order dimensionless coefficient, parameterising the strength of the tidal interaction given the internal structure of the star. $E_{2}$ was calculated for a limited number of stellar models by \citet{1975A&A....41..329Z}. Subsequently, \citet{2002MNRAS.329..897H} fitted a functional form to these values, producing a prescription for $E_{2}$ as a function of mass. We note that \citet{2017MNRAS.467.2146K} significantly improves upon this, however their implementation relies on knowledge of the radial structure of the star - information which is not readily available in the standard BPASS output files, and would again make this process computationally expensive. A full implementation would require recalculating 250,000 individual detailed BPASS stellar evolution models, which is far beyond the scope of this paper. Comparing a handful of $E_{2}$ parameters calculated by \citet{2017MNRAS.467.2146K} to the form assumed by \citet{2002MNRAS.329..897H} gives results that are in broad agreement.

Observations of massive stars, for example the VLT-FLAMES Tarantula survey \citep{2015A&A...580A..92R,2017A&A...600A..81R}, suggest that they are born spinning with initial stellar rotational velocities typically around 30-40 per cent of their critical break-up velocity. \citet{2019arXiv190503359D} construct a probability density function for the rotational velocities of (apparently) single O stars in the Tarantula Nebula (Large Magellanic Cloud, LMC) and NGC 346 (Small Magellanic Cloud) and again find a median rotation around 40 per cent of critical. By contrast, \citet{2018MNRAS.479.4535S} used spectropolarimetry of galactic WO stars to show that their rotational velocities are less than 10 per cent of critical, however these stars are unambiguously at a late stage of evolution, thus suggesting that the spins of these massive stars evolve significantly during their lifetime. However, in low metallicity environments, the spin-down is likely reduced. \citet{2011A&A...536L..10V} and \citet{2017A&A...603A.120V} report the detection of young Galactic and LMC Wolf-Rayet stars which have surface rotations that could be conducive to GRB production.

For primary models, we adopt an initial equatorial rotational velocity that is 0.4 of the critical rotation. This is faster than previously assumed in BPASS and its predecessors. The current prescription, first used by \citet{2000MNRAS.315..543H}, gives early-type stars (specifically, those that end their lives as black holes) initial rotational velocities which are 10-20 per cent of the equatorial break-up velocity. Our ansatz is much faster, but consistent with the observations described above. 

We add the tidally driven change in orbital separation, ${\Delta}a_\mathrm{tides}$, to the orbital change due to stellar evolution processes which have already been accounted for, such as mass loss, so that ${\Delta}a_\mathrm{total} = {\Delta}a_\mathrm{tides} + {\Delta}a_\mathrm{BPASS}$. The BPASS and tidal changes can act against each other. The orbital evolution from this point forward may correspond more closely to that of a BPASS model at the same stellar evolution stage which began life with a slightly different orbital separation. If the new orbit is now closer to a different BPASS model (with the same stellar masses, at the closest matching time step), we switch models and update the parameters, continuing from there. We let the semi-major axis $a$, orbital angular velocity ${\omega}$ and stellar rotational angular velocity ${\Omega}$ vary smoothly and do not update these numbers when a model change occurs. Otherwise, we stay on the same model. 

This process is continued until a significant event happens in the system - this is defined as either a common envelope phase or Roche lobe overflow. At this point, we synchronise the primary spin to the orbital period and maintain this synchronisation until core-collapse. Because tidal forces are strongly dependent on radius, any post-main sequence expansion is likely to result in synchronisation \citep{2002MNRAS.329..897H}. Since we assume that most high mass systems are synchronised before the end of the primary model, we start each secondary model in a synchronised state, where the rotational angular velocity of the secondary equals the orbital angular velocity of the compact object companion. The percentage of model outcomes changed by the influence of tides is around ${\sim}5$ per cent, out of ${\sim}7000$ models considered per metallicity. Most of these changes do not alter the final event categorisation, as many models are shifted onto a slightly different evolutionary track which still results in the same class of transient.

\begin{figure}
\centering
\includegraphics[width=0.95\columnwidth]{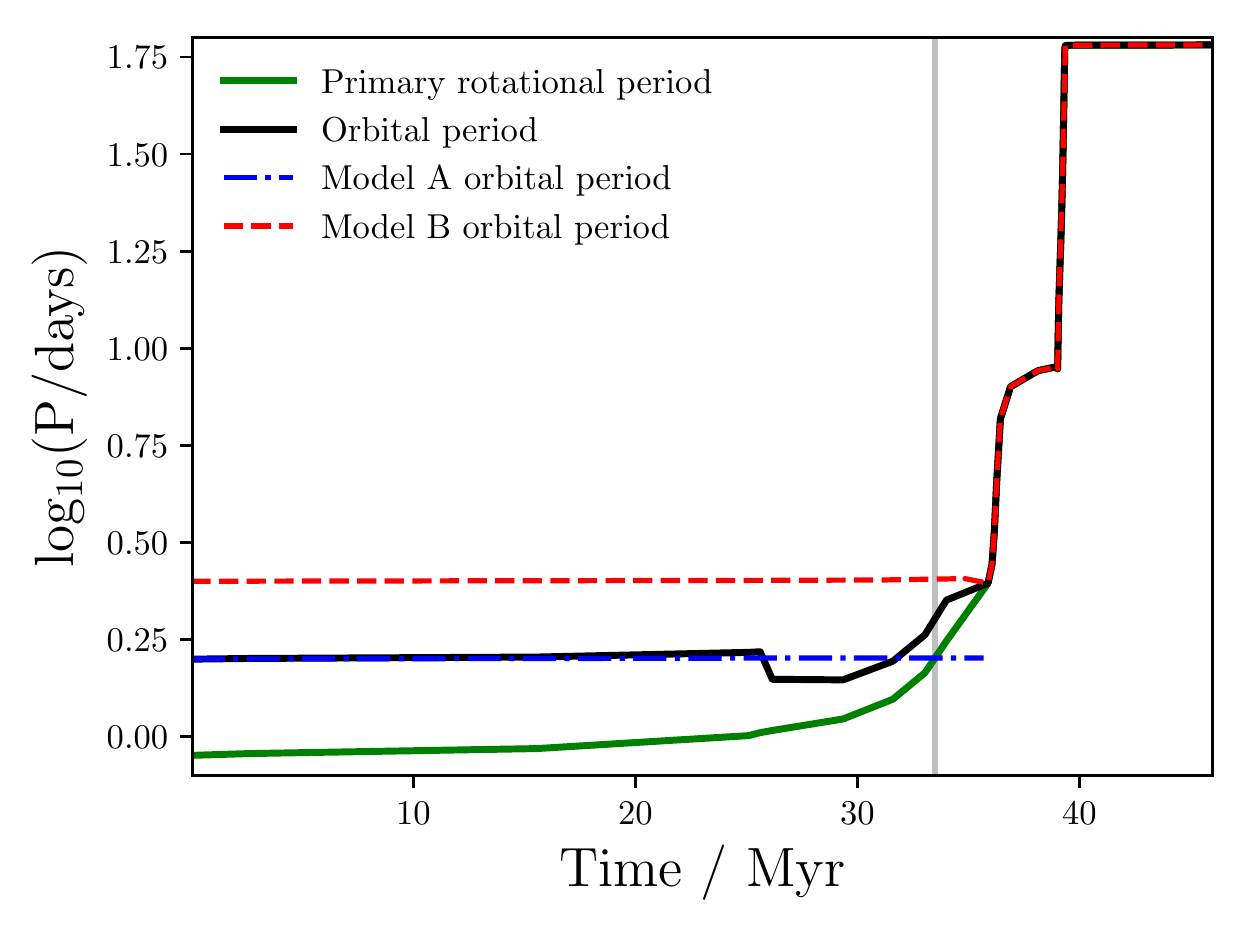}
\caption{An example of our methodology. The blue dashed line indicates the orbital period evolution of the initial model (labelled Model A). The red dashed line shows the orbital period of the model at the time of core-collapse (Model B). Other models can be passed through in between these. The solid green line is the primary star's rotational angular velocity, which starts at 40 per cent of critical, and the solid black line is the orbital angular velocity. A vertical grey line indicates when the model switch occurs. The system in this example is at a metallicity of $Z=0.014$ and consists of a $M_\mathrm{ZAMS}=8$\,$M_{\odot}$ star with a 7.2\,$M_{\odot}$ companion, and an initial orbital period of log$_{10}($P$/$days$)=0.2$. The system starts in a state where the primary is spinning faster than the orbit of the companion, and so the system moves apart as angular momentum is transferred. Tidal forces get stronger as the primary expands, leading to synchronisation at ${\sim}$35\,Myr. In this case, the added orbital angular momentum gained from tidal interactions is sufficient to eject the common envelope, and consequently the binary moves apart. Without tides, the system would retain more mass, shortening the primary lifetime, such that it explodes in $<40$\,Myr without significant evolution in orbital period.}
\label{fig:tideexam}
\end{figure}

\begin{figure}
\centering
\includegraphics[width=0.95\columnwidth]{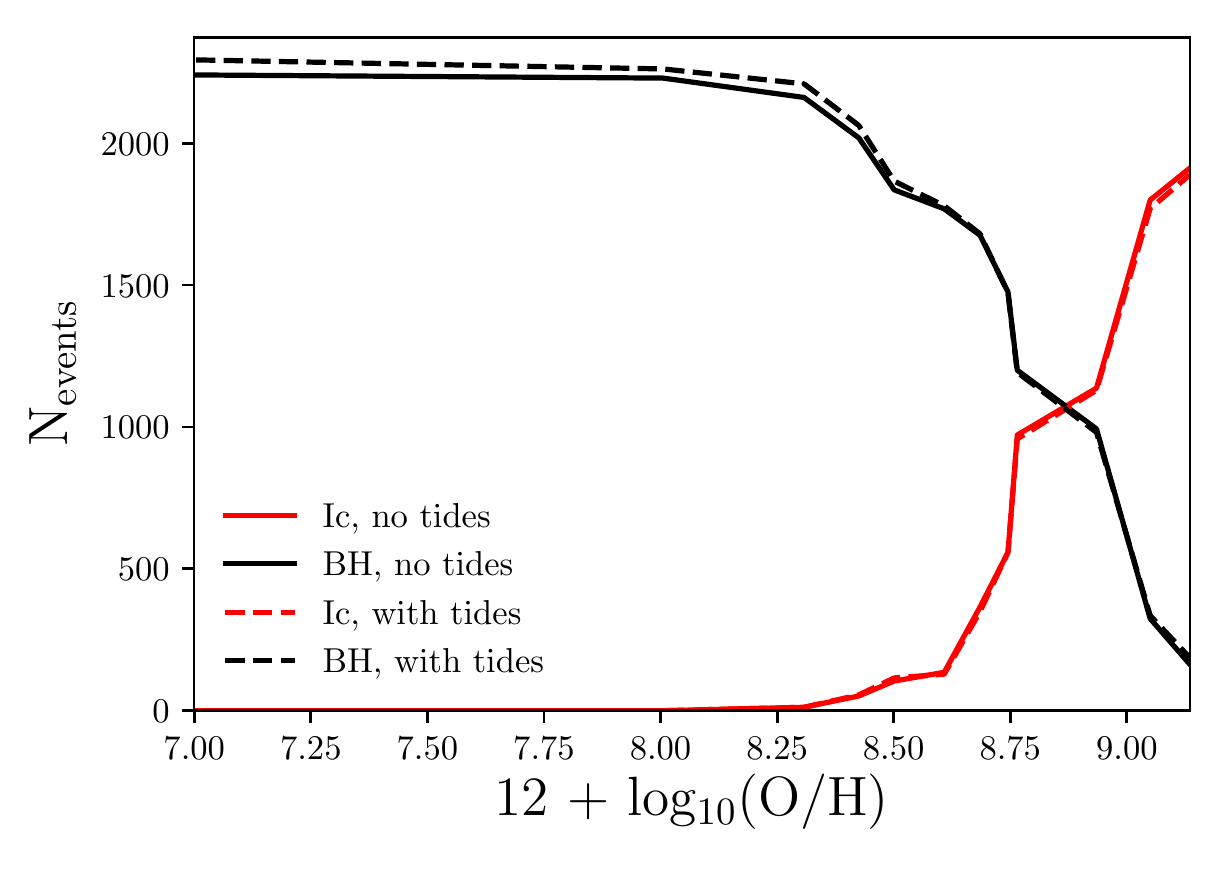}
\caption{The change in the number of Ic SNe and black holes produced per $10^{6}M_{\odot}$ of star formation, as a function of metallicity, due to the inclusion of tidal interactions.}
\label{fig:changes}
\end{figure}

An example of the methodology is shown in Figure \ref{fig:tideexam}, demonstrating the process of jumping between BPASS models when tidal interactions significantly alter the binary orbit. The synchronisation of the binary is also shown when the common envelope phase occurs at $t {\sim} 35$\,Myr.

\subsection{The effect of tides on high-mass stars}
We compare the number of models which now produce either Ic SNe with a neutron star remnant, or direct collapse to a black hole, to the previous numbers from before tides were considered. We find that models are both added to, and removed from, the pool of models which produce Ic SNe and black holes. The overall effect is shown in Figure \ref{fig:changes}. For a more complete breakdown of the changes due to tides, refer to Figure \ref{fig:A1} of the online appendix, which show the number of high-mass stellar death events (Ib, Ic, PISN and vanishing) per $10^{6}M_{\odot}$ of star formation for each metallicity and BPASS model type, both before and after tides. The effect of tides is complex and depends on the specific system in question. In most systems, given our initial rotational velocities, the binary is pushed apart, leading to reduced envelope mass loss from the primary (and a shorter main sequence lifetime). However, this same transfer of angular momentum to the orbit can make the ejection of a common envelope more efficient. These effects are roughly balanced, such that the overall change in the number of stripped envelope progenitors is small, as Figure \ref{fig:changes} shows.

In order for a stripped-envelope star to produce a GRB, it must also launch jets. A key parameter of interest in jet production is the angular momentum of the star at core-collapse. \citet{1993ApJ...405..273W} and \citet{1999A&AS..138..499W} calculated that the specific angular momentum of material just outside the newly formed black hole should be $>{\sim}10^{16}$\,cm$^{2}$s$^{-1}$ in order for accretion to occur, otherwise material directly infalls and energy extraction is inefficient.

\begin{figure}
\centering
\includegraphics[width=0.95\columnwidth]{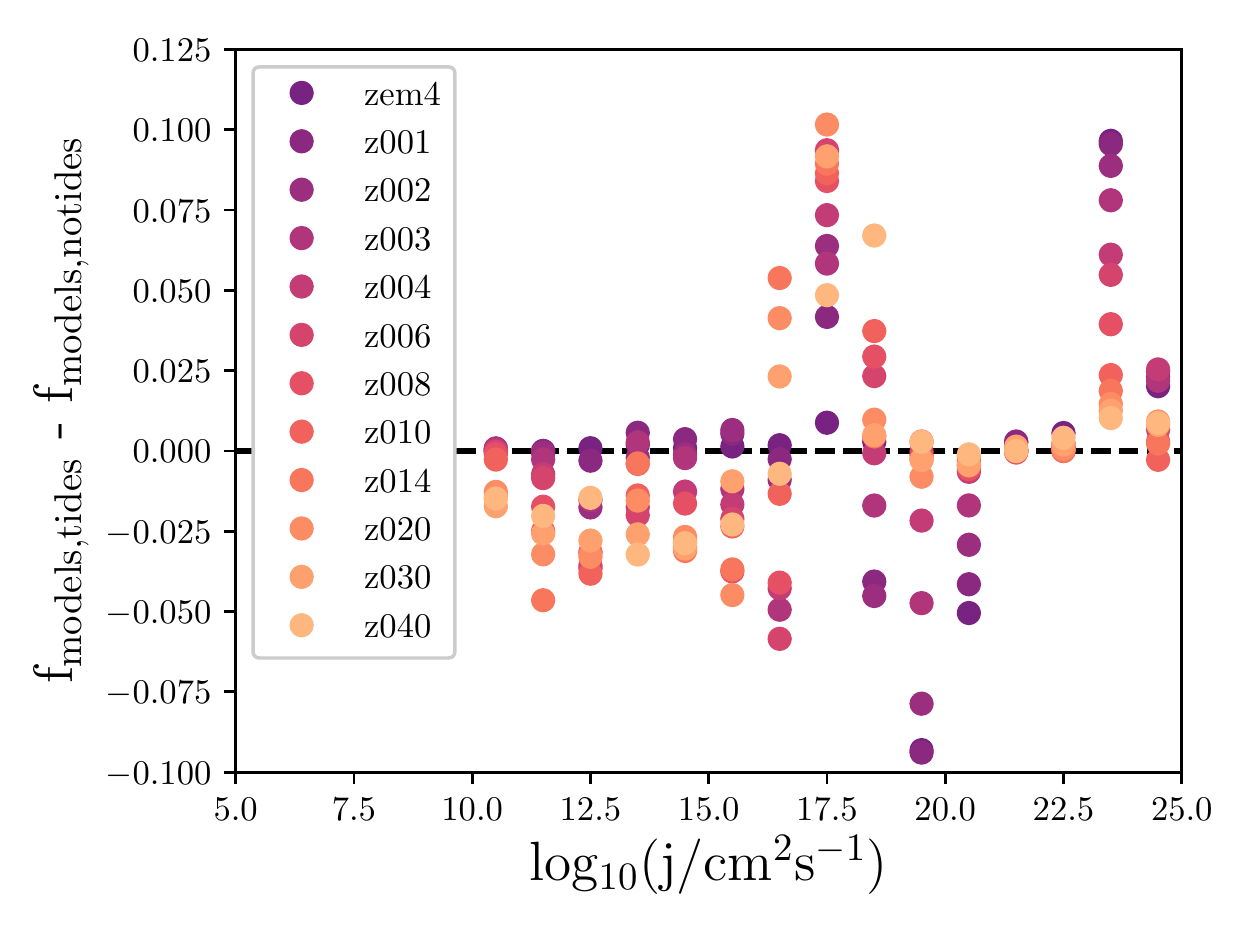}
\caption{The difference in each specific angular momentum bin between the number of stellar models producing Ic-SNe and black holes at core-collapse, before and after tides are considered. Metallicity is indicated by the colour gradient. The $y$-axis represents the bin heights for the number of models after tides, subtracted from the bins heights before tides were considered, where the histograms are normalised so that the enclosed area of each is equal to unity. This therefore provides an indication of changes in the $j$ distribution due to tides.}
\label{fig:jplot}
\end{figure}

In Figure \ref{fig:jplot}, we visualise the effect of tides on the final angular momentum of Ic-SN and black hole progenitors. For the evolution without tides, the spin of the star in question is synchronised to the orbit in the final time-step in order to obtain an estimate of the specific angular momentum $j$. The resultant distributions of $j$, with and without tidal evolution over the stellar lifetime, are binned, and the difference in the normalised fraction contributing to each $j$ bin is shown. Notably, it is the high metallicity models which are predominantly affected by the inclusion of tides. The evolution of these systems was previously dominated by wind-driven mass loss, however tidal interactions are acting to maintain angular momentum in the primary when it would otherwise be lost. Although tides typically push binaries apart over the main sequence lifetime, they tend to end their lives spinning more rapidly due to synchronisation in the mass transfer and common envelope phases.

Figure \ref{fig:jplot} also shows a high-$j$ spike due to tides, which is mostly populated by low metallicity systems. Inspecting the mass distribution of these high angular momentum stars, they are typically very high mass. Our assumed initial mass function (IMF) in this work allows for ZAMS masses up to $300M_{\odot}$. Because tidal forces are extremely radius sensitive (proportional to the eighth power in $R$ in the prescription used), and because low $Z$ stars are more likely to maintain very high masses all the way to core-collapse, the highest $j$ bins are naturally populated by the lowest $Z$ stars.

\begin{figure}
\centering
\includegraphics[width=0.95\columnwidth]{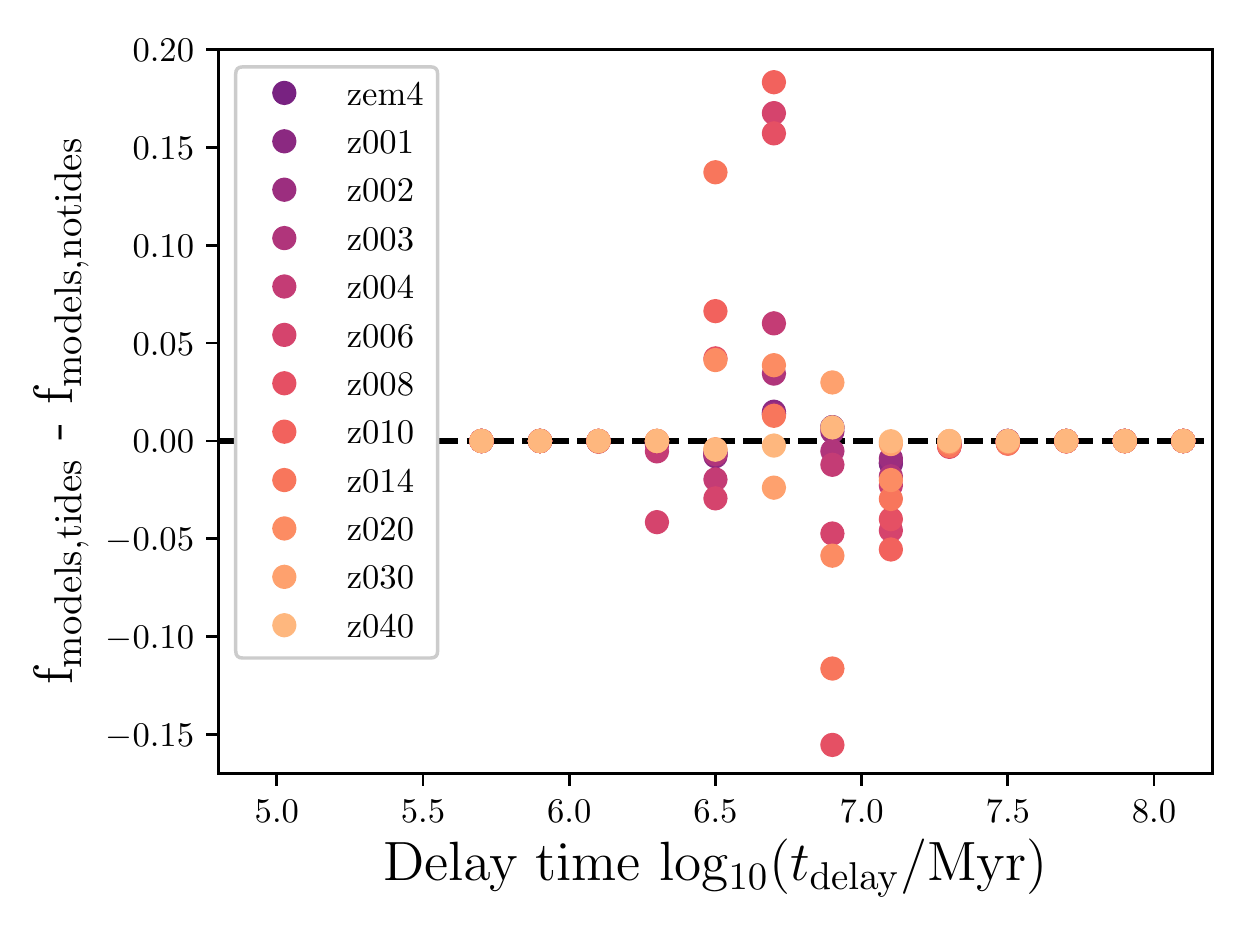}
\caption{The change in the number of models falling in each delay time bin, for black hole and Ic-SN producing models. Delay times are measured from the zero age main sequence. The overall effect of tides on the delay time distribution of massive stars is small, since model lifetimes are both shortened and lengthened by decreased mass transfer and more efficient common envelope ejection respectively. The $y$-axis is calculated in the same way as Figure \ref{fig:jplot}.}
\label{fig:delay}
\end{figure}

Another effect of introducing tides to BPASS is that the delay times, the intervals between star formation and core-collapse, are changed. In general, over the main sequence lifetime, tides act to push binaries apart, therefore decreasing the mass lost to a companion. This means that stars leave the main sequence with higher masses than before tides were considered, which typically shortens lifetimes and therefore delay times. However, if the stars are further apart, the orbital angular momentum is greater and common envelopes are more efficiently ejected, therefore reducing the system mass (and increasing delay times). These two effects, which act to add and remove mass respectively, dominate equally as often. In Figure \ref{fig:delay}, as in Figure \ref{fig:jplot}, we show the change of the contribution to each delay time bin from the Ic-SN and black hole progenitor population, before and after tides are included. The overall effect of tides on the delay time distributions is a small shift to shorter delays.

Because the application of this tidal algorithm changes the frequency with which primary models end in different configurations, the weightings of the secondary models are changed. These are calculated based on the primary model endpoints and follow the evolution of post-supernova systems to later time-steps. Using the IMF and initial binary population parameters of the primary models, and the associated new end points for these binaries, the secondary models are re-weighted to account for the tidal evolution. In subsequent analysis, these new weightings are used.

\section{Calculation of long gamma-ray burst volumetric event rates and progenitor properties}\label{sec:eventrates}
\subsection{Transient population synthesis}
The search for additional, tidally-induced GRB pathways (referred to hereafter as `tidal GRBs') is motivated by the observed metallicity distribution of GRB progenitor environments \citep[see e.g.][]{2015ApJ...802..103T,2019A&A...623A..26P}, but the addition of any such pathway must also be consistent with observed long GRB volumetric event rates. For the existing QHE pathway, any stars with a remnant mass $>3M_{\odot}$, which accrete $>5$ per cent of their initial mass at $Z<0.004$, fulfil our composition and spin requirements for a GRB \citep{2019MNRAS.482..870E}.  We have now further identified the BPASS models which are rapidly spinning at core-collapse, at all metallicities, and which would produce a type Ic-SN if collapse to a black hole did not hamper the escape of emission from an explosion. We expect a currently-undefined subset of these to also launch GRBs and hypothesise that the principle determinant of whether they do so is whether they exceed a threshold in specific angular momentum, $j_\mathrm{cut}$. We will assume that these two pathways are the only contributors to the GRB population and now turn our attention to deciding which of the tidally spun models produce GRBs, and whether we can re-create the observed GRB rate and its evolution over cosmic history using plausible selection criteria. 

We vary the specific angular momentum threshold to define sets of plausible GRB progenitors. The model weightings are used to determine, for a given mass of stars formed at metallicity $Z$, how many GRBs we would expect. We also know the delay times of these models. In order to construct the GRB rate as a function of redshift, we follow the methodology of \citet{2019MNRAS.482..870E} to apply a metallicity-dependent cosmic star formation rate history. The star formation rate density \citep[SFRD,][]{2014ARA&A..52..415M} as a function of redshift is given by,
\begin{equation}
\psi(z) = 0.015\frac{(1+z)^{2.7}}{1+((1+z)/2.9)^{5.6}} M_{\odot}\mathrm{yr}^{-1}\mathrm{Mpc}^{-3},
\end{equation}
and this used in conjunction with the formalism of \citet{2006ApJ...638L..63L}, which decomposes the SFRD into contributions from different metallicities as a function of redshift. The SFRD at or below a metallicity $Z$, at redshift $z$, is given by,
\begin{equation}
\Psi\left(z,\frac{Z}{Z_{\odot}}\right) = \psi(z) \frac{\hat{\Gamma}[0.84,(Z/Z_{\odot})^{2}10^{0.3z}]}{\Gamma(0.84)},
\end{equation}
where $\hat{\Gamma}$ and ${\Gamma}$ are the incomplete and complete Gamma functions. We use 0.020 for the Solar metallicity $Z_{\odot}$, which corresponds to $12+$log$(O/H)=8.93$ in the abundance distribution pattern adopted by BPASS \citep[see][]{2017PASA...34...58E}.

Low metallicities dominate star formation only in the very early Universe, with the peak of Solar-metallicity star formation at around $z{\sim}1$, and the overall peak in star formation at $z{\sim}2$. By applying the GRB progenitor model rate per $10^{6}M_{\odot}$ to the SFR at each redshift and metallicity, and accounting for the delay times, we can construct the intrinsic (and hence estimate the observed) GRB rate. This requires, however, the application of an angular momentum cut-off $j_\mathrm{cut}$ for the tidal GRB pathways, and rate corrections due to jet beaming and the GRB luminosity function. 

\subsection{Bayesian parameter estimation}\label{sec:mcmc}
In order to investigate whether our two-pathway model is plausible, we allow four model parameters to vary, and infer their values by comparison to observational data. The parameters are,
\begin{enumerate}
\item the jet half-opening angle ${\theta}$,
\item the lower limit on the isotropic equivalent energy of GRBs, $E_\mathrm{low}$,
\item the specific angular momentum $j_\mathrm{cut,\odot}$, at Solar metallicity, above which GRBs can occur in our black hole producing, stripped envelope, rapidly rotating models,
\item the index $n$, which allows for a metallicity dependence of this angular momentum cut (as $j_\mathrm{cut}\propto j_\mathrm{cut,\odot}(Z/Z_{\odot})^{n}$).
\end{enumerate}
The first two parameters determine whether a GRB is likely to be seen in the observed sample, while the latter two determine whether an event is likely to launch a jet at all. For observed GRB rates, we use the Swift Gamma-Ray Burst Host Galaxy Legacy Survey (SHOALS) sample \citep{2016ApJ...817....7P}.
This is the largest unbiased sample of GRBs with identified hosts (and hence redshift and metallicity estimates) consisting of bursts with isotropic equivalent energies, E$_\mathrm{iso}$, greater than $10^{51}$\,erg.

Of the four model parameters, two are corrective and change the number density of events equally across redshift (${\theta}$, $E_\mathrm{low}$). The other two ($j_\mathrm{cut}$, $n$) may also affect how the rate varies with redshift. The half-opening angle ${\theta}$ is used to account for the fact that GRBs are strongly beamed, and therefore most events are seen off-axis and not detected, so that the observed rate is much lower than the intrinsic one. This correction is given by $[1-\cos{\theta}]^{-1}$ for bipolar jets, and assumes that within a viewing angle ${\theta}$, we are equally likely to detect the burst whatever the orientation. The second factor corrects for the GRB luminosity function (actually an isotopic-equivalent energy $E_\mathrm{iso}$ function). This is required because the SHOALS comparison data is comprised exclusively of high-energy bursts, above $10^{51}$\,erg, which creates a luminosity-unbiased sample over a wide redshift range. However, we want to know the total number of bursts which are occurring. The assumed GRB luminosity function is taken from \citet{2016A&A...587A..40P}. This has a power law slope of -1.2 below, and -1.92 above, a break energy of $5\times10^{50}$\,erg. The SHOALS data gives us the number of events per redshift bin above $10^{51}$\,erg, to obtain the total number at each redshift for comparison with modelled rates we integrate over the function down to a lower limit $E_\mathrm{low}$. Finally, models are selected as GRB candidates if they have a specific angular momentum at core-collapse greater than $j_\mathrm{cut}$, which is proportional to $(Z/Z_{\odot})^{n}$.

We perform a Markov Chain Monte Carlo (MCMC) analysis to infer the probability density functions of ${\theta}$, $E_\mathrm{low}$, $j_\mathrm{cut}$ and $n$. 
The Python package {\sc emcee} is used to perform our MCMC sampling \citep{2010CAMCS...5...65G,2013PASP..125..306F}. The posterior distribution of the parameters, $P({\theta},L_\mathrm{E},j_\mathrm{cut},n|D,M)$ is given by the product of the log likelihood and log prior, 
\begin{equation}
\begin{split}
& P({\theta},L_\mathrm{E},j_\mathrm{cut},n|D,M) \propto \\
&P(D|{\theta},E_\mathrm{low},j_\mathrm{cut},n,M) \times P({\theta},E_\mathrm{low},j_\mathrm{cut},n)
\end{split}
\end{equation}
where $D$ denotes the data (in this case the SHOALS rates and their quoted uncertainties), and $M$ is the two-pathway model as previously described. The priors assumed, $P({\theta},E_\mathrm{low},j_\mathrm{cut},n)$, are as follows:
\begin{enumerate}
\item For the half opening angle, we simply limit ${\theta}$ to the range $0 < {\theta} < 22.5$, the least informative prior we can use whilst ruling out very weakly beamed (total opening angle $>45$\,deg) or isotropic emission. 
\item We have limited prior knowledge on the lowest possible GRB luminosity. We therefore restrict $E_\mathrm{low}$ to the range $45 < \mathrm{log}_{10}(E_\mathrm{low}/$erg$) < 50.7$, where the upper bound is the break in the assumed luminosity function, and the lower bound is arbitrarily low. Although events are seldom seen at $\mathrm{log}_{10}(E_\mathrm{iso})<50.7$, we are trying to apply a minimal prior and allow for a small number of outliers in $E_\mathrm{iso}$ \citep{2019ApJ...878...52A}.
\item We use a top-hat prior which covers the range $16 < \mathrm{log}_{10}(j_\mathrm{cut}/$cm$^{2}$s$^{-1}) < 19.3$. The lower bound corresponds to the theoretical minimum $j$ required for GRB production by a black hole central engine \citep[e.g.,][]{1993ApJ...405..273W,1999ApJ...524..262M}. The upper bound is simply the maximum value of $j$ obtained from our tidal calculations. 
\item For the index $n$, we again use a flat prior covering all physically realistic values ($0<n<15$). GRBs are rarer at high metallicity, and high stellar envelope opacity could help to impede jet propagation. Because opacity scales approximately as $Z$ \citep[e.g.][]{2001A&A...369..574V}, we might expect $n=1$ to be most probable, however this is not favoured in our prior.
\end{enumerate}

\begin{figure*}
\centering
\includegraphics[width=0.95\textwidth]{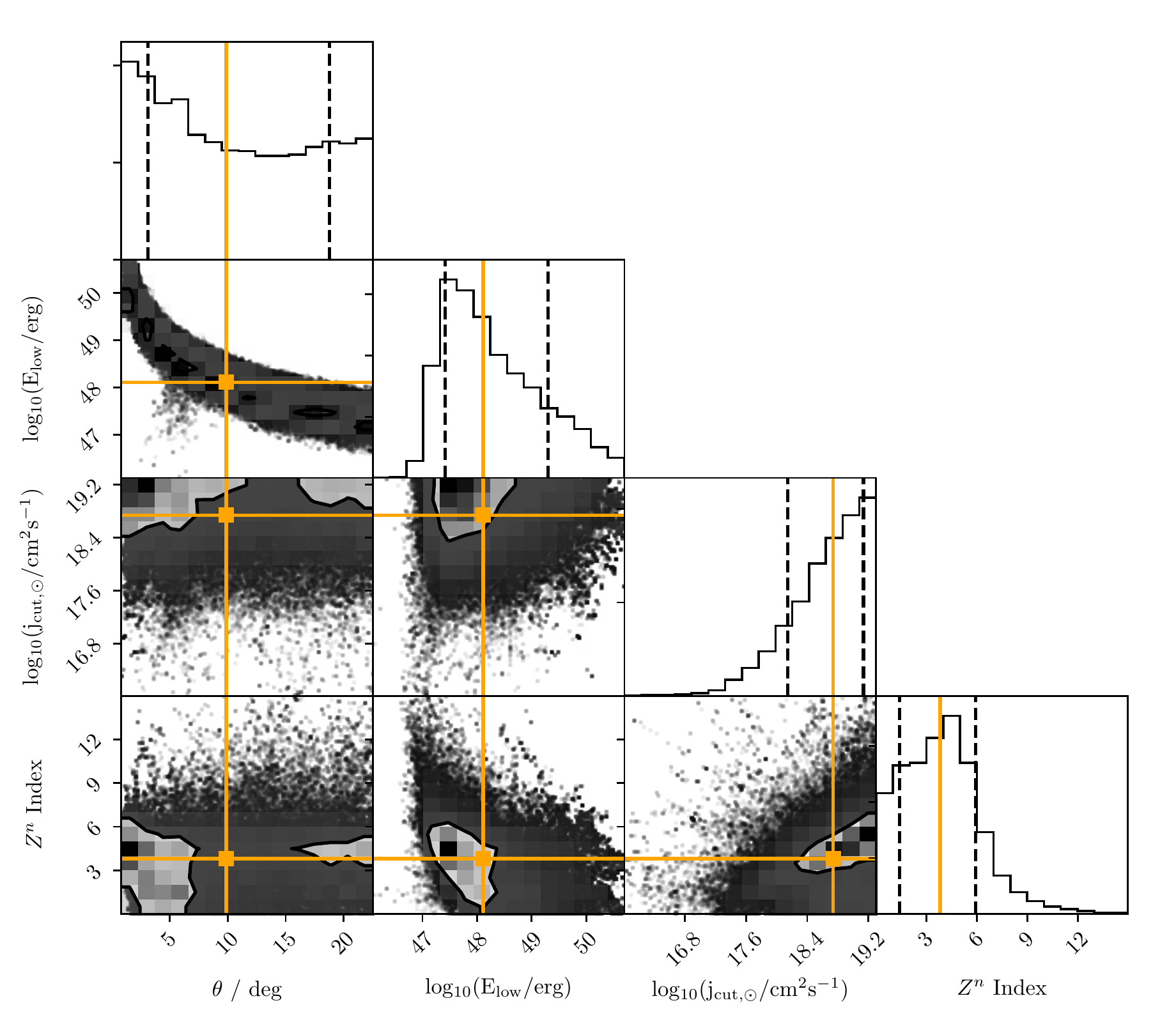}
\caption{Covariances between the four fitted parameters in our two-pathway GRB model, and the marginalised posterior probability density distributions for each. Vertical dashed lines mark the 16$^\mathrm{th}$ and 84$^\mathrm{th}$ percentiles, and the orange crosshairs indicate the distributions medians. 1D histograms at the top of each column represent the marginalised distribution of the parameter indicated on the $x$-axis for that column.}
\label{fig:corner}
\end{figure*}

Initialising the MCMC with 100 walkers, 2000 steps, and a burn-in of 50 steps (checked by visual inspection), we obtain the joint marginalised distributions and correlations shown in the corner plot of Figure \ref{fig:corner}.

\section{Predicted long gamma-ray burst rates and progenitors}\label{sec:results}
\subsection{MCMC results}
The marginalised probability distribution in the half-opening angle ${\theta}$ is similar in form to those found by \citet{2009ApJ...698...43R}, and the well-fit sample of \citet{2015ApJ...799....3R}. The posterior distribution in $E_\mathrm{low}$ favours a critical isotropic equivalent energy of $10^{48.1}$\,erg. This result arises despite a minimal prior which allows for all physically reasonable values below the break in the assumed luminosity function. The cutoff is in good agreement with observations, which have yielded bursts with log$_{10}(E_\mathrm{iso}) < 48$ on only a few occasions, despite such events being theoretically detectable at low redshift \citep[e.g.][]{2009ApJ...698...43R,2019ApJ...878...52A}. The posterior distribution of the angular momentum cutoff log$_{10}(j_\mathrm{cut,\odot})$ has its median at 18.74. Finally, the distribution of the power-law index $n$ drops away at very high values, with a peak and median at ${\sim}$4. The mean, median, and percentiles of all four posterior distributions are listed in Table \ref{tab:post}.

\begin{table}
\centering 
\caption{Properties of the parameter posterior distributions, obtained from the MCMC analysis carried out in Section \ref{sec:results}.}
\begin{tabular}{l p{0.9cm} p{1.2cm} p{0.8cm} p{1.2cm}} 
\hline 
Parameter & Mean & 16$^\mathrm{th}$ Percentile & Median & 84$^\mathrm{th}$ Percentile \\ 
\hline 
${\theta}$ / deg & 10.62 & 3.12 & 9.88 & 18.75 \\
log$_{10}$(E$_\mathrm{low}$/erg) & 48.31 & 47.42 & 48.11 & 49.30 \\
log$_{10}$(j$_\mathrm{cut,\odot}$/cm$^{2}$s$^{-1}$) & 18.64 & 18.14 & 18.74 & 19.13 \\
$Z^{n}\mathrm{index}$ & 3.87 & 1.39 & 3.81 & 5.94 \\
\end{tabular}
\label{tab:post}
\end{table}

\begin{figure*}
\centering
\includegraphics[width=0.95\textwidth]{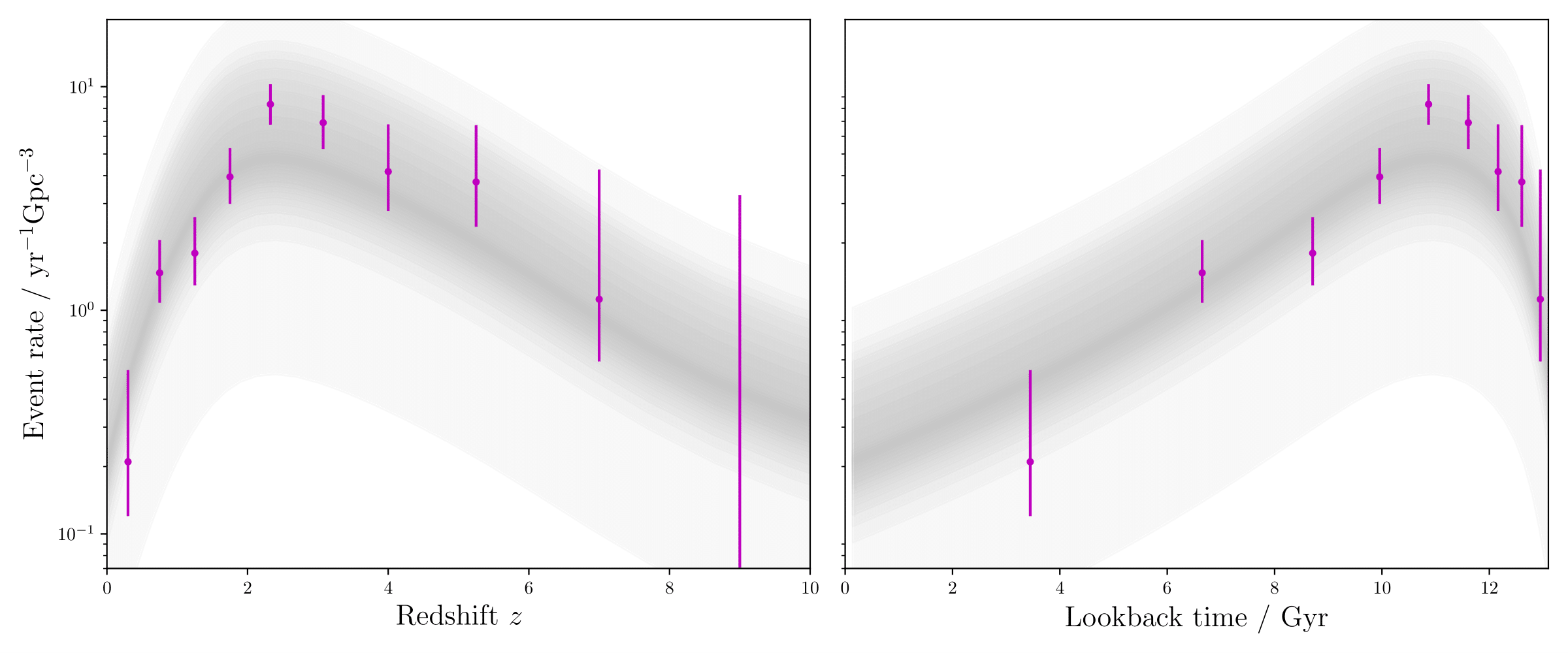}
\caption{The distribution of observed GRB rates, as a function of redshift (or lookback time). Darker shading represents higher probability density. The observed SHOALS rates are shown with their uncertainties. BPASS produces intrinsic rates, these have been corrected using the distributions of ${\theta}$ and $E_\mathrm{low}$ which are output as posteriors from the MCMC run.}
\label{fig:posterior}
\end{figure*}

\begin{figure*}
\centering
\includegraphics[width=0.95\textwidth]{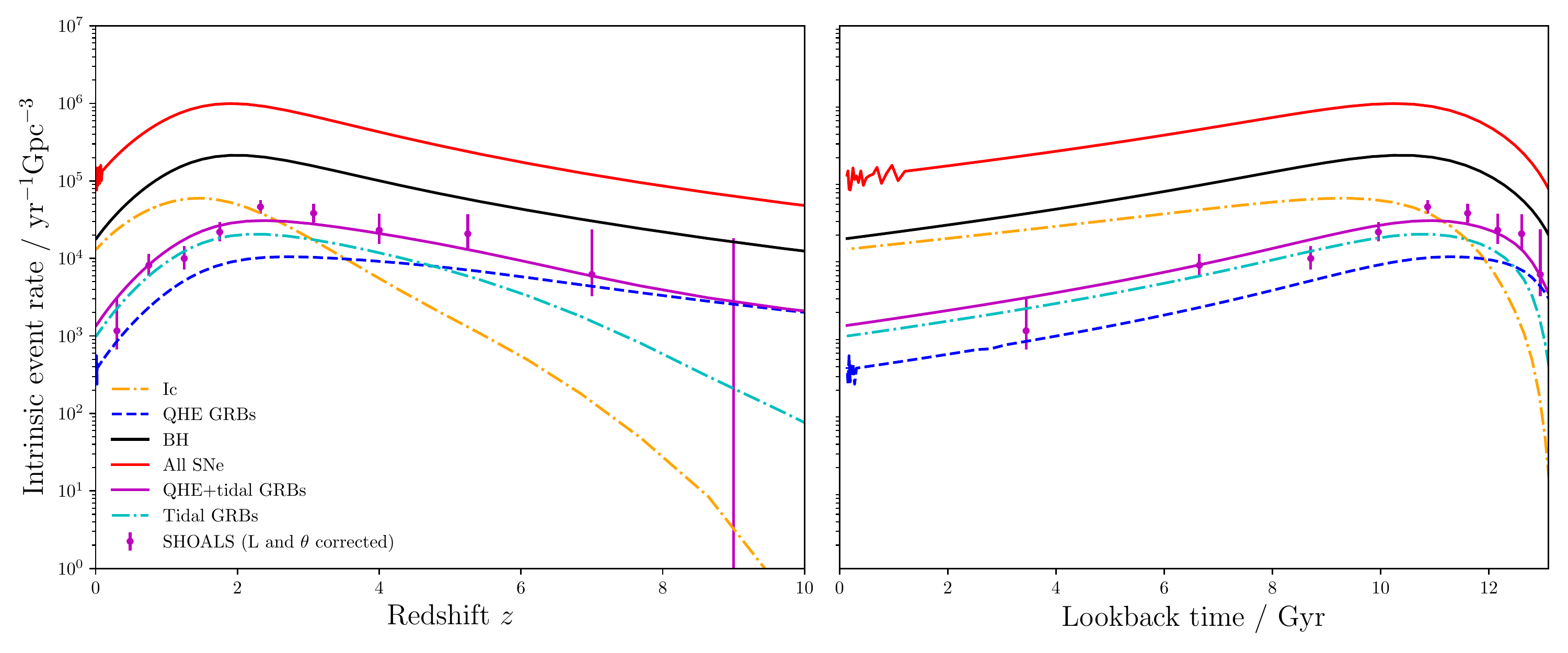}
\caption{The BPASS prediction for the intrinsic volumetric rate of GRBs arising from both QHE and tidal pathways, shown by the solid purple line. The dashed dark blue line below is the contribution from QHE progenitors, and the dashed orange line represents bursts from the tidal influence pathway. These rates were obtained by selecting black hole-producing, stripped envelope progenitors with a specific angular momentum cut that best reproduces the evolution of the rate over cosmic time. The SHOALS rates have then been corrected for beaming and a luminosity function, using the medians of the ${\theta}$ and $E_\mathrm{low}$ MCMC posterior distributions.}
\label{fig:grb_rate}
\end{figure*}

In Figure \ref{fig:posterior} we show the posterior probability density for our model, compared with the observed $E>10^{51}$\,erg SHOALS rates. The model (as a function of $j_\mathrm{cut}$ and $n$) predicts an intrinsic rate, which is converted to an observed rate by the ${\theta}$ and $E_\mathrm{low}$ parameters as described above. The probability density shown in Figure \ref{fig:posterior} therefore represents the posterior distribution for the correction to the intrinsic rate, which is a convolution of the corrections arising from the opening angle and lower luminosity limit. 

For best estimates of the parameters values, we use the posterior medians and 68 per cent credible intervals given by the 16$^\mathrm{th}$ and 84$^\mathrm{th}$ percentiles. This gives ${\theta}=9.9^{+8.9}_{-6.8}$ degrees, log$_{10}($E$_\mathrm{low}$/erg$)=48.1^{+1.2}_{-0.7}$, log$_{10}$(j$_\mathrm{cut,\odot}$/cm$^2$\,s$^{-1}$)=$18.7^{+0.4}_{-0.6}$ and $n=3.8^{+2.1}_{-2.4}$. Using the median values gives the fit shown in Figure \ref{fig:grb_rate} (we do not fit to the highest redshift SHOALS point, which gives an event rate consistent with zero). We note that the fitting just the QHE pathway can produce similarly good results given different assumptions for ${\theta}$ and $E_\mathrm{low}$, as demonstrated by \citet{2019MNRAS.482..870E}, but we know that GRBs occur above $0.2Z_{\odot}$ metallicity, and therefore the QHE pathway cannot be the sole contributor.

\subsection{Metallicity distribution}
\begin{figure}
\centering
\includegraphics[width=0.95\columnwidth]{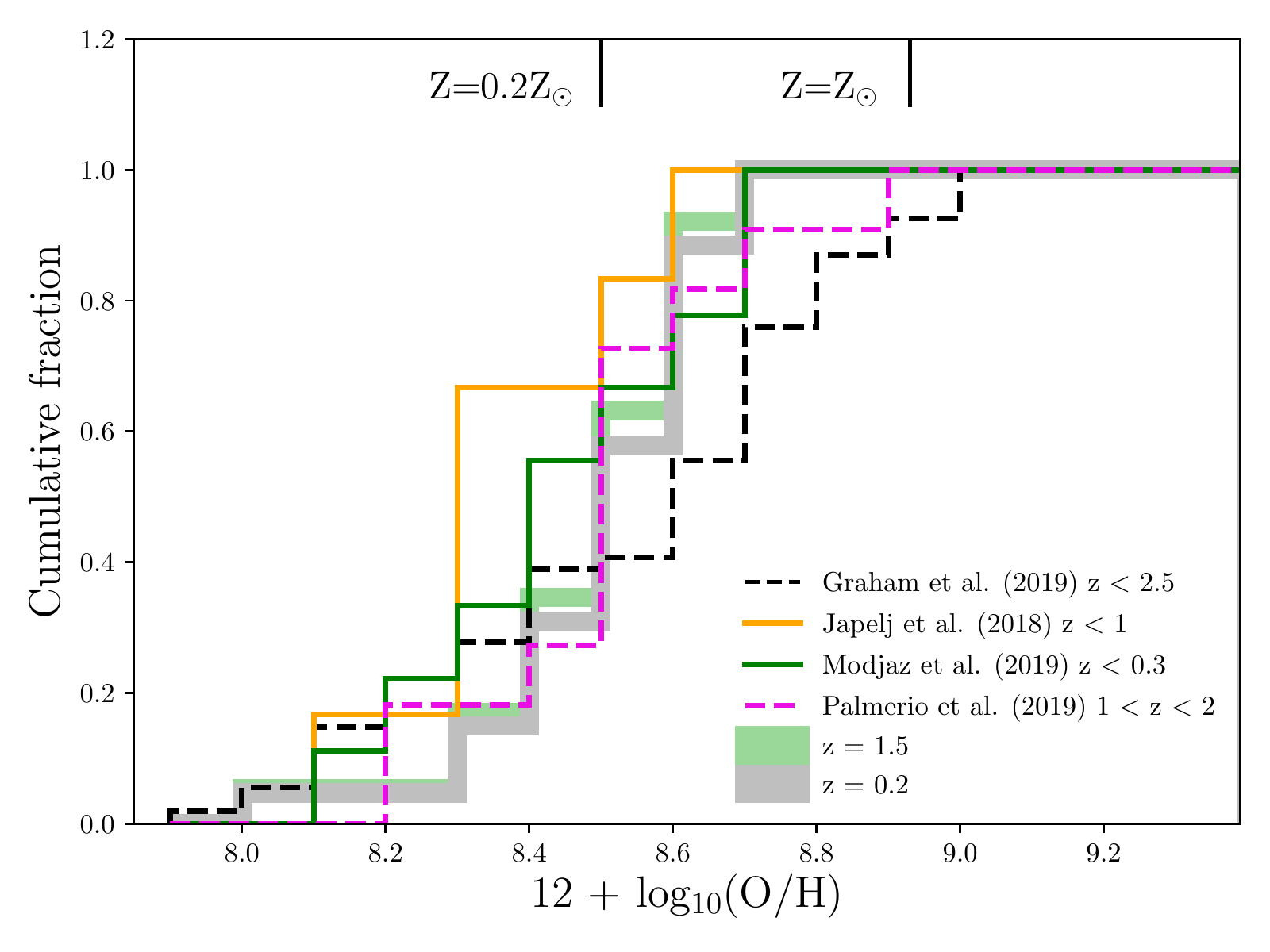}
\caption{Cumulative metallicity distributions of our GRB progenitors at $z=0.2$ and $z=1.5$, marked by shaded grey and green lines, compared with observed host galaxy distributions from \citet{2018A&A...617A.105J}, \citet{2019arXiv190100872M}, \citet{2019A&A...623A..26P} and \citet{2019arXiv190402673G}. The comparison data has been shifted so that at the mass fraction used to define Solar metallicity in their scale, 12$+$log(O$/$H) is the same as the BPASS value at that mass fraction. Marked on the plot are $0.2Z_{\odot}$ and $Z_{\odot}$ metallicities in the (modified) BPASS scaling.}
\label{fig:metals}
\end{figure}

An independent test is to compare the metallicity distribution predicted for our tidal and QHE GRB progenitors, to that of observed host galaxies. In Figure \ref{fig:metals}, we show the synthetic metallicity distribution of our GRB-producing stars at two redshifts, $z=0.2$ and $z=1.5$. Also shown are host galaxy metallicity distributions from \citet{2018A&A...617A.105J}, \citet{2019arXiv190100872M}, \citet{2019arXiv190402673G} and \citet{2019A&A...623A..26P}.

To draw comparisons between metallicity distributions, we need to ensure that the scales being used do not have significant offsets. \citet{2019arXiv190402673G} used the metallicity diagnostic and scaling of \citet{2004ApJ...617..240K}, with Solar metallicity defined at 12$+$log(O$/$H)$=8.69$ \citep{2001ApJ...556L..63A}. This corresponds in their scale to a metal mass fraction of 0.014. In our BPASS scaling, a mass of fraction of 0.014 corresponds to 12$+$log(O$/$H)$=8.76$. To reconcile this with \citet{2004ApJ...617..240K}, 0.07 dex is added to each value in the \citet{2019arXiv190402673G} distribution.

The other three comparison data sets use a \citet{2008A&A...488..463M} scaling (\citet{2019arXiv190100872M} provide a variety, we choose the same scaling for consistency), where Solar is again at 12$+$log(O$/$H)$=8.69$, but this now corresponds to 0.0134 by mass fraction \citep{2009ARA&A..47..481A}. Again, these data sets are shifted by 0.07, which brings the 0.0134 Solar value into agreement with BPASS at 12$+$log(O$/$H)$=8.76$. For a discussion of these issues, and their impact within BPASS, we refer the reader to \citet{2017PASA...34...58E} and \citet{2018MNRAS.477..904X}.

Interestingly, there appears to be little evolution of the metallicity distribution in GRBs predicted by the BPASS models with redshift, with an overall shift to lower values of only ${\sim}$0.2 dex between redshifts 0 and 5. The data similarly shows a lack of variation in the observed fraction of high metallicity bursts out to redshift 2.5 \citep{2019arXiv190402673G}. The observed samples nonetheless span a wide redshift range, and we compare each to the closer of the two redshift curves shown on Figure \ref{fig:metals}. Anderson-Darling tests between the BPASS results and the data fail to reject the null hypothesis (i.e. $p>0.05$) that they are drawn from the same distribution, in every case except for \citet{2019arXiv190402673G}.

\begin{figure}
\centering
\includegraphics[width=0.95\columnwidth]{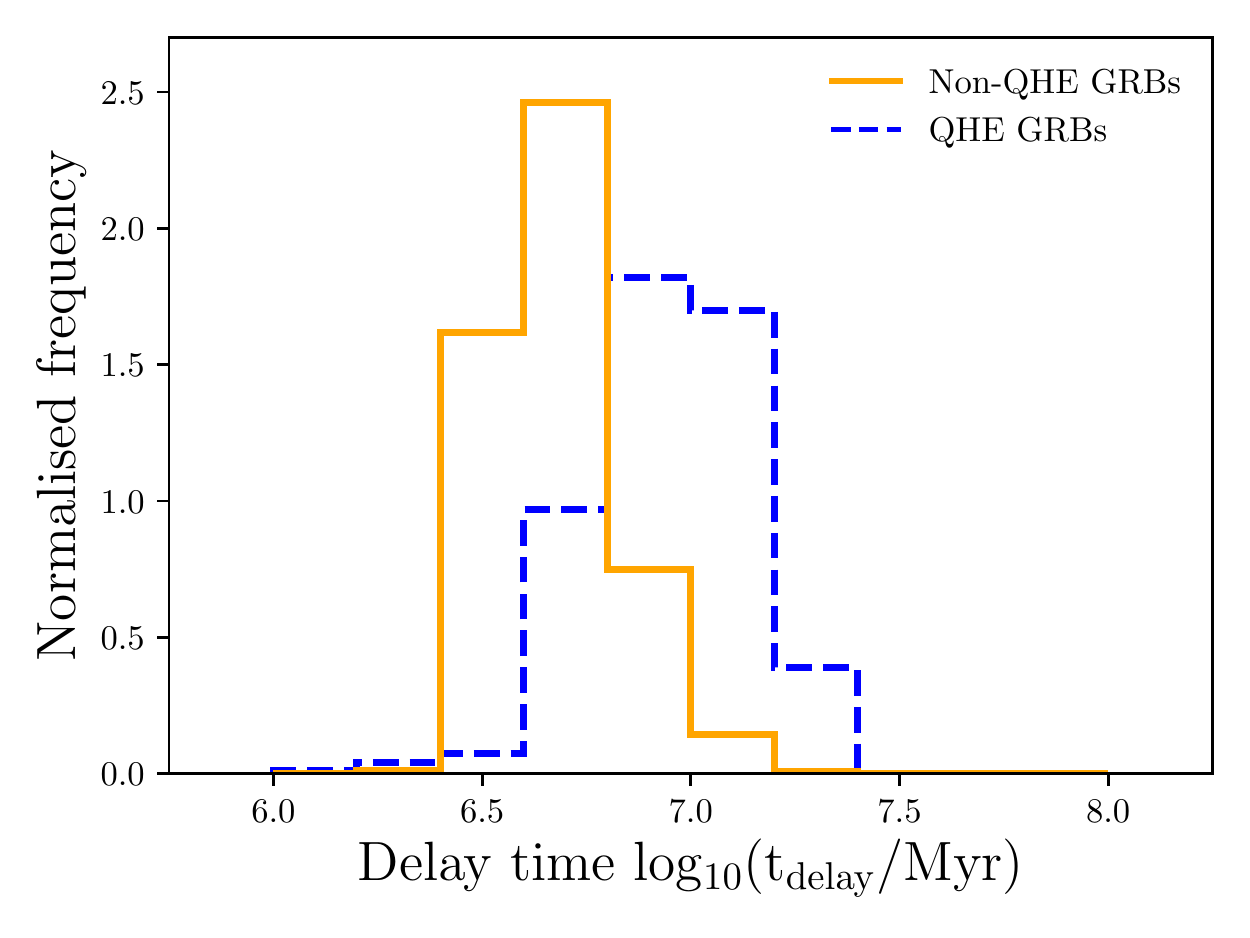}
\caption{Delay time distributions (DTDs) for QHE and tidal GRB progenitors, across all metallicities considered in this work. QHE pathways are limited to $Z<0.2Z_{\odot}$ by construction. The contribution from each model is weighted according to the (tide adjusted) BPASS weightings, and delays are measured from zero-age main sequence. The total area of each distribution is normalised to unity.}
\label{fig:GRBdelay}
\end{figure}

\subsection{Delay-time distribution}
In Figure \ref{fig:GRBdelay}, the delay-time distribution for the QHE and tidal progenitors are shown. In the tidal case this is for all metallicities considered, whereas the QHE distribution is limited to $Z<0.2Z_{\odot}$ by construction. The new tidal pathways have shorter delay times, decreasing the mean temporal offset between star formation and GRB events. We note that these times are technically only until the end of core carbon burning. However, the final stages of core burning before core collapse occupy $<<$1\,Myr \citep[e.g.][]{2014A&A...564A..30G}.
The GRB progenitors have among the shortest delay times of any stars, particularly so for the tidal GRBs. This implies that they will be also be among the most luminous main sequence stars. Unlike the QHE pathway, in which it is lower mass secondary stars that produce GRBs, in the tidal pathway the progenitor will usually not have been subject to a supernova kick, or have received a smaller kick. The progenitor stars would therefore be preferentially formed in the brightest regions of their host galaxies, and stay there, leading to a GRB distribution within their hosts that traces the host light - a trend which has been previously been observed \citep[e.g.][]{2006Natur.441..463F,2011MNRAS.414.3501E,2017MNRAS.467.1795L}.

\subsection{Progenitor systems}
Table \ref{tab:prog_table} shows distribution statistics for the tidal progenitor systems in mass ratio, orbital period, initial mass, final mass and delay time parameter space, over the metallicity range $Z=0.008-0.020$. Stars with final masses of ${\sim}20M_{\odot}$, in tight binaries with large mass ratios, are the most frequent progenitors. A corner plot showing this information, and covariances between parameters, is available in Figure \ref{fig:B1} of the online appendix. There are signs of an excess in the number of twin systems (i.e. systems with mass-ratio near unity) which give rise to GRBs, however this is simply reflecting the increased likelihood of twin systems in all massive binaries, and is not specific to GRB progenitors.  

\begin{table}
\centering 
\caption{Properties of the tidal GRB progenitors over the metallicity range 0.008-0.020. For GRBs arising from the primary star (the majority), mass ratio $q$, initial mass and log($P$) are all given at ZAMS. For GRBs arising from the secondary star in a binary, values are given immediately after the supernova of its primary companion. We list the minimum, maximum, mean and standard deviation ${\sigma}$ of the parameter distributions for the population.}
\begin{tabular}{l c c c c} 
\hline 
Property & Min & Max & Mean & ${\sigma}$ \\
\hline 
Mass ratio $q$ & 0.03 & 0.90 & 0.48 & 0.28 \\
log$_{10}$(P$/$days) & 0.00 & 2.20 & 0.45 & 0.50 \\
Initial mass $/$ $M_{\odot}$ & 15.0 & 300.0 & 84.7 & 67.6 \\
Final mass $/$ $M_{\odot}$ & 8.5 & 46.6 & 18.7 & 8.6 \\
Delay time $/$ Myr & 2.5 & 16.0 & 4.5 & 2.1 \\
\end{tabular}
\label{tab:prog_table}
\end{table}

\begin{figure}
\centering
\includegraphics[width=0.95\columnwidth]{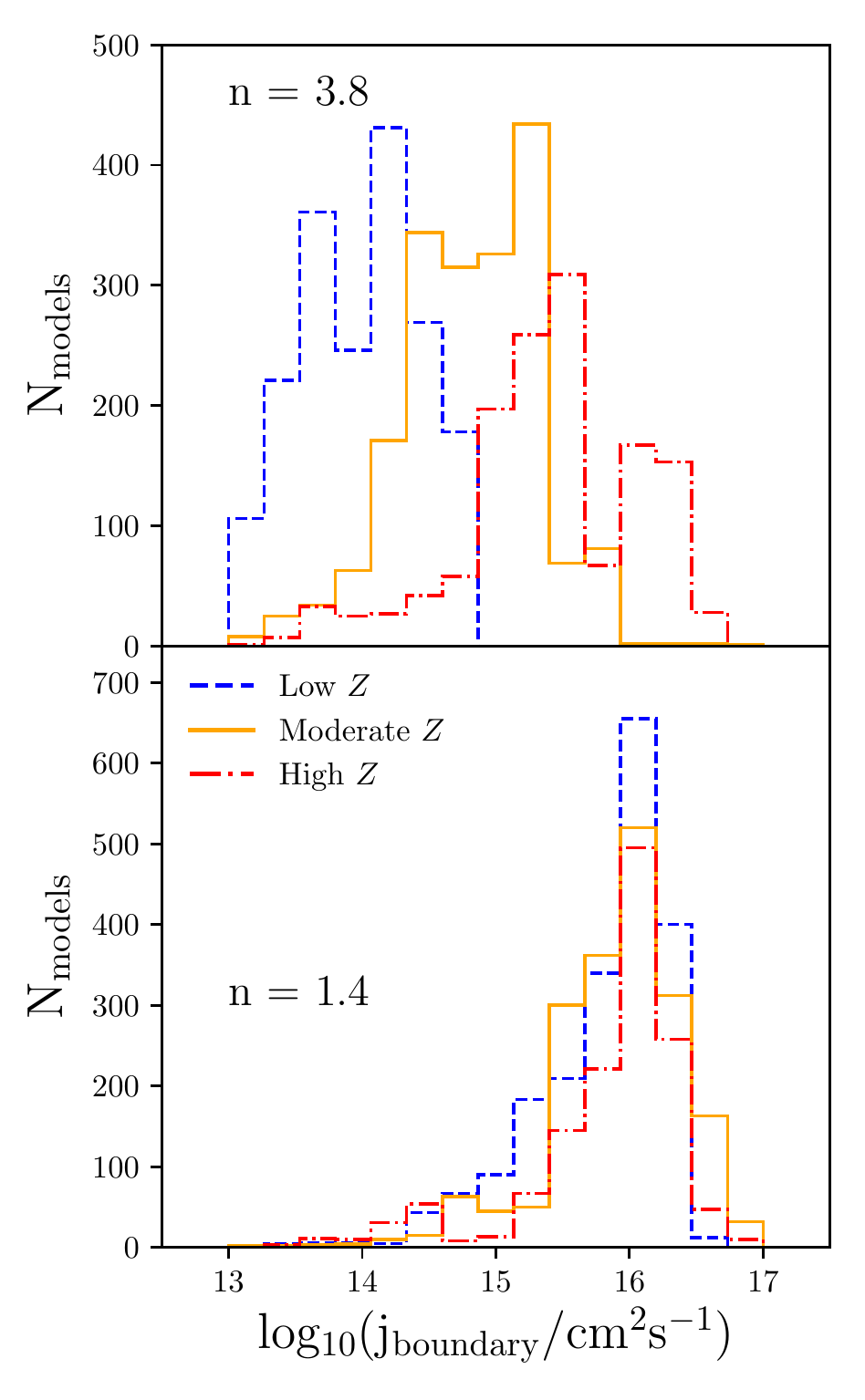}
\caption{The minimum specific angular momenta required for a GRB, measured at the remnant-ejecta boundary of each model at the point of collapse. These correspond to collapsing stars with minimum surface specific angular momenta $j_\mathrm{cut}$. The upper panel assumes a $Z^{3.8}$ metallicity dependence, and the lower assumes that $j_\mathrm{cut} \propto Z^{1.4}$. Detailed stellar interior models are used to calculate $j_\mathrm{boundary}$ as discussed in the text. The models, which include primary and secondary stars, are binned into three metallicity ranges: low ($10\times10^{-4} \leq Z \leq 0.004$), moderate ($0.006 \leq Z \leq 0.008$) and high ($0.010 \leq Z \leq 0.040$). The typical boundary momenta required are in the range log$_{10}(j_\mathrm{boundary} / \mathrm{cm}^{2}\mathrm{s}^{-1}){\sim}13-17$ for $n=3.8$, and $14.5-16.5$ for $n=1.4$, although these are minimum values and could easily be ${\sim}$100 times greater (see discussion in Section \ref{sec:grbproduction}).}
\label{fig:jcore}
\end{figure}

\subsection{Core angular momentum}
Finally, we turn our attention to the angular momentum distribution of the tidally spun models selected as GRB progenitors. The angular momentum cut applies to $j = {\Omega} \times R^{2}$, where $R$ is the radius of the star, however the quantity of interest for GRBs is the interior specific angular momentum. Theoretical modelling \citep{1993ApJ...405..273W,1999ApJ...524..262M} suggests that to launch a jet, $\gtrsim 16\,\mathrm{cm}^{2}\,\mathrm{s}^{-1}$ is required at the innermost stable orbit around the newly formed black hole. To estimate these $j$ values in our tidal GRB models, we make two assumptions. First, that the star has a constant rotational angular velocity ${\Omega}$ throughout its structure at the end point of our models. Secondly, that initial radius of the material which will form the accretion disk of the nascent black hole is at $R_\mathrm{boundary}$ - the radius in the pre-collapse star which encloses the post-collapse remnant mass. This material is assumed to retain its specific angular momentum during core-collapse. 

To calculate $R_\mathrm{boundary}$, we use the files output directly from the BPASS version of the STARS code, which includes information on the radial structure of the stellar models\footnote{Note: these detailed output files do not form part of the standard BPASS stellar model data release due to their data volume and technical complexity.}.  The boundary radius is calculated by summing shells of mass until the remnant mass is enclosed, from $r=0$ to $r=R_\mathrm{boundary}$. The specific angular momentum at this radius is then given by,
\begin{equation}
j_\mathrm{boundary} = \frac{J_\mathrm{shell}}{M_\mathrm{shell}} = \frac{2}{3} \Omega R_\mathrm{boundary}^{2}
\end{equation}
where the angular momentum and mass of the shell just outside the boundary radius are $J_\mathrm{shell}$ and $M_\mathrm{shell}$.

In Figure \ref{fig:jcore}, we show the specific angular momenta, evaluated at the remnant-ejecta boundary $r=R_\mathrm{boundary}$, for the tidal GRB models immediately before core-collapse in three metallicity bins. The angular velocity assumed in all cases is $j_\mathrm{cut}$/$R^{2}$, the minimum value allowed by our metallicity-dependent surface momentum cut. We show two cases, the first has a $Z^{3.8}$ dependence on this cut (the favoured value from our MCMC run), the other has $n=1.4$ (this is the lower, 1${\sigma}$ equivalent bound). We might expect no strong metallicity dependence on the angular momentum that a new-forming black hole requires to launch a jet, although an index of ${\sim}$1 may be expected for jet escape if envelope opacity is solely responsible for suppressing jets and hence dominating any $Z$ dependence \citep{2001A&A...369..574V,2005A&A...442..587V}.

The distributions shown in Figure \ref{fig:jcore} represents lower limits in that the angular momentum of any specific GRB progenitor may exceed the fitted cut level for the population. The distributions show scatter around $10^{13} - 10^{17}$cm\,s$^{-1}$ (a wide spread, $n=3.8$) and $10^{14.5} - 10^{16.5}$cm\,s$^{-1}$ (more peaked, $n=1.4$). We note that using a $n{\sim}1$ metallicity dependence preferentially shifts the low $Z$ models to greater specific angular momenta, and brings the three metallicity bins into good agreement.

\section{Discussion}\label{sec:discussion}
\subsection{The production of GRBs}\label{sec:grbproduction}
\begin{figure}
\centering
\includegraphics[width=0.95\columnwidth]{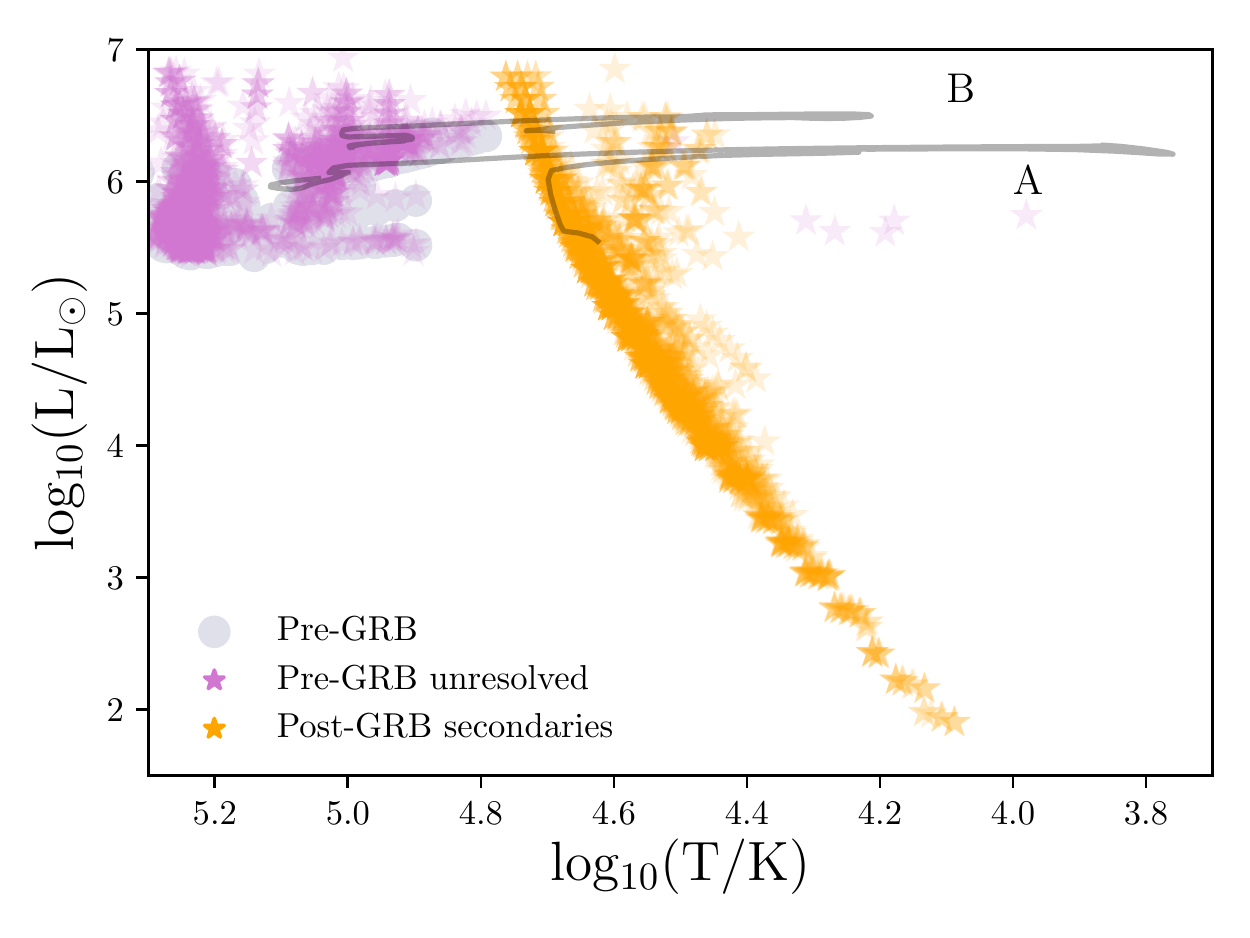}
\caption{Every model (primary and secondary) which produces a GRB via the tidal pathway, in the metallicity range $Z=0.008$ to $Z=0.020$, shown on the HR diagram. The model values for the individual pre core-collapse stars that go GRB are marked by grey circles. Also marked are the secondary stars left behind after a primary goes GRB (orange stars) and the unresolved systems that would be seen in pre-primary explosion imaging (purple stars). The shading represents the number density of models. Two example evolutionary tracks are overlaid. Track (A) is for a 50\,M$_{\odot}$ primary with a 45\,M$_{\odot}$ companion, starting with an orbital period of 0.2 days at a metallicity of 0.5\,Z$_{\odot}$. Track (B) follows a 150\,M$_{\odot}$ secondary star with an 11\,M$_{\odot}$ black hole companion, starting with an orbital period of 1.4 days, at 0.4\,Z$_{\odot}$.}
\label{fig:evolve}
\end{figure}

We have identified the subset of probability weighted stellar evolution models which are likely to generate a GRB through either quasihomogenous evolution or the results of tidal interactions modifying the angular momentum of the progenitor star. The location of the tidal GRB progenitors on the Hertzsprung-Russell diagram is shown in Figure \ref{fig:evolve}, for all models in the metallicity range $Z=0.008$ to $Z=0.020$. Purple stars indicate the predicted total optical light from the binary (i.e. both primary and secondary star, or secondary plus remnant in rare cases) immediately before the supernova explosion, where the more luminous component is assumed to dominate the temperature measurement. The yellow stars indicate the properties of the surviving binary companion expected to be observable after the GRB has faded. Grey circles represent the properties of individual progenitor stars immediately before core-collapse. In some cases, this progenitor is the secondary in the original ZAMS binary. 

The GRB progenitor binary component is often not responsible for all the light coming from an observed progenitor system; this is particularly true for the twin systems identified above, in which the secondary is likely to be very nearly as bright and evolved as the primary. Immediately prior to core-collapse, 5 per cent of progenitor systems between $0.4Z_{\odot}$ and $Z_{\odot}$ metallicity have a secondary star that is more luminous than the pre-explosion primary.

Progenitors end in the hot and bright region on the upper left, as luminous or more luminous than typical Wolf-Rayet stars seen in the Local Group \citep{2019arXiv190806238N}. Figure \ref{fig:evolve} and Table \ref{tab:prog_table} indicate that main sequence stars are the most frequent companions left behind after a primary star goes GRB in the models considered here. This is consistent with earlier findings. \citet{2017ApJ...842..125Z} predict the companions expected for stripped envelope (type IIb, Ib and Ic) supernovae using the {\sc binary\_c} rapid population synthesis code. They found that at $0.3Z_{\odot}$ metallicity, given their assumed IMF and binary parameters, ${\sim}$68 per cent of the progenitors should have a main sequence companion at the point of explosion, and the remainder have compact object companions. 

The modelled intrinsic GRB rate shown in Figure \ref{fig:grb_rate} is around ${\sim}10$ per cent of the type Ic SN rate expected in the local Universe ($z < 1$). This is consistent with previous estimates of this fraction \citep[][]{2007PASP..119.1211F} from observational data, corrected for selection effects. 

The key property in our model determining whether or not a massive stripped envelope star produces a GRB is its internal specific angular momentum. Our estimates (using $n=3.8$) for the minimum boundary specific angular momentum required to launch a GRB, log$_{10}(j_\mathrm{boundary} / \mathrm{cm}^{2}\mathrm{s}^{-1}) {\sim}13-17$, drops well below that expected from theory. This is also lower than the values found from detailed modelling of massive star interiors \citep{2000ApJ...528..368H,2012A&A...542A.113Y,2019EPJA...55..132F}. The BPASS models rapidly evolve through their final stages, and do not track the end phases of evolution in detail. The core properties we have used are therefore more representative of stellar structure at the end of core carbon burning, rather than at collapse, and stellar cores can contract further by around two orders of magnitude in density after carbon burning \citep[][]{2019arXiv190807762E}, potentially raising the specific angular momentum of a given shell. During this time, the assumption of solid body rotation almost certainly breaks down, but circumstances in which the envelope rotates faster than the core are highly unlikely. The $j$ values shown in Figure \ref{fig:jcore} are already lower limits since we have adopted our log$_{10}(j_\mathrm{cut})$ minimum value. Actual model values extend up to log$_{10}(j){\sim}19$.
If a further core contraction occurs, as expected, the true values may be higher still by several orders of magnitude \citep[e.g.][]{2000ApJ...528..368H}. Even a modest cumulative increase of 2 dex would push the distribution towards log$_{10}(j_\mathrm{boundary} / \mathrm{cm}^{2}\mathrm{s}^{-1})$ ${\sim}15-19$, in better agreement with theoretical predictions \citep{1993ApJ...405..273W,1999A&AS..138..499W}. This is the first time this critical threshold has been derived from observational data, albeit through fitting with stellar models. It arises naturally, without fine tuning, from our two-pathway model and associated assumptions.

\subsection{The metallicity dependence of GRBs}
\begin{figure}
\centering
\includegraphics[width=0.95\columnwidth]{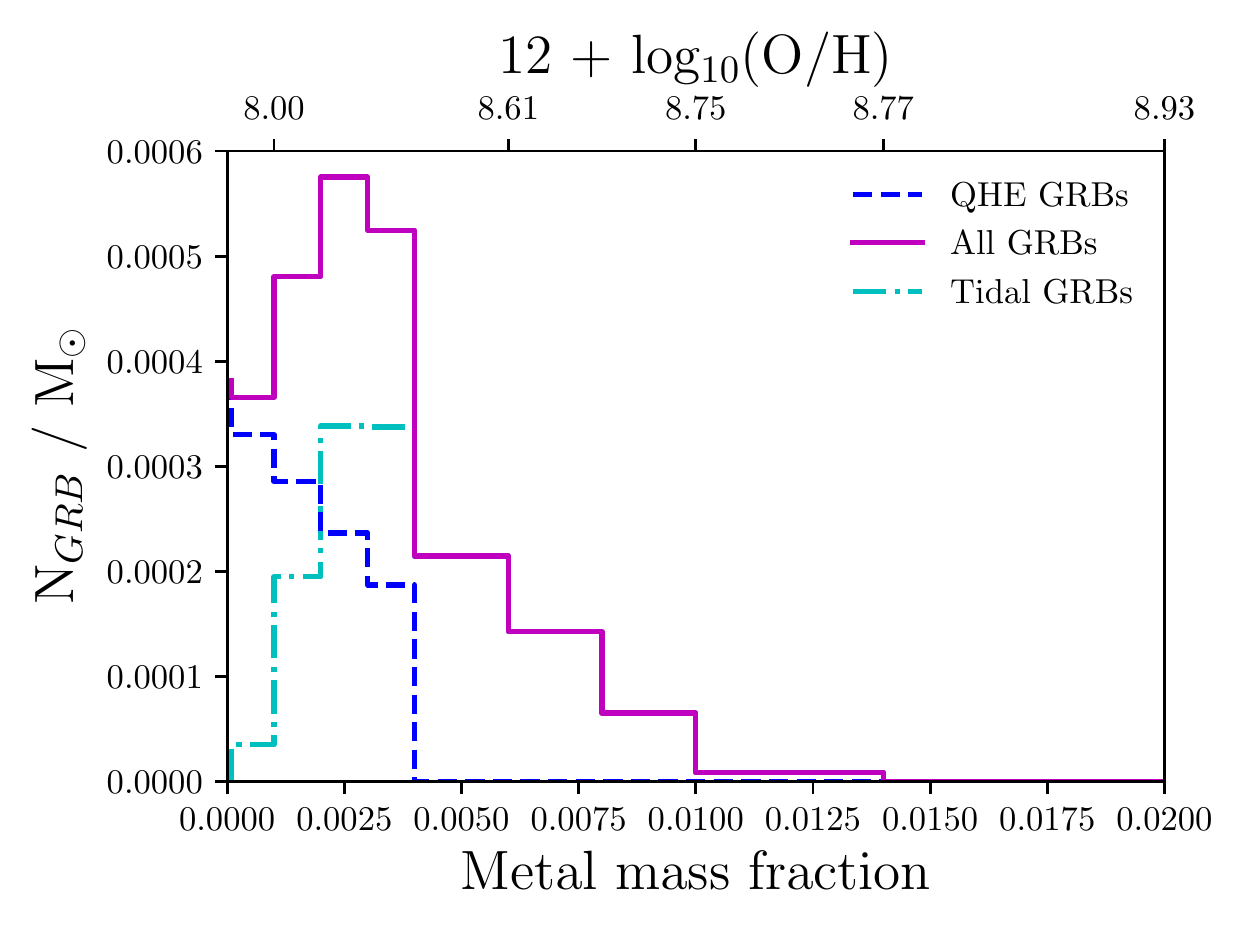}
\caption{The efficiency of GRB progenitor production across metallicity. For a given mass of gas at a single metallicity, we would expect $N_\mathrm{GRB}$ progenitors to be formed per Solar mass of gas. We show this number for all GRB progenitors, and separately for the QHE pathway. QHE progenitors show a smooth decline in occurrence rate as metallicity increases, before the 0.2\,Z$_{\odot}$ cut-off is reached. The tidally-induced pathways, however, peak around this metallicity.}
\label{fig:efficiency}
\end{figure}

GRBs show a deficit with respect to the cosmic volume-averaged star formation rate density at low-redshift, but occur in proportion to it at $z{\sim}3$ and above \citep{2019arXiv190807762E}. This is reflecting the well documented high-metallicity aversion of GRBs \citep[e.g.][]{2006Natur.441..463F,2015ApJ...809...76G,2019arXiv190100872M}, which is demonstrated further in Figure \ref{fig:efficiency}. Here we show the GRB progenitor production efficiency, defined as GRB progenitor fraction per unit star formation at each metallicity, given our assumed IMF and binary parameters. There is a clear aversion to high metallicity. The QHE pathway becomes increasingly scarce as metallicity increases, until the $Z=0.004$ mass fraction cut-off implemented by BPASS is reached. Tidal pathways, on the other hand, peak at around 20 per cent of Solar. This is likely due to a trade-off between tides being more effective when the stellar envelope is extended, and increasing angular momentum loss through winds.

We can further test this by examining the metallicity distribution of predicted progenitors.  In Figure \ref{fig:metals} we compared our synthetic GRB  metallicity distributions to those from observed host galaxies, making the assumption that the metallicity of the host stellar population changes by a negligible amount over the lifetime of a GRB progenitor. The resultant distribution accounts for the cosmic star formation history and model weightings, and is broadly consistent with the observed host galaxy population.

Given that our predicted metallicity distribution and volumetric rate evolution is in good agreement with observations, using our best-fit Bayesian analysis values, thought must be given to the strong metallicity dependence suggested by that analysis, $Z\propto Z^n$ where  $n=3.8^{+2.1}_{-2.4}$. The uncertainties on this parameter are large, and at their upper end suggest a far stronger metallicity dependence for GRBs than seen in earlier work \citep[e.g.][]{2015ApJ...802..103T}. 

We have chosen in this analysis to place the metallicity dependence as a modifier of the angular momentum threshold. However, we expect that the minimum angular momentum threshold at the remnant-ejecta boundary (i.e. at the point at which a jet is launched) is independent of metallicity. Thus the effective threshold encodes information not on jet launching but rather on the metallicity dependence of whether the jet, once launched, can escape the stellar envelope or is stifled before producing a GRB.

Successful jet break-out relies on a number of different properties of the collapsing star. In principle, it will depend on the column density of the material through which the jet must tunnel (i.e. the density and thickness of the stellar envelope at point of collapse), the probability of photons interacting with that material (i.e. the envelope opacity), and also the timescale for jet escape (i.e. if the central engine deactivates before breakout is achieved there will no visible event). 

We calculate the envelope column density for each GRB progenitor model by summing the density of mass shells from the core-envelope boundary out to the surface, multiplying by the shell thickness at each stage. The detailed STARS outputs are again used. We find that although high metallicity stars do have some of the highest columns, there is no distinct trend across the full metallicity range. We therefore cannot confidently attribute the high $Z$ dependence suggested by our model fitting analysis to a column density effect.

The opacity of the ejecta which must be tunnelled through is also dependent on the metal content of the stellar material. For each photon, its probability of interaction scales with the number of possible electron energy level transitions which it may be able to excite. Heavier metallic elements, with their extended electron shells, dominate this probability, and so the opacity scales broadly linearly with the abundance of iron group elements. For higher opacities, more energy is dissipated from the jet in exciting electrons and so an initially more relativistic, more collimated jet is required. This corresponds to a greater reservoir of angular momentum also required to successfully tunnel through the envelope. Opacity is roughly proportional to metal mass fraction $Z$  \citep{2001A&A...369..574V,2005A&A...442..587V}, and would lead to $n\sim1$ in our formalism. It therefore cannot reproduce the $n=3.8$ dependence that results as a best fit from our Bayesian inference.

If instead we assume the lower 1\,$\sigma$ bound from our analysis ($n=1.4$), the resultant $j$ distribution in Figure \ref{fig:jcore} is good agreement with theory and removes any clear metallicity dependence at the point where the jet launches. A lower $n$ index also introduces bursts at $Z_{\odot}$ and above, improving the metallicity distribution agreement with observations. The question then is why $Z^{3.8}$ was favoured by the Bayesian inference. In Figure \ref{fig:grb_rate}, a model which tracks the assumed cosmic star formation history overestimates the GRB rate at low redshift and underestimates it at high redshift. Larger powers of $Z$ rectify this, allowing more bursts to occur at low metallicity (fewer at high values) - boosting the high redshift rate (diminishing it at low redshift). However, a $Z^n$ dependence that is too strong begins to remove too many bursts and makes it difficult to reproduce the observed population numbers with plausible $E_\mathrm{iso}$ and ${\theta}$ corrections. The $n=3.8^{+2.1}_{-2.4}$ best fit arises from the combination of these factors, and inherits all their uncertainties. In particular we have not considered the substantial uncertainty on the cosmic volume-averaged metallicity evolution and its scatter as a function of redshift, nor uncertainties in the somewhat better constrained cosmic star formation rate density history. The inclusion of other progenitor pathways, which could have a different occurrence rate over cosmic history, may also affect the best fit index $n$. Thus while we clearly favour a non-zero metallicity dependence, we caution against over-interpretation of the best-fit value, and note that an $n\sim1$ dependence is both permitted and explained by a physically plausible mechanism.

We note that the derived metallicity dependence arises from a fit to the evolution in the inferred GRB production rate. The fraction of very low metallicity (i.e $Z<0.001$) models contributing to the volume averaged rate is very low and so the derived $Z^n$ dependence is likely poorly constrained or not applicable at these very low metallicities. Indeed, it is possible that below a minimum metallicity threshold non-iron-dominated opacities begin to dominate and the metallicity-dependence will plateau. There is very little difference in the stellar atmosphere opacity tables used by BPASS in this regime \citep{2017PASA...34...58E}.

\subsection{Uncertainties in the stellar modelling}
Throughout this paper, we have made  assumptions and simplifications with regards to the tidal evolution of binary systems, and the structure of high mass stars. For each model an initial weighting was selected from the IMF and binary parameter distributions, and assigned a value for an initial spin. The default BPASS IMF incorporates stars up to $300M_{\odot}$ at ZAMS, in order to accommodate the rare but important Very Massive Star population seen in the Local Group. However the occurrence of such stars and the distribution of binary parameters may well be metallicity-dependent. This is unconstrained by observations and so not accounted for in the current version of BPASS \citep[v2.2.1,][]{2018MNRAS.479...75S,2019A&A...621A.105S}.

The initial stellar angular velocity chosen for every primary model was 40 per cent of the Keplarian value at the equator \citep[a typical value for OB stars][]{2019arXiv190503359D}. While a distribution of velocities could be sampled, the exact choice of ${\Omega}$ does not have a significant impact on tidal evolution \citep[also seen by][]{2017A&A...601A..29Z},  as there is typically much more angular momentum stored in the orbit than in stellar spins. We also assume that the binaries start in circular orbits with their spin vectors aligned, so that only $\dot{a}$ and $\dot{{\Omega}}$ need be considered.

Given that most significant changes due to tides occur with the onset of mass transfer \citep{2002MNRAS.329..897H}, changes to evolution before that stage of factor a few are unlikely to have a major impact on our results. This is relevant to our assumptions concerning stellar structure. We take the value of the apsidal motion constant, $k$ (which dependes on internal structure) to be 0.05 for every model. The theoretical modelled range of values is 0.01-0.1 \citep{1975A&A....41..329Z}. ${\Delta}{\Omega}$ and ${\Delta}a$ both scale with $k^{-1}$. Varying $k$ by a factor of a few will have similarly small effects on these parameters.
For the system shown in Figure \ref{fig:tideexam}, varying $k$ from 0.01 to 0.1 makes no difference to the size or timing of the model jump made by our tidal algorithm, and the system ends in the same state independent of $k$. 

We note that the secondary star in our tidal prescription is treated as a point mass. For evolved binaries, where this object is a compact remnant, the approximation is reasonable. Otherwise, tidal distortion of the secondary may play a role in the system evolution, even if it remains inside its Roche lobe. One possible outcome of spin-orbit interactions is a merger. For this to occur, the total system angular momentum - the sum of the orbital and spin components - must be less than a critical value. Angular momentum is then transferred from the orbit to the spin, but the stars merge before a synchronisation can occur \citep{2019ses..book.....E}. Given our starting velocity of $0.4{\Omega}_\mathrm{crit}$, mergers solely due to tides are rare in our model set, only occurring in very tight binaries. 

Finally, a major assumption in this analysis is that most black hole producing core-collapse events do not produce a visible supernova. The picture is more complex than there being a simple mass cutoff, as \citet{2012ApJ...754...76D} discuss in their work on black hole production in GRBs, with other factors including the rotationally-driven magnetic field and envelope structure. Core compactness is also important criteria for successful supernovae \citep{2016ApJ...821...38S,2019arXiv191001641E}, leading to so-called `islands of explodability'. We address the core compactness by assuming that the radius of interest within the pre-collapse star is that which encloses the eventual remnant mass. This is reasonable approximation, provided that there is negligible fallback accretion. We also assume that the core and envelope co-rotate, with a flat angular velocity profile throughout the entire star. As previously discussed, any deviation from this will likely result in the core spinning faster than assumed. Such an increase in core spin would increase the number of rapidly spinning cores, and therefore increase the inferred $j_\mathrm{cut,\odot}$ threshold, improving agreement with collapsar theory.

\subsection{Magnetars as GRB central engines}
Although we have assumed in this paper that GRBs operate under the collapsar mechanism, which requires black hole formation upon core-collapse, it is possible that newly formed magnetars might be able to launch jets too. Indeed, there is some evidence from GRB-SN energetics that this is the case \citep{2014MNRAS.443...67M}. Therefore, we relaxed our remnant mass constraint to 1.4M$_{\odot}$, allowing for the possibility that neutron stars are formed in core-collapse GRBs, and re-performed the MCMC analysis as described in section \ref{sec:mcmc}. We can see from Figure \ref{fig:changes} that while black holes form readily at low $Z$, type Ic SNe with neutron star remnants are rare. By including neutron star producing Ic events, we are preferentially adding models at high metallicity. Because these end their lives spinning slower, most of those added are then rejected by the angular momentum cut. Therefore, neutron star forming events can contribute, but are not an important pathway if black hole forming GRBs are also considered. The small difference their inclusion makes is demonstrated in Figure \ref{fig:C1} of the online appendix.

We then took the extreme case that GRBs can {\it only} occur if a neutron star is formed in core-collapse. An MCMC run under this assumption, as described in section \ref{sec:mcmc}, yielded ${\theta}=12.8^{+6.4}_{-4.6}$ degrees, log$_{10}($E$_\mathrm{low}$/erg$)=49.6^{+0.6}_{-0.5}$, log$_{10}$(j$_\mathrm{cut,\odot}$/cm$^2$\,s$^{-1}$)=$18.8^{+0.3}_{-0.4}$ and $n=2.2^{+1.0}_{-1.3}$. Because stripped envelope, fast-spinning, neutron star forming progenitors are rarer (than the equivalent black hole forming events), the best-fit parameter values are naturally quite different, as shown in Figure \ref{fig:C1} of the appendix. The most notable among them is E$_\mathrm{low}$ - the minimum isotropic equivalent energy is nearly two orders of magnitude higher than the faintest GRBs observed. Therefore, barring strong co-variances between the four model parameters which have not been considered, or a very different luminosity function than that assumed, this appears to disfavour magnetars as the sole central engines capable of launching GRBs. Furthermore, the metallicity distribution assuming just magnetar engines is inconsistent with the host galaxy data (see Figure \ref{fig:C2} of the online appendix). There is narrow range of metallicity in which stars lose enough mass to produce neutron star remnants, but retain enough to not fully spin-down. Anderson-Darling $p$-values between the synthetic distribution and the hosts data sets are all $<0.05$, rejecting the null hypothesis that they are drawn from the same distribution at the 2${\sigma}$ level.

\subsection{Future possibilities}
In Figure \ref{fig:evolve} we predicted that the progenitors of GRBs induced through tidal evolution are amongst the most luminous stripped envelope stars. They are also extremely hot. Thus despite their high luminosities they may be challenging to observe, requiring ultraviolet, or very blue-sensitive optical, detectors. Supernova 2017ein, the only SN Ic to have a progenitor system identified so far, occurred at a redshift ${\sim}$0.0027, with the candidate identified in archival {\it Hubble Space Telescope (HST)} imaging \citep{2018ApJ...860...90V,2018MNRAS.480.2072K}. The star, if single, had an absolute F555W magnitude of $-7.5$. The nearest GRB confirmed so far is GRB\,980425, associated with SN\,1998bw. This occurred at $z{\sim}0.0085$ \citep{1998Natur.395..670G}, corresponding to a luminosity distance which is approximately three times greater than SN\,2017ein. If SN\,2017ein had occurred at the distance of SN\,1998bw, the progenitor (or unresolved progenitor system) would have had an apparent magnitude of ${\sim}25.5$ in this band.

It is also interesting to consider the possibility of surviving companions being seen once the supernova and GRB afterglow have faded. We have identified these companions as most-likely main sequence stars, with a slightly preference for large mass ratios or twin systems. Such stars are faint, but not beyond the realms of possibility for {\it HST} or upcoming facilities such as the {\it European Extremely Large Telescope (E-ELT)} or the {\it James Webb Space Telescope (JWST)}. The low metallicity QHE pathway is restricted to high redshifts where these kind of observations are unfeasible. However, the tidal GRB pathways we have identified, where tides maintain spin in a stripped envelope progenitor, are a plausible target. 

Although we can reproduce the observed GRB rate using mass-transfer QHE and tidally spun-up pathways, it is worth considering other, rarer possibilities. A number of such pathways have been identified in the literature, including the merger of helium stars with a companion \citep{2005ApJ...623..302F,2008A&A...484..831D,2013ApJ...764..166D,2014ApJ...782....7D}, binary-driven hypernovae \citep{2019arXiv190506050R} and white dwarf-compact object mergers \citep{1999ApJ...520..650F,2009A&A...498..501C}. 

Another scenario we have not considered is that tidally spun-up stars could enter the QHE regime \citep{2013ApJ...764..166D,2016A&A...585A.120S,2016MNRAS.460.3545D,2016MNRAS.458.2634M,2016A&A...585A.120S,2016A&A...588A..50M,2017A&A...604A..55M} - in our formalism the only way to induce QHE is through mass transfer at very low metallicity. Tidal interactions may in some cases lead to increased mass transfer and QHE, and there is an overall slight increase in the frequency of QHE occurring when tides are considered (see Figure \ref{fig:A1} of the online appendix). This pathway has received significant attention in the literature as a route to producing massive black hole binaries. However, because the low-metallicity, low mass-loss GRB requirement holds even if QHE is tidally induced, such events will not significantly contribute to the overall observed GRB rate. We note that our analysis above has produced a good match to the observed volumetric rates and properties of known GRBs without assuming exotic or extremely rare progenitor pathways. We therefore leave a full BPASS study of the viability and occurrence rates of rarer alternative pathways for future consideration.

Finally, an extension of the BPASS modelling project is the synthesis of SN lightcurves from the stellar structure models at the point of explosion \citep[CURVEPOPS,][]{2018PASA...35...49E,2019arXiv190807762E}. In a similar fashion, the lightcurves arising from the type Ic supernovae of our candidate GRB progenitors could be synthesised, and a comparison between the CURVEPOPS lightcurves and GRB-SN lightcurves made. This will be explored in future work.

\section{Conclusions}\label{sec:conc}
We have used the binary stellar evolution models of BPASS to investigate the nature of supernova and core-collapse gamma-ray burst progenitors. By considering tidal interactions, and applying a prescription for the cosmic star formation rate history, we have shown that two binary pathways can explain the bulk of the observed GRB population. The first involves secondary stars spun up by accretion into a quasi-homogeneous state, and the second occurs when massive stars in binaries have their envelope removed while tides maintain the required angular momentum for jet production. This two pathway model can reproduce the rates as a function of redshift and the observed host metallicity spread, using reasonable parameter distributions for the jet opening angle and isotropic equivalent energy. The model predicts minimum angular momentum requirements which are in general agreement with collapsar GRB theory. Finally, predictions are made for the nature of the progenitor systems and their companions, which may be testable with next generation facilities. The exploratory work presented here does not represent a definitive or unique solution to the GRB progenitor problem, and there are several areas which need to be further explored or developed in future work. None the less, our approach can simultaneously reproduce several aspects of long GRB theory and observation, demonstrating the utility of population synthesis models in transient progenitor studies.

\section*{Acknowledgements}
AAC is supported by Science and Technology Facilities Council (STFC) grant 1763016. AAC also thanks the William Edwards educational charity. ERS is supported by STFC consolidated grant ST/P000495/1. JJE acknowledges support from the University of Auckland and also the Royal Society Te Ap\={a}rangi of New Zealand under Marsden Fund grant UOA1818.

We acknowledge use of Python algorithms including the {\sc corner} module \citep{2016JOSS....1...24F}, the excellent Python-based introduction to Bayesianism by \citet{2014arXiv1411.5018V}, and the use of Ned Wright and James Schombert's Python cosmology calculator \citep{2006PASP..118.1711W}.




\bibliographystyle{mnras}
\bibliography{grbs_bpass} 

\begin{thebibliography}{}
\makeatletter
\relax
\def\mn@urlcharsother{\let\do\@makeother \do\$\do\&\do\#\do\^\do\_\do\%\do\~}
\def\mn@doi{\begingroup\mn@urlcharsother \@ifnextchar [ {\mn@doi@}
  {\mn@doi@[]}}
\def\mn@doi@[#1]#2{\def\@tempa{#1}\ifx\@tempa\@empty \href
  {http://dx.doi.org/#2} {doi:#2}\else \href {http://dx.doi.org/#2} {#1}\fi
  \endgroup}
\def\mn@eprint#1#2{\mn@eprint@#1:#2::\@nil}
\def\mn@eprint@arXiv#1{\href {http://arxiv.org/abs/#1} {{\tt arXiv:#1}}}
\def\mn@eprint@dblp#1{\href {http://dblp.uni-trier.de/rec/bibtex/#1.xml}
  {dblp:#1}}
\def\mn@eprint@#1:#2:#3:#4\@nil{\def\@tempa {#1}\def\@tempb {#2}\def\@tempc
  {#3}\ifx \@tempc \@empty \let \@tempc \@tempb \let \@tempb \@tempa \fi \ifx
  \@tempb \@empty \def\@tempb {arXiv}\fi \@ifundefined
  {mn@eprint@\@tempb}{\@tempb:\@tempc}{\expandafter \expandafter \csname
  mn@eprint@\@tempb\endcsname \expandafter{\@tempc}}}

\bibitem[\protect\citeauthoryear{{Adams}, {Kochanek}, {Gerke}, {Stanek}  \&
  {Dai}}{{Adams} et~al.}{2017}]{2017MNRAS.468.4968A}
{Adams} S.~M.,  {Kochanek} C.~S.,  {Gerke} J.~R.,  {Stanek} K.~Z.,   {Dai} X.,
  2017, \mn@doi [\mnras] {10.1093/mnras/stx816}, \href
  {http://adsabs.harvard.edu/abs/2017MNRAS.468.4968A} {468, 4968}

\bibitem[\protect\citeauthoryear{{Ajello} et~al.,}{{Ajello}
  et~al.}{2019}]{2019ApJ...878...52A}
{Ajello} M.,  et~al., 2019, \mn@doi [\apj] {10.3847/1538-4357/ab1d4e}, \href
  {https://ui.adsabs.harvard.edu/abs/2019ApJ...878...52A} {878, 52}

\bibitem[\protect\citeauthoryear{{Aldering}, {Humphreys}  \&
  {Richmond}}{{Aldering} et~al.}{1994}]{1994AJ....107..662A}
{Aldering} G.,  {Humphreys} R.~M.,   {Richmond} M.,  1994, \mn@doi [\aj]
  {10.1086/116886}, \href
  {https://ui.adsabs.harvard.edu/abs/1994AJ....107..662A} {107, 662}

\bibitem[\protect\citeauthoryear{{Allende Prieto}, {Lambert}  \&
  {Asplund}}{{Allende Prieto} et~al.}{2001}]{2001ApJ...556L..63A}
{Allende Prieto} C.,  {Lambert} D.~L.,   {Asplund} M.,  2001, \mn@doi [\apjl]
  {10.1086/322874}, \href
  {https://ui.adsabs.harvard.edu/abs/2001ApJ...556L..63A} {556, L63}

\bibitem[\protect\citeauthoryear{{Asplund}, {Grevesse}, {Sauval}  \&
  {Scott}}{{Asplund} et~al.}{2009}]{2009ARA&A..47..481A}
{Asplund} M.,  {Grevesse} N.,  {Sauval} A.~J.,   {Scott} P.,  2009, \mn@doi
  [\araa] {10.1146/annurev.astro.46.060407.145222}, \href
  {https://ui.adsabs.harvard.edu/abs/2009ARA&A..47..481A} {47, 481}

\bibitem[\protect\citeauthoryear{{Auchettl}, {Lopez}, {Badenes},
  {Ramirez-Ruiz}, {Beacom}  \& {Holland -Ashford}}{{Auchettl}
  et~al.}{2019}]{2019ApJ...871...64A}
{Auchettl} K.,  {Lopez} L.~A.,  {Badenes} C.,  {Ramirez-Ruiz} E.,  {Beacom}
  J.~F.,   {Holland -Ashford} T.,  2019, \mn@doi [\apj]
  {10.3847/1538-4357/aaf395}, \href
  {https://ui.adsabs.harvard.edu/abs/2019ApJ...871...64A} {871, 64}

\bibitem[\protect\citeauthoryear{{Barnes}, {Duffell}, {Liu}, {Modjaz},
  {Bianco}, {Kasen}  \& {MacFadyen}}{{Barnes}
  et~al.}{2018}]{2018ApJ...860...38B}
{Barnes} J.,  {Duffell} P.~C.,  {Liu} Y.,  {Modjaz} M.,  {Bianco} F.~B.,
  {Kasen} D.,   {MacFadyen} A.~I.,  2018, \mn@doi [\apj]
  {10.3847/1538-4357/aabf84}, \href
  {https://ui.adsabs.harvard.edu/abs/2018ApJ...860...38B} {860, 38}

\bibitem[\protect\citeauthoryear{{Caito}, {Bernardini}, {Bianco}, {Dainotti},
  {Guida}  \& {Ruffini}}{{Caito} et~al.}{2009}]{2009A&A...498..501C}
{Caito} L.,  {Bernardini} M.~G.,  {Bianco} C.~L.,  {Dainotti} M.~G.,  {Guida}
  R.,   {Ruffini} R.,  2009, \mn@doi [\aap] {10.1051/0004-6361/200810676},
  \href {https://ui.adsabs.harvard.edu/abs/2009A&A...498..501C} {498, 501}

\bibitem[\protect\citeauthoryear{{Cano}, {Wang}, {Dai}  \& {Wu}}{{Cano}
  et~al.}{2017}]{2017AdAst2017E...5C}
{Cano} Z.,  {Wang} S.-Q.,  {Dai} Z.-G.,   {Wu} X.-F.,  2017, \mn@doi [Advances
  in Astronomy] {10.1155/2017/8929054}, \href
  {https://ui.adsabs.harvard.edu/abs/2017AdAst2017E...5C} {2017, 8929054}

\bibitem[\protect\citeauthoryear{{Cantiello}, {Yoon}, {Langer}  \&
  {Livio}}{{Cantiello} et~al.}{2007}]{2007A&A...465L..29C}
{Cantiello} M.,  {Yoon} S.~C.,  {Langer} N.,   {Livio} M.,  2007, \mn@doi
  [\aap] {10.1051/0004-6361:20077115}, \href
  {https://ui.adsabs.harvard.edu/abs/2007A&A...465L..29C} {465, L29}

\bibitem[\protect\citeauthoryear{{Dessart}, {Hillier}, {Livne}, {Yoon},
  {Woosley}, {Waldman}  \& {Langer}}{{Dessart}
  et~al.}{2011}]{2011MNRAS.414.2985D}
{Dessart} L.,  {Hillier} D.~J.,  {Livne} E.,  {Yoon} S.-C.,  {Woosley} S.,
  {Waldman} R.,   {Langer} N.,  2011, \mn@doi [\mnras]
  {10.1111/j.1365-2966.2011.18598.x}, \href
  {https://ui.adsabs.harvard.edu/abs/2011MNRAS.414.2985D} {414, 2985}

\bibitem[\protect\citeauthoryear{{Dessart}, {Hillier}, {Li}  \&
  {Woosley}}{{Dessart} et~al.}{2012a}]{2012MNRAS.424.2139D}
{Dessart} L.,  {Hillier} D.~J.,  {Li} C.,   {Woosley} S.,  2012a, \mn@doi
  [\mnras] {10.1111/j.1365-2966.2012.21374.x}, \href
  {https://ui.adsabs.harvard.edu/abs/2012MNRAS.424.2139D} {424, 2139}

\bibitem[\protect\citeauthoryear{{Dessart}, {O'Connor}  \& {Ott}}{{Dessart}
  et~al.}{2012b}]{2012ApJ...754...76D}
{Dessart} L.,  {O'Connor} E.,   {Ott} C.~D.,  2012b, \mn@doi [\apj]
  {10.1088/0004-637X/754/1/76}, \href
  {https://ui.adsabs.harvard.edu/abs/2012ApJ...754...76D} {754, 76}

\bibitem[\protect\citeauthoryear{{Detmers}, {Langer}, {Podsiadlowski}  \&
  {Izzard}}{{Detmers} et~al.}{2008}]{2008A&A...484..831D}
{Detmers} R.~G.,  {Langer} N.,  {Podsiadlowski} P.,   {Izzard} R.~G.,  2008,
  \mn@doi [\aap] {10.1051/0004-6361:200809371}, \href
  {https://ui.adsabs.harvard.edu/abs/2008A&A...484..831D} {484, 831}

\bibitem[\protect\citeauthoryear{{Dufton}, {Evans}, {Hunter}, {Lennon}  \&
  {Schneider}}{{Dufton} et~al.}{2019}]{2019arXiv190503359D}
{Dufton} P.~L.,  {Evans} C.~J.,  {Hunter} I.,  {Lennon} D.~J.,   {Schneider}
  F.~R.~N.,  2019, arXiv e-prints, \href
  {https://ui.adsabs.harvard.edu/abs/2019arXiv190503359D} {p. arXiv:1905.03359}

\bibitem[\protect\citeauthoryear{{Eldridge} \& {Maund}}{{Eldridge} \&
  {Maund}}{2016}]{2016MNRAS.461L.117E}
{Eldridge} J.~J.,  {Maund} J.~R.,  2016, \mn@doi [\mnras]
  {10.1093/mnrasl/slw099}, \href
  {https://ui.adsabs.harvard.edu/abs/2016MNRAS.461L.117E} {461, L117}

\bibitem[\protect\citeauthoryear{{Eldridge} \& {Tout}}{{Eldridge} \&
  {Tout}}{2004}]{2004MNRAS.353...87E}
{Eldridge} J.~J.,  {Tout} C.~A.,  2004, \mn@doi [\mnras]
  {10.1111/j.1365-2966.2004.08041.x}, \href
  {http://adsabs.harvard.edu/abs/2004MNRAS.353...87E} {353, 87}

\bibitem[\protect\citeauthoryear{{Eldridge} \& {Tout}}{{Eldridge} \&
  {Tout}}{2019}]{2019ses..book.....E}
{Eldridge} J.~J.,  {Tout} C.~A.,  2019, {The Structure and Evolution of Stars}.
World Scientific, \mn@doi{10.1142/p974}

\bibitem[\protect\citeauthoryear{{Eldridge} \& {Xiao}}{{Eldridge} \&
  {Xiao}}{2019}]{2019MNRAS.485L..58E}
{Eldridge} J.~J.,  {Xiao} L.,  2019, \mn@doi [\mnras] {10.1093/mnrasl/slz030},
  \href {https://ui.adsabs.harvard.edu/abs/2019MNRAS.485L..58E} {485, L58}

\bibitem[\protect\citeauthoryear{{Eldridge}, {Langer}  \& {Tout}}{{Eldridge}
  et~al.}{2011}]{2011MNRAS.414.3501E}
{Eldridge} J.~J.,  {Langer} N.,   {Tout} C.~A.,  2011, \mn@doi [\mnras]
  {10.1111/j.1365-2966.2011.18650.x}, \href
  {https://ui.adsabs.harvard.edu/abs/2011MNRAS.414.3501E} {414, 3501}

\bibitem[\protect\citeauthoryear{{Eldridge}, {Fraser}, {Smartt}, {Maund}  \&
  {Crockett}}{{Eldridge} et~al.}{2013}]{2013MNRAS.436..774E}
{Eldridge} J.~J.,  {Fraser} M.,  {Smartt} S.~J.,  {Maund} J.~R.,   {Crockett}
  R.~M.,  2013, \mn@doi [\mnras] {10.1093/mnras/stt1612}, \href
  {http://adsabs.harvard.edu/abs/2013MNRAS.436..774E} {436, 774}

\bibitem[\protect\citeauthoryear{{Eldridge}, {Stanway}, {Xiao}, {McClelland},
  {Taylor}, {Ng}, {Greis}  \& {Bray}}{{Eldridge}
  et~al.}{2017}]{2017PASA...34...58E}
{Eldridge} J.~J.,  {Stanway} E.~R.,  {Xiao} L.,  {McClelland} L.~A.~S.,
  {Taylor} G.,  {Ng} M.,  {Greis} S.~M.~L.,   {Bray} J.~C.,  2017, \mn@doi
  [\pasa] {10.1017/pasa.2017.51}, \href
  {http://adsabs.harvard.edu/abs/2017PASA...34...58E} {34, e058}

\bibitem[\protect\citeauthoryear{{Eldridge}, {Xiao}, {Stanway}, {Rodrigues}  \&
  {Guo}}{{Eldridge} et~al.}{2018}]{2018PASA...35...49E}
{Eldridge} J.~J.,  {Xiao} L.,  {Stanway} E.~R.,  {Rodrigues} N.,   {Guo} N.~Y.,
   2018, \mn@doi [\pasa] {10.1017/pasa.2018.47}, \href
  {https://ui.adsabs.harvard.edu/abs/2018PASA...35...49E} {35, 49}

\bibitem[\protect\citeauthoryear{{Eldridge}, {Guo}, {Rodrigues}, {Stanway}  \&
  {Xiao}}{{Eldridge} et~al.}{2019a}]{2019arXiv190807762E}
{Eldridge} J.~J.,  {Guo} N.~Y.,  {Rodrigues} N.,  {Stanway} E.~R.,   {Xiao} L.,
   2019a, arXiv e-prints, \href
  {https://ui.adsabs.harvard.edu/abs/2019arXiv190807762E} {p. arXiv:1908.07762}

\bibitem[\protect\citeauthoryear{{Eldridge}, {Stanway}  \& {Tang}}{{Eldridge}
  et~al.}{2019b}]{2019MNRAS.482..870E}
{Eldridge} J.~J.,  {Stanway} E.~R.,   {Tang} P.~N.,  2019b, \mn@doi [\mnras]
  {10.1093/mnras/sty2714}, \href
  {https://ui.adsabs.harvard.edu/abs/2019MNRAS.482..870E} {482, 870}

\bibitem[\protect\citeauthoryear{{Ertl}, {Woosley}, {Sukhbold}  \&
  {Janka}}{{Ertl} et~al.}{2019}]{2019arXiv191001641E}
{Ertl} T.,  {Woosley} S.~E.,  {Sukhbold} T.,   {Janka} H.~T.,  2019, arXiv
  e-prints, \href {https://ui.adsabs.harvard.edu/abs/2019arXiv191001641E} {p.
  arXiv:1910.01641}

\bibitem[\protect\citeauthoryear{Filippenko}{Filippenko}{1997}]{doi:10.1146/annurev.astro.35.1.309}
Filippenko A.~V.,  1997, \mn@doi [Annual Review of Astronomy and Astrophysics]
  {10.1146/annurev.astro.35.1.309}, 35, 309

\bibitem[\protect\citeauthoryear{{Foreman-Mackey}}{{Foreman-Mackey}}{2016}]{2016JOSS....1...24F}
{Foreman-Mackey} D.,  2016, \mn@doi [The Journal of Open Source Software]
  {10.21105/joss.00024}, \href
  {https://ui.adsabs.harvard.edu/abs/2016JOSS....1...24F} {1, 24}

\bibitem[\protect\citeauthoryear{{Foreman-Mackey}, {Hogg}, {Lang}  \&
  {Goodman}}{{Foreman-Mackey} et~al.}{2013}]{2013PASP..125..306F}
{Foreman-Mackey} D.,  {Hogg} D.~W.,  {Lang} D.,   {Goodman} J.,  2013, \mn@doi
  [\pasp] {10.1086/670067}, \href
  {https://ui.adsabs.harvard.edu/abs/2013PASP..125..306F} {125, 306}

\bibitem[\protect\citeauthoryear{{Fruchter} et~al.,}{{Fruchter}
  et~al.}{2006}]{2006Natur.441..463F}
{Fruchter} A.~S.,  et~al., 2006, \mn@doi [\nat] {10.1038/nature04787}, \href
  {https://ui.adsabs.harvard.edu/abs/2006Natur.441..463F} {441, 463}

\bibitem[\protect\citeauthoryear{{Fryer} \& {Heger}}{{Fryer} \&
  {Heger}}{2005}]{2005ApJ...623..302F}
{Fryer} C.~L.,  {Heger} A.,  2005, \mn@doi [\apj] {10.1086/428379}, \href
  {https://ui.adsabs.harvard.edu/abs/2005ApJ...623..302F} {623, 302}

\bibitem[\protect\citeauthoryear{{Fryer}, {Woosley}, {Herant}  \&
  {Davies}}{{Fryer} et~al.}{1999}]{1999ApJ...520..650F}
{Fryer} C.~L.,  {Woosley} S.~E.,  {Herant} M.,   {Davies} M.~B.,  1999, \mn@doi
  [\apj] {10.1086/307467}, \href
  {https://ui.adsabs.harvard.edu/abs/1999ApJ...520..650F} {520, 650}

\bibitem[\protect\citeauthoryear{{Fryer} et~al.,}{{Fryer}
  et~al.}{2007}]{2007PASP..119.1211F}
{Fryer} C.~L.,  et~al., 2007, \mn@doi [\pasp] {10.1086/523768}, \href
  {https://ui.adsabs.harvard.edu/abs/2007PASP..119.1211F} {119, 1211}

\bibitem[\protect\citeauthoryear{{Fryer}, {Lloyd-Ronning}, {Wollaeger},
  {Wiggins}, {Miller}, {Dolence}, {Ryan}  \& {Fields}}{{Fryer}
  et~al.}{2019}]{2019EPJA...55..132F}
{Fryer} C.~L.,  {Lloyd-Ronning} N.,  {Wollaeger} R.,  {Wiggins} B.,  {Miller}
  J.,  {Dolence} J.,  {Ryan} B.,   {Fields} C.~E.,  2019, \mn@doi [European
  Physical Journal A] {10.1140/epja/i2019-12818-y}, \href
  {https://ui.adsabs.harvard.edu/abs/2019EPJA...55..132F} {55, 132}

\bibitem[\protect\citeauthoryear{{Galama} et~al.,}{{Galama}
  et~al.}{1998}]{1998Natur.395..670G}
{Galama} T.~J.,  et~al., 1998, \mn@doi [\nat] {10.1038/27150}, \href
  {https://ui.adsabs.harvard.edu/abs/1998Natur.395..670G} {395, 670}

\bibitem[\protect\citeauthoryear{{Goldreich} \& {Nicholson}}{{Goldreich} \&
  {Nicholson}}{1989}]{1989ApJ...342.1079G}
{Goldreich} P.,  {Nicholson} P.~D.,  1989, \mn@doi [\apj] {10.1086/167665},
  \href {https://ui.adsabs.harvard.edu/abs/1989ApJ...342.1079G} {342, 1079}

\bibitem[\protect\citeauthoryear{{Goodman} \& {Weare}}{{Goodman} \&
  {Weare}}{2010}]{2010CAMCS...5...65G}
{Goodman} J.,  {Weare} J.,  2010, \mn@doi [Communications in Applied
  Mathematics and Computational Science] {10.2140/camcos.2010.5.65}, \href
  {https://ui.adsabs.harvard.edu/abs/2010CAMCS...5...65G} {5, 65}

\bibitem[\protect\citeauthoryear{{Graham}, {Fruchter}, {Kewley}, {Levesque},
  {Levan}, {Tanvir}, {Reichart}  \& {Nysewander}}{{Graham}
  et~al.}{2009}]{2009AIPC.1133..269G}
{Graham} J.~F.,  {Fruchter} A.~S.,  {Kewley} L.~J.,  {Levesque} E.~M.,  {Levan}
  A.~J.,  {Tanvir} N.~R.,  {Reichart} D.~E.,   {Nysewander} M.,  2009, in
  {Meegan} C.,  {Kouveliotou} C.,   {Gehrels} N.,  eds,  American Institute of
  Physics Conference Series Vol. 1133, American Institute of Physics Conference
  Series. pp 269--272 (\mn@eprint {arXiv} {0903.5544}),
  \mn@doi{10.1063/1.3155900}

\bibitem[\protect\citeauthoryear{{Graham}, {Schady}  \& {Fruchter}}{{Graham}
  et~al.}{2019}]{2019arXiv190402673G}
{Graham} J.~F.,  {Schady} P.,   {Fruchter} A.~S.,  2019, arXiv e-prints, \href
  {https://ui.adsabs.harvard.edu/abs/2019arXiv190402673G} {p. arXiv:1904.02673}

\bibitem[\protect\citeauthoryear{{Greiner} et~al.,}{{Greiner}
  et~al.}{2015}]{2015ApJ...809...76G}
{Greiner} J.,  et~al., 2015, \mn@doi [\apj] {10.1088/0004-637X/809/1/76}, \href
  {https://ui.adsabs.harvard.edu/abs/2015ApJ...809...76G} {809, 76}

\bibitem[\protect\citeauthoryear{{Groh}, {Meynet}, {Ekstr{\"o}m}  \&
  {Georgy}}{{Groh} et~al.}{2014}]{2014A&A...564A..30G}
{Groh} J.~H.,  {Meynet} G.,  {Ekstr{\"o}m} S.,   {Georgy} C.,  2014, \mn@doi
  [\aap] {10.1051/0004-6361/201322573}, \href
  {https://ui.adsabs.harvard.edu/abs/2014A&A...564A..30G} {564, A30}

\bibitem[\protect\citeauthoryear{{Heger} \& {Woosley}}{{Heger} \&
  {Woosley}}{2002}]{2002ApJ...567..532H}
{Heger} A.,  {Woosley} S.~E.,  2002, \mn@doi [\apj] {10.1086/338487}, \href
  {https://ui.adsabs.harvard.edu/abs/2002ApJ...567..532H} {567, 532}

\bibitem[\protect\citeauthoryear{{Heger}, {Langer}  \& {Woosley}}{{Heger}
  et~al.}{2000}]{2000ApJ...528..368H}
{Heger} A.,  {Langer} N.,   {Woosley} S.~E.,  2000, \mn@doi [\apj]
  {10.1086/308158}, \href
  {https://ui.adsabs.harvard.edu/abs/2000ApJ...528..368H} {528, 368}

\bibitem[\protect\citeauthoryear{{Hirschi}, {Meynet}  \& {Maeder}}{{Hirschi}
  et~al.}{2005}]{2005A&A...443..581H}
{Hirschi} R.,  {Meynet} G.,   {Maeder} A.,  2005, \mn@doi [\aap]
  {10.1051/0004-6361:20053329}, \href
  {https://ui.adsabs.harvard.edu/abs/2005A&A...443..581H} {443, 581}

\bibitem[\protect\citeauthoryear{{Hjorth}}{{Hjorth}}{2013}]{2013RSPTA.37120275H}
{Hjorth} J.,  2013, \mn@doi [Philosophical Transactions of the Royal Society of
  London Series A] {10.1098/rsta.2012.0275}, \href
  {https://ui.adsabs.harvard.edu/abs/2013RSPTA.37120275H} {371, 20120275}

\bibitem[\protect\citeauthoryear{{Hjorth} et~al.,}{{Hjorth}
  et~al.}{2003}]{2003Natur.423..847H}
{Hjorth} J.,  et~al., 2003, \mn@doi [\nat] {10.1038/nature01750}, \href
  {https://ui.adsabs.harvard.edu/abs/2003Natur.423..847H} {423, 847}

\bibitem[\protect\citeauthoryear{{Horiuchi}, {Beacom}, {Kochanek}, {Prieto},
  {Stanek}  \& {Thompson}}{{Horiuchi} et~al.}{2011}]{2011ApJ...738..154H}
{Horiuchi} S.,  {Beacom} J.~F.,  {Kochanek} C.~S.,  {Prieto} J.~L.,  {Stanek}
  K.~Z.,   {Thompson} T.~A.,  2011, \mn@doi [\apj]
  {10.1088/0004-637X/738/2/154}, \href
  {http://adsabs.harvard.edu/abs/2011ApJ...738..154H} {738, 154}

\bibitem[\protect\citeauthoryear{{Hurley}, {Pols}  \& {Tout}}{{Hurley}
  et~al.}{2000}]{2000MNRAS.315..543H}
{Hurley} J.~R.,  {Pols} O.~R.,   {Tout} C.~A.,  2000, \mn@doi [\mnras]
  {10.1046/j.1365-8711.2000.03426.x}, \href
  {https://ui.adsabs.harvard.edu/abs/2000MNRAS.315..543H} {315, 543}

\bibitem[\protect\citeauthoryear{{Hurley}, {Tout}  \& {Pols}}{{Hurley}
  et~al.}{2002}]{2002MNRAS.329..897H}
{Hurley} J.~R.,  {Tout} C.~A.,   {Pols} O.~R.,  2002, \mn@doi [\mnras]
  {10.1046/j.1365-8711.2002.05038.x}, \href
  {https://ui.adsabs.harvard.edu/abs/2002MNRAS.329..897H} {329, 897}

\bibitem[\protect\citeauthoryear{{Hut}}{{Hut}}{1981}]{1981A&A....99..126H}
{Hut} P.,  1981, \aap, \href
  {https://ui.adsabs.harvard.edu/abs/1981A&A....99..126H} {99, 126}

\bibitem[\protect\citeauthoryear{{Iwamoto} et~al.,}{{Iwamoto}
  et~al.}{1998}]{1998Natur.395..672I}
{Iwamoto} K.,  et~al., 1998, \mn@doi [\nat] {10.1038/27155}, \href
  {https://ui.adsabs.harvard.edu/abs/1998Natur.395..672I} {395, 672}

\bibitem[\protect\citeauthoryear{{Izzard}, {Ramirez-Ruiz}  \& {Tout}}{{Izzard}
  et~al.}{2004}]{2004MNRAS.348.1215I}
{Izzard} R.~G.,  {Ramirez-Ruiz} E.,   {Tout} C.~A.,  2004, \mn@doi [\mnras]
  {10.1111/j.1365-2966.2004.07436.x}, \href
  {https://ui.adsabs.harvard.edu/abs/2004MNRAS.348.1215I} {348, 1215}

\bibitem[\protect\citeauthoryear{{Izzo} et~al.,}{{Izzo}
  et~al.}{2019}]{2019Natur.565..324I}
{Izzo} L.,  et~al., 2019, \mn@doi [\nat] {10.1038/s41586-018-0826-3}, \href
  {https://ui.adsabs.harvard.edu/abs/2019Natur.565..324I} {565, 324}

\bibitem[\protect\citeauthoryear{{Japelj}, {Vergani}, {Salvaterra}, {Renzo},
  {Zapartas}, {de Mink}, {Kaper}  \& {Zibetti}}{{Japelj}
  et~al.}{2018}]{2018A&A...617A.105J}
{Japelj} J.,  {Vergani} S.~D.,  {Salvaterra} R.,  {Renzo} M.,  {Zapartas} E.,
  {de Mink} S.~E.,  {Kaper} L.,   {Zibetti} S.,  2018, \mn@doi [\aap]
  {10.1051/0004-6361/201833209}, \href
  {https://ui.adsabs.harvard.edu/abs/2018A&A...617A.105J} {617, A105}

\bibitem[\protect\citeauthoryear{{Kilpatrick} et~al.,}{{Kilpatrick}
  et~al.}{2018}]{2018MNRAS.480.2072K}
{Kilpatrick} C.~D.,  et~al., 2018, \mn@doi [\mnras] {10.1093/mnras/sty2022},
  \href {http://adsabs.harvard.edu/abs/2018MNRAS.480.2072K} {480, 2072}

\bibitem[\protect\citeauthoryear{{Kobulnicky} \& {Kewley}}{{Kobulnicky} \&
  {Kewley}}{2004}]{2004ApJ...617..240K}
{Kobulnicky} H.~A.,  {Kewley} L.~J.,  2004, \mn@doi [\apj] {10.1086/425299},
  \href {https://ui.adsabs.harvard.edu/abs/2004ApJ...617..240K} {617, 240}

\bibitem[\protect\citeauthoryear{{Kushnir}, {Zaldarriaga}, {Kollmeier}  \&
  {Waldman}}{{Kushnir} et~al.}{2017}]{2017MNRAS.467.2146K}
{Kushnir} D.,  {Zaldarriaga} M.,  {Kollmeier} J.~A.,   {Waldman} R.,  2017,
  \mn@doi [\mnras] {10.1093/mnras/stx255}, \href
  {https://ui.adsabs.harvard.edu/abs/2017MNRAS.467.2146K} {467, 2146}

\bibitem[\protect\citeauthoryear{{Langer} \& {Norman}}{{Langer} \&
  {Norman}}{2006}]{2006ApJ...638L..63L}
{Langer} N.,  {Norman} C.~A.,  2006, \mn@doi [\apj] {10.1086/500363}, \href
  {https://ui.adsabs.harvard.edu/abs/2006ApJ...638L..63L} {638, L63}

\bibitem[\protect\citeauthoryear{{Levan}, {Crowther}, {de Grijs}, {Langer},
  {Xu}  \& {Yoon}}{{Levan} et~al.}{2016}]{2016SSRv..202...33L}
{Levan} A.,  {Crowther} P.,  {de Grijs} R.,  {Langer} N.,  {Xu} D.,   {Yoon}
  S.-C.,  2016, \mn@doi [\ssr] {10.1007/s11214-016-0312-x}, \href
  {https://ui.adsabs.harvard.edu/abs/2016SSRv..202...33L} {202, 33}

\bibitem[\protect\citeauthoryear{{Levesque}, {Kewley}, {Berger}  \&
  {Zahid}}{{Levesque} et~al.}{2010a}]{2010AJ....140.1557L}
{Levesque} E.~M.,  {Kewley} L.~J.,  {Berger} E.,   {Zahid} H.~J.,  2010a,
  \mn@doi [\aj] {10.1088/0004-6256/140/5/1557}, \href
  {https://ui.adsabs.harvard.edu/abs/2010AJ....140.1557L} {140, 1557}

\bibitem[\protect\citeauthoryear{{Levesque}, {Kewley}, {Graham}  \&
  {Fruchter}}{{Levesque} et~al.}{2010b}]{2010ApJ...712L..26L}
{Levesque} E.~M.,  {Kewley} L.~J.,  {Graham} J.~F.,   {Fruchter} A.~S.,  2010b,
  \mn@doi [\apjl] {10.1088/2041-8205/712/1/L26}, \href
  {https://ui.adsabs.harvard.edu/abs/2010ApJ...712L..26L} {712, L26}

\bibitem[\protect\citeauthoryear{{Lyman} et~al.,}{{Lyman}
  et~al.}{2017}]{2017MNRAS.467.1795L}
{Lyman} J.~D.,  et~al., 2017, \mn@doi [\mnras] {10.1093/mnras/stx220}, \href
  {https://ui.adsabs.harvard.edu/abs/2017MNRAS.467.1795L} {467, 1795}

\bibitem[\protect\citeauthoryear{{MacFadyen} \& {Woosley}}{{MacFadyen} \&
  {Woosley}}{1999}]{1999ApJ...524..262M}
{MacFadyen} A.~I.,  {Woosley} S.~E.,  1999, \mn@doi [\apj] {10.1086/307790},
  \href {https://ui.adsabs.harvard.edu/abs/1999ApJ...524..262M} {524, 262}

\bibitem[\protect\citeauthoryear{{MacFadyen}, {Woosley}  \&
  {Heger}}{{MacFadyen} et~al.}{2001}]{2001ApJ...550..410M}
{MacFadyen} A.~I.,  {Woosley} S.~E.,   {Heger} A.,  2001, \mn@doi [\apj]
  {10.1086/319698}, \href
  {https://ui.adsabs.harvard.edu/abs/2001ApJ...550..410M} {550, 410}

\bibitem[\protect\citeauthoryear{{Madau} \& {Dickinson}}{{Madau} \&
  {Dickinson}}{2014}]{2014ARA&A..52..415M}
{Madau} P.,  {Dickinson} M.,  2014, \mn@doi [\araa]
  {10.1146/annurev-astro-081811-125615}, \href
  {https://ui.adsabs.harvard.edu/abs/2014ARA&A..52..415M} {52, 415}

\bibitem[\protect\citeauthoryear{{Maiolino} et~al.,}{{Maiolino}
  et~al.}{2008}]{2008A&A...488..463M}
{Maiolino} R.,  et~al., 2008, \mn@doi [\aap] {10.1051/0004-6361:200809678},
  \href {https://ui.adsabs.harvard.edu/abs/2008A&A...488..463M} {488, 463}

\bibitem[\protect\citeauthoryear{{Mandel} \& {de Mink}}{{Mandel} \& {de
  Mink}}{2016}]{2016MNRAS.458.2634M}
{Mandel} I.,  {de Mink} S.~E.,  2016, \mn@doi [\mnras] {10.1093/mnras/stw379},
  \href {https://ui.adsabs.harvard.edu/abs/2016MNRAS.458.2634M} {458, 2634}

\bibitem[\protect\citeauthoryear{{Mapelli}, {Spera}, {Montanari}, {Limongi},
  {Chieffi}, {Giacobbo}  \& {Bressan}}{{Mapelli}
  et~al.}{2019}]{2019arXiv190901371M}
{Mapelli} M.,  {Spera} M.,  {Montanari} E.,  {Limongi} M.,  {Chieffi} A.,
  {Giacobbo} N.,   {Bressan} A.,  2019, arXiv e-prints, \href
  {https://ui.adsabs.harvard.edu/abs/2019arXiv190901371M} {p. arXiv:1909.01371}

\bibitem[\protect\citeauthoryear{{Marchant}, {Langer}, {Podsiadlowski},
  {Tauris}  \& {Moriya}}{{Marchant} et~al.}{2016}]{2016A&A...588A..50M}
{Marchant} P.,  {Langer} N.,  {Podsiadlowski} P.,  {Tauris} T.~M.,   {Moriya}
  T.~J.,  2016, \mn@doi [\aap] {10.1051/0004-6361/201628133}, \href
  {https://ui.adsabs.harvard.edu/abs/2016A&A...588A..50M} {588, A50}

\bibitem[\protect\citeauthoryear{{Marchant}, {Langer}, {Podsiadlowski},
  {Tauris}, {de Mink}, {Mandel}  \& {Moriya}}{{Marchant}
  et~al.}{2017}]{2017A&A...604A..55M}
{Marchant} P.,  {Langer} N.,  {Podsiadlowski} P.,  {Tauris} T.~M.,  {de Mink}
  S.,  {Mandel} I.,   {Moriya} T.~J.,  2017, \mn@doi [\aap]
  {10.1051/0004-6361/201630188}, \href
  {https://ui.adsabs.harvard.edu/abs/2017A&A...604A..55M} {604, A55}

\bibitem[\protect\citeauthoryear{{Mazzali} et~al.,}{{Mazzali}
  et~al.}{2008}]{2008Sci...321.1185M}
{Mazzali} P.~A.,  et~al., 2008, \mn@doi [Science] {10.1126/science.1158088},
  \href {https://ui.adsabs.harvard.edu/abs/2008Sci...321.1185M} {321, 1185}

\bibitem[\protect\citeauthoryear{{Mazzali}, {McFadyen}, {Woosley}, {Pian}  \&
  {Tanaka}}{{Mazzali} et~al.}{2014}]{2014MNRAS.443...67M}
{Mazzali} P.~A.,  {McFadyen} A.~I.,  {Woosley} S.~E.,  {Pian} E.,   {Tanaka}
  M.,  2014, \mn@doi [\mnras] {10.1093/mnras/stu1124}, \href
  {https://ui.adsabs.harvard.edu/abs/2014MNRAS.443...67M} {443, 67}

\bibitem[\protect\citeauthoryear{{Micha{\l}owski} et~al.,}{{Micha{\l}owski}
  et~al.}{2018}]{2018A&A...616A.169M}
{Micha{\l}owski} M.~J.,  et~al., 2018, \mn@doi [\aap]
  {10.1051/0004-6361/201629942}, \href
  {https://ui.adsabs.harvard.edu/abs/2018A&A...616A.169M} {616, A169}

\bibitem[\protect\citeauthoryear{{Modjaz} et~al.,}{{Modjaz}
  et~al.}{2008}]{2008AJ....135.1136M}
{Modjaz} M.,  et~al., 2008, \mn@doi [\aj] {10.1088/0004-6256/135/4/1136}, \href
  {http://adsabs.harvard.edu/abs/2008AJ....135.1136M} {135, 1136}

\bibitem[\protect\citeauthoryear{{Modjaz}, {Liu}, {Bianco}  \&
  {Graur}}{{Modjaz} et~al.}{2016}]{2016ApJ...832..108M}
{Modjaz} M.,  {Liu} Y.~Q.,  {Bianco} F.~B.,   {Graur} O.,  2016, \mn@doi [\apj]
  {10.3847/0004-637X/832/2/108}, \href
  {http://adsabs.harvard.edu/abs/2016ApJ...832..108M} {832, 108}

\bibitem[\protect\citeauthoryear{{Modjaz} et~al.,}{{Modjaz}
  et~al.}{2019}]{2019arXiv190100872M}
{Modjaz} M.,  et~al., 2019, arXiv e-prints, \href
  {https://ui.adsabs.harvard.edu/abs/2019arXiv190100872M} {p. arXiv:1901.00872}

\bibitem[\protect\citeauthoryear{{Moe} \& {Di Stefano}}{{Moe} \& {Di
  Stefano}}{2017}]{2017ApJS..230...15M}
{Moe} M.,  {Di Stefano} R.,  2017, \mn@doi [\apjs] {10.3847/1538-4365/aa6fb6},
  \href {http://adsabs.harvard.edu/abs/2017ApJS..230...15M} {230, 15}

\bibitem[\protect\citeauthoryear{{Moriya} \& {Eldridge}}{{Moriya} \&
  {Eldridge}}{2016}]{2016MNRAS.461.2155M}
{Moriya} T.~J.,  {Eldridge} J.~J.,  2016, \mn@doi [\mnras]
  {10.1093/mnras/stw1471}, \href
  {https://ui.adsabs.harvard.edu/abs/2016MNRAS.461.2155M} {461, 2155}

\bibitem[\protect\citeauthoryear{{Nakamura}, {Mazzali}, {Nomoto}  \&
  {Iwamoto}}{{Nakamura} et~al.}{2001}]{2001ApJ...550..991N}
{Nakamura} T.,  {Mazzali} P.~A.,  {Nomoto} K.,   {Iwamoto} K.,  2001, \mn@doi
  [\apj] {10.1086/319784}, \href
  {https://ui.adsabs.harvard.edu/abs/2001ApJ...550..991N} {550, 991}

\bibitem[\protect\citeauthoryear{{Neugent} \& {Massey}}{{Neugent} \&
  {Massey}}{2019}]{2019arXiv190806238N}
{Neugent} K.~F.,  {Massey} P.,  2019, arXiv e-prints, \href
  {https://ui.adsabs.harvard.edu/abs/2019arXiv190806238N} {p. arXiv:1908.06238}

\bibitem[\protect\citeauthoryear{{Nomoto}, {Maeda}, {Tominaga}, {Ohkubo},
  {Umeda}, {Deng}  \& {Mazzali}}{{Nomoto} et~al.}{2004a}]{2004PThPS.155..299N}
{Nomoto} K.,  {Maeda} K.,  {Tominaga} N.,  {Ohkubo} T.,  {Umeda} H.,  {Deng}
  J.,   {Mazzali} P.~A.,  2004a, \mn@doi [Progress of Theoretical Physics
  Supplement] {10.1143/PTPS.155.299}, \href
  {https://ui.adsabs.harvard.edu/abs/2004PThPS.155..299N} {155, 299}

\bibitem[\protect\citeauthoryear{{Nomoto}, {Maeda}, {Mazzali}, {Umeda}, {Deng}
  \& {Iwamoto}}{{Nomoto} et~al.}{2004b}]{2004ASSL..302..277N}
{Nomoto} K.,  {Maeda} K.,  {Mazzali} P.~A.,  {Umeda} H.,  {Deng} J.,
  {Iwamoto} K.,  2004b, in {Fryer} C.~L.,  ed.,  Astrophysics and Space Science
  Library Vol. 302, Astrophysics and Space Science Library. pp 277--325
  (\mn@eprint {arXiv} {astro-ph/0308136}),
  \mn@doi{10.1007/978-0-306-48599-2_10}

\bibitem[\protect\citeauthoryear{{Palmerio} et~al.,}{{Palmerio}
  et~al.}{2019}]{2019A&A...623A..26P}
{Palmerio} J.~T.,  et~al., 2019, \mn@doi [\aap] {10.1051/0004-6361/201834179},
  \href {https://ui.adsabs.harvard.edu/abs/2019A&A...623A..26P} {623, A26}

\bibitem[\protect\citeauthoryear{{Perley} et~al.,}{{Perley}
  et~al.}{2016a}]{2016ApJ...817....7P}
{Perley} D.~A.,  et~al., 2016a, \mn@doi [\apj] {10.3847/0004-637X/817/1/7},
  \href {https://ui.adsabs.harvard.edu/abs/2016ApJ...817....7P} {817, 7}

\bibitem[\protect\citeauthoryear{{Perley} et~al.,}{{Perley}
  et~al.}{2016b}]{2016ApJ...817....8P}
{Perley} D.~A.,  et~al., 2016b, \mn@doi [\apj] {10.3847/0004-637X/817/1/8},
  \href {https://ui.adsabs.harvard.edu/abs/2016ApJ...817....8P} {817, 8}

\bibitem[\protect\citeauthoryear{{Pescalli} et~al.,}{{Pescalli}
  et~al.}{2016}]{2016A&A...587A..40P}
{Pescalli} A.,  et~al., 2016, \mn@doi [\aap] {10.1051/0004-6361/201526760},
  \href {https://ui.adsabs.harvard.edu/abs/2016A&A...587A..40P} {587, A40}

\bibitem[\protect\citeauthoryear{{Petrovic}, {Langer}, {Yoon}  \&
  {Heger}}{{Petrovic} et~al.}{2005}]{2005A&A...435..247P}
{Petrovic} J.,  {Langer} N.,  {Yoon} S.~C.,   {Heger} A.,  2005, \mn@doi [\aap]
  {10.1051/0004-6361:20042545}, \href
  {https://ui.adsabs.harvard.edu/abs/2005A&A...435..247P} {435, 247}

\bibitem[\protect\citeauthoryear{{Piran}, {Bromberg}, {Nakar}  \&
  {Sari}}{{Piran} et~al.}{2013}]{2013RSPTA.37120273P}
{Piran} T.,  {Bromberg} O.,  {Nakar} E.,   {Sari} R.,  2013, \mn@doi
  [Philosophical Transactions of the Royal Society of London Series A]
  {10.1098/rsta.2012.0273}, \href
  {https://ui.adsabs.harvard.edu/abs/2013RSPTA.37120273P} {371, 20120273}

\bibitem[\protect\citeauthoryear{{Piran}, {Nakar}, {Mazzali}  \&
  {Pian}}{{Piran} et~al.}{2019}]{2019ApJ...871L..25P}
{Piran} T.,  {Nakar} E.,  {Mazzali} P.,   {Pian} E.,  2019, \mn@doi [\apjl]
  {10.3847/2041-8213/aaffce}, \href
  {https://ui.adsabs.harvard.edu/abs/2019ApJ...871L..25P} {871, L25}

\bibitem[\protect\citeauthoryear{{Podsiadlowski}, {Mazzali}, {Nomoto},
  {Lazzati}  \& {Cappellaro}}{{Podsiadlowski}
  et~al.}{2004}]{2004ApJ...607L..17P}
{Podsiadlowski} P.,  {Mazzali} P.~A.,  {Nomoto} K.,  {Lazzati} D.,
  {Cappellaro} E.,  2004, \mn@doi [\apjl] {10.1086/421347}, \href
  {https://ui.adsabs.harvard.edu/abs/2004ApJ...607L..17P} {607, L17}

\bibitem[\protect\citeauthoryear{{Racusin} et~al.,}{{Racusin}
  et~al.}{2009}]{2009ApJ...698...43R}
{Racusin} J.~L.,  et~al., 2009, \mn@doi [\apj] {10.1088/0004-637X/698/1/43},
  \href {https://ui.adsabs.harvard.edu/abs/2009ApJ...698...43R} {698, 43}

\bibitem[\protect\citeauthoryear{{Ram{\'\i}rez-Agudelo}
  et~al.,}{{Ram{\'\i}rez-Agudelo} et~al.}{2015}]{2015A&A...580A..92R}
{Ram{\'\i}rez-Agudelo} O.~H.,  et~al., 2015, \mn@doi [\aap]
  {10.1051/0004-6361/201425424}, \href
  {https://ui.adsabs.harvard.edu/abs/2015A&A...580A..92R} {580, A92}

\bibitem[\protect\citeauthoryear{{Ram{\'\i}rez-Agudelo}
  et~al.,}{{Ram{\'\i}rez-Agudelo} et~al.}{2017}]{2017A&A...600A..81R}
{Ram{\'\i}rez-Agudelo} O.~H.,  et~al., 2017, \mn@doi [\aap]
  {10.1051/0004-6361/201628914}, \href
  {https://ui.adsabs.harvard.edu/abs/2017A&A...600A..81R} {600, A81}

\bibitem[\protect\citeauthoryear{{Rueda}, {Ruffini}  \& {Wang}}{{Rueda}
  et~al.}{2019}]{2019arXiv190506050R}
{Rueda} J.~A.,  {Ruffini} R.,   {Wang} Y.,  2019, arXiv e-prints, \href
  {https://ui.adsabs.harvard.edu/abs/2019arXiv190506050R} {p. arXiv:1905.06050}

\bibitem[\protect\citeauthoryear{{Ryan}, {van Eerten}, {MacFadyen}  \&
  {Zhang}}{{Ryan} et~al.}{2015}]{2015ApJ...799....3R}
{Ryan} G.,  {van Eerten} H.,  {MacFadyen} A.,   {Zhang} B.-B.,  2015, \mn@doi
  [\apj] {10.1088/0004-637X/799/1/3}, \href
  {https://ui.adsabs.harvard.edu/abs/2015ApJ...799....3R} {799, 3}

\bibitem[\protect\citeauthoryear{{Savaglio}, {Glazebrook}  \& {Le
  Borgne}}{{Savaglio} et~al.}{2009}]{2009ApJ...691..182S}
{Savaglio} S.,  {Glazebrook} K.,   {Le Borgne} D.,  2009, \mn@doi [\apj]
  {10.1088/0004-637X/691/1/182}, \href
  {https://ui.adsabs.harvard.edu/abs/2009ApJ...691..182S} {691, 182}

\bibitem[\protect\citeauthoryear{{Schady} et~al.,}{{Schady}
  et~al.}{2015}]{2015A&A...579A.126S}
{Schady} P.,  et~al., 2015, \mn@doi [\aap] {10.1051/0004-6361/201526060}, \href
  {https://ui.adsabs.harvard.edu/abs/2015A&A...579A.126S} {579, A126}

\bibitem[\protect\citeauthoryear{{Shivvers} et~al.,}{{Shivvers}
  et~al.}{2017}]{2017PASP..129e4201S}
{Shivvers} I.,  et~al., 2017, \mn@doi [\pasp] {10.1088/1538-3873/aa54a6}, \href
  {http://adsabs.harvard.edu/abs/2017PASP..129e4201S} {129, 054201}

\bibitem[\protect\citeauthoryear{{Smartt}}{{Smartt}}{2009}]{2009ARA&A..47...63S}
{Smartt} S.~J.,  2009, \mn@doi [\araa] {10.1146/annurev-astro-082708-101737},
  \href {http://adsabs.harvard.edu/abs/2009ARA%26A..47...63S} {47, 63}

\bibitem[\protect\citeauthoryear{{Smartt}}{{Smartt}}{2015}]{2015PASA...32...16S}
{Smartt} S.~J.,  2015, \mn@doi [\pasa] {10.1017/pasa.2015.17}, \href
  {http://adsabs.harvard.edu/abs/2015PASA...32...16S} {32, e016}

\bibitem[\protect\citeauthoryear{{Sobacchi}, {Granot}, {Bromberg}  \&
  {Sormani}}{{Sobacchi} et~al.}{2017}]{2017MNRAS.472..616S}
{Sobacchi} E.,  {Granot} J.,  {Bromberg} O.,   {Sormani} M.~C.,  2017, \mn@doi
  [\mnras] {10.1093/mnras/stx2083}, \href
  {https://ui.adsabs.harvard.edu/abs/2017MNRAS.472..616S} {472, 616}

\bibitem[\protect\citeauthoryear{{Song}, {Meynet}, {Maeder}, {Ekstr{\"o}m}  \&
  {Eggenberger}}{{Song} et~al.}{2016}]{2016A&A...585A.120S}
{Song} H.~F.,  {Meynet} G.,  {Maeder} A.,  {Ekstr{\"o}m} S.,   {Eggenberger}
  P.,  2016, \mn@doi [\aap] {10.1051/0004-6361/201526074}, \href
  {https://ui.adsabs.harvard.edu/abs/2016A&A...585A.120S} {585, A120}

\bibitem[\protect\citeauthoryear{{Stanek} et~al.,}{{Stanek}
  et~al.}{2003}]{2003ApJ...591L..17S}
{Stanek} K.~Z.,  et~al., 2003, \mn@doi [The Astrophysical Journal]
  {10.1086/376976}, \href
  {https://ui.adsabs.harvard.edu/abs/2003ApJ...591L..17S} {591, L17}

\bibitem[\protect\citeauthoryear{{Stanway} \& {Eldridge}}{{Stanway} \&
  {Eldridge}}{2018}]{2018MNRAS.479...75S}
{Stanway} E.~R.,  {Eldridge} J.~J.,  2018, \mn@doi [\mnras]
  {10.1093/mnras/sty1353}, \href
  {https://ui.adsabs.harvard.edu/abs/2018MNRAS.479...75S} {479, 75}

\bibitem[\protect\citeauthoryear{{Stanway} \& {Eldridge}}{{Stanway} \&
  {Eldridge}}{2019}]{2019A&A...621A.105S}
{Stanway} E.~R.,  {Eldridge} J.~J.,  2019, \mn@doi [\aap]
  {10.1051/0004-6361/201834359}, \href
  {https://ui.adsabs.harvard.edu/abs/2019A&A...621A.105S} {621, A105}

\bibitem[\protect\citeauthoryear{{Stevance}, {Ignace}, {Crowther}, {Maund},
  {Davies}  \& {Rate}}{{Stevance} et~al.}{2018}]{2018MNRAS.479.4535S}
{Stevance} H.~F.,  {Ignace} R.,  {Crowther} P.~A.,  {Maund} J.~R.,  {Davies}
  B.,   {Rate} G.,  2018, \mn@doi [\mnras] {10.1093/mnras/sty1827}, \href
  {https://ui.adsabs.harvard.edu/abs/2018MNRAS.479.4535S} {479, 4535}

\bibitem[\protect\citeauthoryear{{Sukhbold} \& {Adams}}{{Sukhbold} \&
  {Adams}}{2019}]{2019arXiv190500474S}
{Sukhbold} T.,  {Adams} S.,  2019, arXiv e-prints, \href
  {https://ui.adsabs.harvard.edu/abs/2019arXiv190500474S} {p. arXiv:1905.00474}

\bibitem[\protect\citeauthoryear{{Sukhbold}, {Ertl}, {Woosley}, {Brown}  \&
  {Janka}}{{Sukhbold} et~al.}{2016}]{2016ApJ...821...38S}
{Sukhbold} T.,  {Ertl} T.,  {Woosley} S.~E.,  {Brown} J.~M.,   {Janka} H.-T.,
  2016, \mn@doi [\apj] {10.3847/0004-637X/821/1/38}, \href
  {http://adsabs.harvard.edu/abs/2016ApJ...821...38S} {821, 38}

\bibitem[\protect\citeauthoryear{{Trenti}, {Perna}  \& {Jimenez}}{{Trenti}
  et~al.}{2015}]{2015ApJ...802..103T}
{Trenti} M.,  {Perna} R.,   {Jimenez} R.,  2015, \mn@doi [\apj]
  {10.1088/0004-637X/802/2/103}, \href
  {https://ui.adsabs.harvard.edu/abs/2015ApJ...802..103T} {802, 103}

\bibitem[\protect\citeauthoryear{{Van Dyk} et~al.,}{{Van Dyk}
  et~al.}{2018}]{2018ApJ...860...90V}
{Van Dyk} S.~D.,  et~al., 2018, \mn@doi [\apj] {10.3847/1538-4357/aac32c},
  \href {https://ui.adsabs.harvard.edu/abs/2018ApJ...860...90V} {860, 90}

\bibitem[\protect\citeauthoryear{{VanderPlas}}{{VanderPlas}}{2014}]{2014arXiv1411.5018V}
{VanderPlas} J.,  2014, arXiv e-prints, \href
  {https://ui.adsabs.harvard.edu/abs/2014arXiv1411.5018V} {p. arXiv:1411.5018}

\bibitem[\protect\citeauthoryear{{Vink} \& {Harries}}{{Vink} \&
  {Harries}}{2017}]{2017A&A...603A.120V}
{Vink} J.~S.,  {Harries} T.~J.,  2017, \mn@doi [\aap]
  {10.1051/0004-6361/201730503}, \href
  {https://ui.adsabs.harvard.edu/abs/2017A&A...603A.120V} {603, A120}

\bibitem[\protect\citeauthoryear{{Vink} \& {de Koter}}{{Vink} \& {de
  Koter}}{2005}]{2005A&A...442..587V}
{Vink} J.~S.,  {de Koter} A.,  2005, \mn@doi [\aap]
  {10.1051/0004-6361:20052862}, \href
  {https://ui.adsabs.harvard.edu/abs/2005A&A...442..587V} {442, 587}

\bibitem[\protect\citeauthoryear{{Vink}, {de Koter}  \& {Lamers}}{{Vink}
  et~al.}{2001}]{2001A&A...369..574V}
{Vink} J.~S.,  {de Koter} A.,   {Lamers} H.~J.~G.~L.~M.,  2001, \mn@doi [\aap]
  {10.1051/0004-6361:20010127}, \href
  {https://ui.adsabs.harvard.edu/abs/2001A&A...369..574V} {369, 574}

\bibitem[\protect\citeauthoryear{{Vink}, {Gr{\"a}fener}  \& {Harries}}{{Vink}
  et~al.}{2011}]{2011A&A...536L..10V}
{Vink} J.~S.,  {Gr{\"a}fener} G.,   {Harries} T.~J.,  2011, \mn@doi [\aap]
  {10.1051/0004-6361/201118197}, \href
  {https://ui.adsabs.harvard.edu/abs/2011A&A...536L..10V} {536, L10}

\bibitem[\protect\citeauthoryear{{Walborn}, {Lasker}, {Laidler}  \&
  {Chu}}{{Walborn} et~al.}{1987}]{1987ApJ...321L..41W}
{Walborn} N.~R.,  {Lasker} B.~M.,  {Laidler} V.~G.,   {Chu} Y.-H.,  1987,
  \mn@doi [\apj] {10.1086/185002}, \href
  {https://ui.adsabs.harvard.edu/abs/1987ApJ...321L..41W} {321, L41}

\bibitem[\protect\citeauthoryear{{Woosley}}{{Woosley}}{1993}]{1993ApJ...405..273W}
{Woosley} S.~E.,  1993, \mn@doi [\apj] {10.1086/172359}, \href
  {http://adsabs.harvard.edu/abs/1993ApJ...405..273W} {405, 273}

\bibitem[\protect\citeauthoryear{{Woosley}}{{Woosley}}{2019}]{2019arXiv190100215W}
{Woosley} S.~E.,  2019, arXiv e-prints, \href
  {http://adsabs.harvard.edu/abs/2019arXiv190100215W} {}

\bibitem[\protect\citeauthoryear{{Woosley} \& {Bloom}}{{Woosley} \&
  {Bloom}}{2006}]{2006ARA&A..44..507W}
{Woosley} S.~E.,  {Bloom} J.~S.,  2006, \mn@doi [\araa]
  {10.1146/annurev.astro.43.072103.150558}, \href
  {https://ui.adsabs.harvard.edu/abs/2006ARA&A..44..507W} {44, 507}

\bibitem[\protect\citeauthoryear{{Woosley} \& {MacFadyen}}{{Woosley} \&
  {MacFadyen}}{1999}]{1999A&AS..138..499W}
{Woosley} S.~E.,  {MacFadyen} A.~I.,  1999, \mn@doi [\aaps]
  {10.1051/aas:1999325}, \href
  {http://adsabs.harvard.edu/abs/1999A%26AS..138..499W} {138, 499}

\bibitem[\protect\citeauthoryear{{Woosley}, {Eastman}  \& {Schmidt}}{{Woosley}
  et~al.}{1999}]{1999ApJ...516..788W}
{Woosley} S.~E.,  {Eastman} R.~G.,   {Schmidt} B.~P.,  1999, \mn@doi [\apj]
  {10.1086/307131}, \href {http://adsabs.harvard.edu/abs/1999ApJ...516..788W}
  {516, 788}

\bibitem[\protect\citeauthoryear{{Woosley}, {Heger}  \& {Weaver}}{{Woosley}
  et~al.}{2002}]{2002RvMP...74.1015W}
{Woosley} S.~E.,  {Heger} A.,   {Weaver} T.~A.,  2002, \mn@doi [Reviews of
  Modern Physics] {10.1103/RevModPhys.74.1015}, \href
  {https://ui.adsabs.harvard.edu/abs/2002RvMP...74.1015W} {74, 1015}

\bibitem[\protect\citeauthoryear{{Wright}}{{Wright}}{2006}]{2006PASP..118.1711W}
{Wright} E.~L.,  2006, \mn@doi [\pasp] {10.1086/510102}, \href
  {https://ui.adsabs.harvard.edu/abs/2006PASP..118.1711W} {118, 1711}

\bibitem[\protect\citeauthoryear{{Xiao}, {Stanway}  \& {Eldridge}}{{Xiao}
  et~al.}{2018}]{2018MNRAS.477..904X}
{Xiao} L.,  {Stanway} E.~R.,   {Eldridge} J.~J.,  2018, \mn@doi [\mnras]
  {10.1093/mnras/sty646}, \href
  {https://ui.adsabs.harvard.edu/abs/2018MNRAS.477..904X} {477, 904}

\bibitem[\protect\citeauthoryear{{Yoon}, {Langer}  \& {Norman}}{{Yoon}
  et~al.}{2006}]{2006A&A...460..199Y}
{Yoon} S.~C.,  {Langer} N.,   {Norman} C.,  2006, \mn@doi [\aap]
  {10.1051/0004-6361:20065912}, \href
  {https://ui.adsabs.harvard.edu/abs/2006A&A...460..199Y} {460, 199}

\bibitem[\protect\citeauthoryear{{Yoon}, {Dierks}  \& {Langer}}{{Yoon}
  et~al.}{2012}]{2012A&A...542A.113Y}
{Yoon} S.~C.,  {Dierks} A.,   {Langer} N.,  2012, \mn@doi [\aap]
  {10.1051/0004-6361/201117769}, \href
  {https://ui.adsabs.harvard.edu/abs/2012A&A...542A.113Y} {542, A113}

\bibitem[\protect\citeauthoryear{{Zahn}}{{Zahn}}{1975}]{1975A&A....41..329Z}
{Zahn} J.~P.,  1975, \aap, \href
  {https://ui.adsabs.harvard.edu/abs/1975A&A....41..329Z} {41, 329}

\bibitem[\protect\citeauthoryear{{Zahn}}{{Zahn}}{1977}]{1977A&A....57..383Z}
{Zahn} J.~P.,  1977, \aap, \href
  {https://ui.adsabs.harvard.edu/abs/1977A&A....57..383Z} {500, 121}

\bibitem[\protect\citeauthoryear{{Zapartas} et~al.,}{{Zapartas}
  et~al.}{2017a}]{2017A&A...601A..29Z}
{Zapartas} E.,  et~al., 2017a, \mn@doi [\aap] {10.1051/0004-6361/201629685},
  \href {https://ui.adsabs.harvard.edu/abs/2017A&A...601A..29Z} {601, A29}

\bibitem[\protect\citeauthoryear{{Zapartas} et~al.,}{{Zapartas}
  et~al.}{2017b}]{2017ApJ...842..125Z}
{Zapartas} E.,  et~al., 2017b, \mn@doi [\apj] {10.3847/1538-4357/aa7467}, \href
  {https://ui.adsabs.harvard.edu/abs/2017ApJ...842..125Z} {842, 125}

\bibitem[\protect\citeauthoryear{{de Mink} \& {Mandel}}{{de Mink} \&
  {Mandel}}{2016}]{2016MNRAS.460.3545D}
{de Mink} S.~E.,  {Mandel} I.,  2016, \mn@doi [\mnras] {10.1093/mnras/stw1219},
  \href {https://ui.adsabs.harvard.edu/abs/2016MNRAS.460.3545D} {460, 3545}

\bibitem[\protect\citeauthoryear{{de Mink}, {Cantiello}, {Langer}, {Pols},
  {Brott}  \& {Yoon}}{{de Mink} et~al.}{2009}]{2009A&A...497..243D}
{de Mink} S.~E.,  {Cantiello} M.,  {Langer} N.,  {Pols} O.~R.,  {Brott} I.,
  {Yoon} S.~C.,  2009, \mn@doi [\aap] {10.1051/0004-6361/200811439}, \href
  {https://ui.adsabs.harvard.edu/abs/2009A&A...497..243D} {497, 243}

\bibitem[\protect\citeauthoryear{{de Mink}, {Langer}, {Izzard}, {Sana}  \& {de
  Koter}}{{de Mink} et~al.}{2013}]{2013ApJ...764..166D}
{de Mink} S.~E.,  {Langer} N.,  {Izzard} R.~G.,  {Sana} H.,   {de Koter} A.,
  2013, \mn@doi [\apj] {10.1088/0004-637X/764/2/166}, \href
  {https://ui.adsabs.harvard.edu/abs/2013ApJ...764..166D} {764, 166}

\bibitem[\protect\citeauthoryear{{de Mink}, {Sana}, {Langer}, {Izzard}  \&
  {Schneider}}{{de Mink} et~al.}{2014}]{2014ApJ...782....7D}
{de Mink} S.~E.,  {Sana} H.,  {Langer} N.,  {Izzard} R.~G.,   {Schneider}
  F.~R.~N.,  2014, \mn@doi [\apj] {10.1088/0004-637X/782/1/7}, \href
  {https://ui.adsabs.harvard.edu/abs/2014ApJ...782....7D} {782, 7}

\makeatother
\end{thebibliography}

\bsp	


\appendix\label{sec:appendix}

\section{Supplementary information}

\begin{figure*}
\begin{minipage}[c]{\textwidth}
\includegraphics[width=0.8\textwidth]{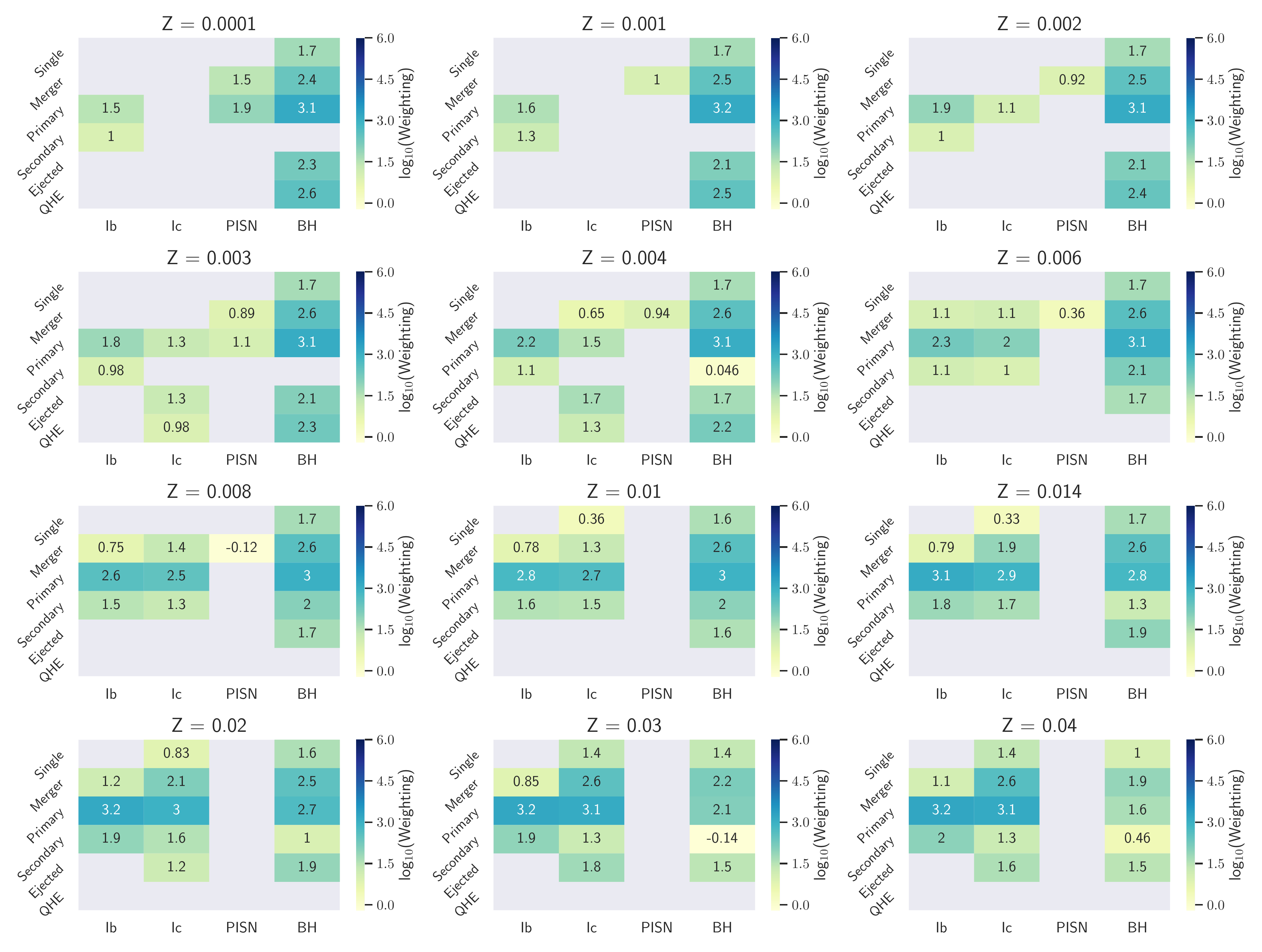}
\includegraphics[width=0.8\textwidth]{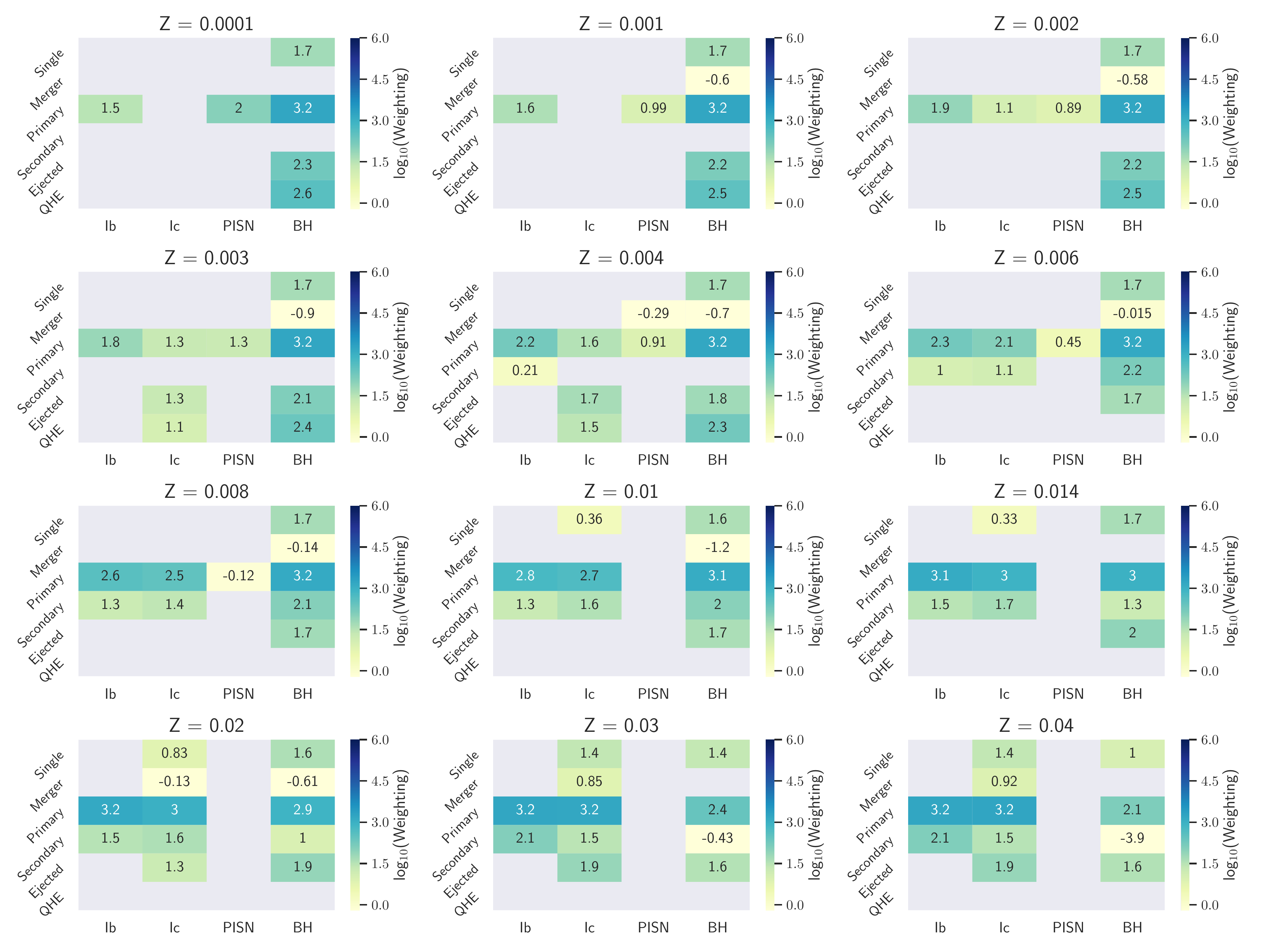}
\centering
\caption{Upper panel: The number of high-mass progenitor core-collapse events arising from each BPASS model type. Higher BPASS weightings indicate that the model occurs more frequently. This version shows the weightings before tides were added. Lower panel: this version shows the weightings including the effects of tidal evolution.}
\label{fig:A1}
\end{minipage}
\end{figure*}

\paragraph*{Changes to the relative frequency of high-mass core-collapse events due to tides\\}
In Figure \ref{fig:A1} we show the change in model frequency of core-collapse events as a function of event type and progenitor evolution pathway, and at each metallicity. Numbers and colour coding correspond to the logarithmic change in number of such events for every $10^6$\,M$_\odot$ of star formation. Initial distributions in mass, mass ratio, period and binary fraction are drawn from BPASS v2.2.1.

\begin{figure*}
\centering
\begin{minipage}[c]{\textwidth}
\includegraphics[width=0.95\textwidth]{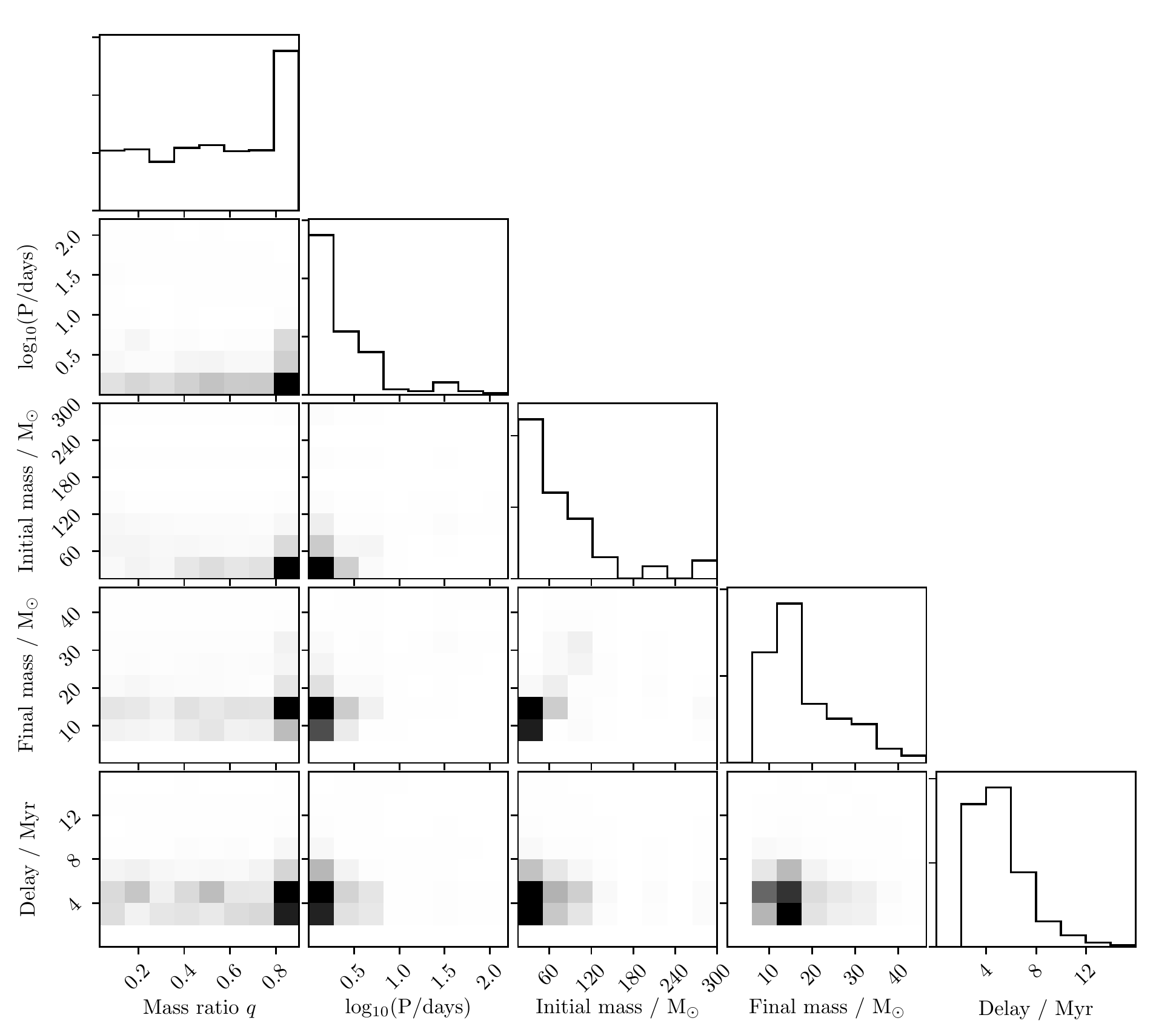}
\caption{A corner plot showing the relative occurrence of the predicted non-QHE progenitors over the metallicity range $Z=0.008$ to $0.020$, for various system parameters. Distribution statistics are available in Table \ref{tab:prog_table}.}
\label{fig:B1}
\end{minipage}
\end{figure*}

\paragraph*{Predicted properties of tidal pathway GRB progenitors\\}
Figure \ref{fig:B1} presents the detailed distribution of tidal GRB progenitors in our formalism and the correlations in their initial and final stellar model properties.

\begin{figure*}
\begin{minipage}[c]{\textwidth}
\centering
\includegraphics[width=0.9\textwidth]{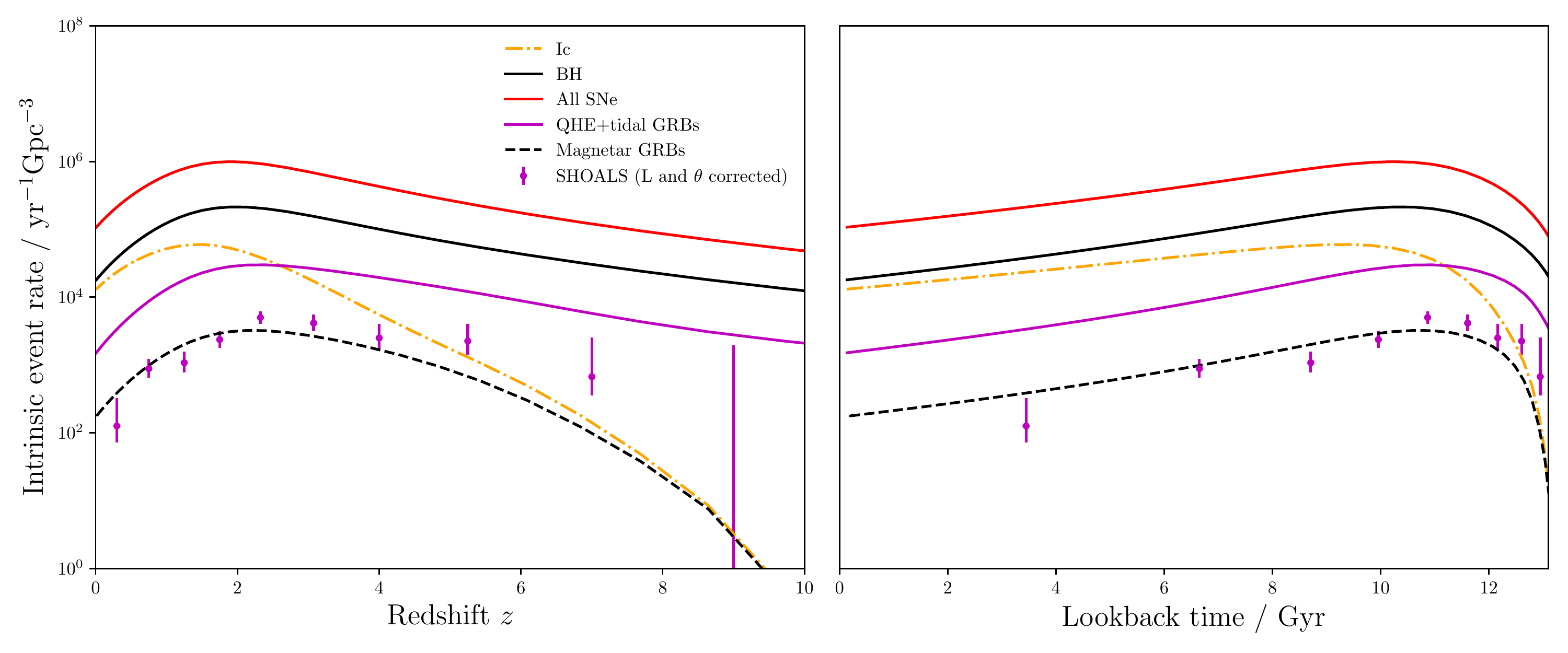}
\caption{The intrinsic GRB rate, if only magnetar central engines are allowed, is shown by the black dashed line (to which the SHOALS data have been fitted). Although magnetars can be used to recreate the GRB rate, the inferred minimum GRB energy that results is ${\sim}$100 greater than the faintest bursts observed.}
\label{fig:C1}
\end{minipage}
\end{figure*}

\begin{figure*}
\begin{minipage}[c]{\textwidth}
\centering
\includegraphics[width=0.9\textwidth]{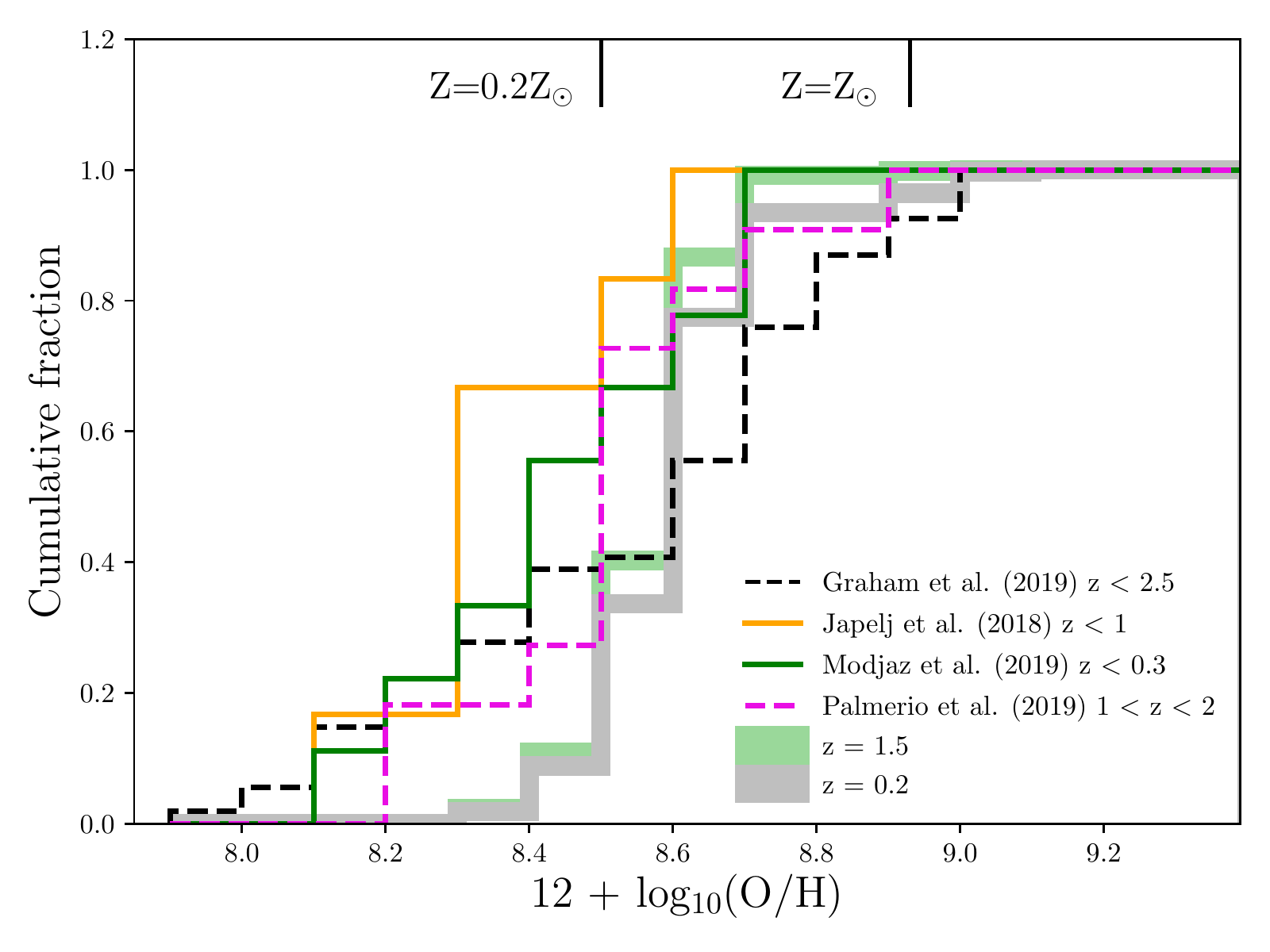}
\caption{The synthesised host metallicity distribution, if GRBs are limited to magnetar central engines only. None of the comparison data sets are consistent (at the 2${\sigma}$ level, according to Anderson-Darling tests).}
\label{fig:C2}
\end{minipage}
\end{figure*}

\paragraph*{The effect of assuming magnetar central engines\\}
Figures \ref{fig:C1} and \ref{fig:C2} are versions of Figures 10 and 11 respectively, where GRBs have been restricted to occurring only in supernovae which produce neutron star remnants.

\clearpage

\label{lastpage}
\end{document}